\documentclass[acmsmall]{acmart}

\AtBeginDocument{%
  \providecommand\BibTeX{{%
    \normalfont B\kern-0.5em{\scshape i\kern-0.25em b}\kern-0.8em\TeX}}}

\copyrightyear{2022}
\acmYear{2022}

\acmConference[]{ACM Trans. Comput.-Hum. Interact.}{}{}
\acmBooktitle{ACM Trans. Comput.-Hum. Interact.}

\usepackage{color}
\usepackage{textcomp}
\usepackage{multirow}
\usepackage{subcaption}
\usepackage{bm}
\usepackage{siunitx}
\usepackage{wrapfig}
\usepackage{soul}
\usepackage{colortbl}

\newcommand{\new}[1]{\textcolor{black}{#1}}
\newcommand\sbullet[1][.5]{\mathbin{\vcenter{\hbox{\scalebox{#1}{$\bullet$}}}}}
	
\definecolor{lgrey}{gray}{0.9}
\definecolor{green}{RGB}{71, 148, 69}

\newcommand*\cirnum[1]{\raisebox{.5pt}{\textcircled{\raisebox{-.9pt} {#1}}}}

\newcommand{\p}[2]{$\text{#1}_{\textit{#2}}$}

\newcolumntype{L}{>{\raggedright\arraybackslash}X}
\newcommand{\xhdr}[1]{\vspace{1mm}\noindent{{\bf #1.}}} 
\renewcommand{\vec}[1]{\mathbf{#1}}

\newcommand\fnum[1]{#1}

\setcopyright{rightsretained}
\acmJournal{TOCHI}
\acmYear{2022} \acmVolume{1} \acmNumber{1} \acmArticle{1} \acmMonth{1} \acmPrice{}\acmDOI{10.1145/3530013}

\begin{document}

\title{Augmenting Scientific Creativity with an Analogical Search Engine}

\author{Hyeonsu B. Kang}
\orcid{0000-0002-1990-2050}
\affiliation{%
  \institution{Carnegie Mellon University}
  \streetaddress{5000 Forbes Ave}
  \city{Pittsburgh}
  \state{PA}
  \country{USA}
}
\email{hyeonsuk@cs.cmu.edu}

\author{Xin Qian}
\email{xinq@umd.edu}
\affiliation{%
  \institution{University of Maryland, College Park}
  \city{College Park}
  \state{MD}
  \country{USA}
  \postcode{20742}
}

\author{Tom Hope}
\email{tomh@allenai.org}
\affiliation{%
  \institution{Allen Institute for AI and The University of Washington}
  \city{Seattle}
  \state{WA}
  \country{USA}
}

\author{Dafna Shahaf}
\email{dshahaf@cs.huji.ac.il}
\affiliation{%
  \institution{Hebrew University of Jerusalem}
  \country{Israel}
}

\author{Joel Chan}
\email{joelchan@umd.edu}
\affiliation{%
  \institution{University of Maryland, College Park}
  \city{College Park}
  \state{MD}
  \country{USA}
  \postcode{20742}
}

\author{Aniket Kittur}
\orcid{0000-0003-4192-9302}
\affiliation{%
  \institution{Carnegie Mellon University}
  \streetaddress{5000 Forbes Ave}
  \city{Pittsburgh} 
  \state{PA} 
  \postcode{15213}
  \country{USA}
}
\email{nkittur@cs.cmu.edu}

\begin{abstract}
Analogies have been central to creative problem-solving throughout the history of science and technology. As the number of scientific papers continues to increase exponentially, there is a growing opportunity for finding diverse solutions to existing problems. However, realizing this potential requires the development of a means for searching through a large corpus that goes beyond surface matches and simple keywords. Here we contribute the first end-to-end system for analogical search on scientific papers and evaluate its effectiveness with scientists' own problems. Using a human-in-the-loop AI system as a probe we find that our system facilitates creative ideation, and that ideation success is mediated by \new{an intermediate level of matching on the problem abstraction (i.e., high versus low).} We also demonstrate a fully automated AI search engine that achieves a similar accuracy with the human-in-the-loop system. We conclude with design implications for enabling automated analogical inspiration engines to accelerate scientific innovation.
\end{abstract}


\maketitle



\section{Introduction}
Analogical reasoning has been central to creative problem solving throughout the history of science and technology~\cite{gentnerAnalogyCreativityWorks1997,gruberDarwinManPsychological1974,holyoakAnalogicalScientist1996,dunbarHowScientistsThink1997,hesseModelsAnalogiesScience1966,oppenheimerAnalogyScience1956}. Many important scientific discoveries were driven by analogies: the Greek philosopher Chrysippus made a connection between observable water waves and sound waves; an analogy between bacteria and slot machines helped Salvador Luria advance the theory of bacterial mutation; a pioneering chemist Joseph Priestly suggested charges attract or repel each other with an inverse square force by an analogy to gravity. 

Today the potential for finding analogies to accelerate innovation in science and engineering is greater than ever before. As of 2009 fifty million scientific papers had been published, and the number continues to grow at an exceedingly fast rate~\cite{noorden_2014,de_Solla_Price510,bornmann2015growth,jinha2010article}. These papers represent a potential treasure trove for finding inspirations from distant domains and generating creative solutions to challenging problems.

However, searching analogical inspirations in a large corpus of papers remains a longstanding challenge~\cite{ontology_based_expertise_finding,analogical_reminding,ackerman_expertise_recommender,automatic_representation_semantic_structure}. Previous systems for retrieving analogies have largely focused on modeling analogical relations in non-scientific domains and/or in limited scopes (e.g., structure-mapping~\cite{gentner1983structure,forbus1994incremental,forbus2001exploring,forbus2017extending,turney2008latent}, multiconstraint-based~\cite{holyoak1989analogical,eliasmith2001integrating,hummel2003symbolic}, connectionist~\cite{hofstadter1995copycat}, rule-based reasoning~\cite{ashley1991reasoning,carbonell1985derivational,carbonell1983learning,veloso1993derivational} systems), and the prohibitive costs of creating highly structured representations prevented hand-crafted systems (e.g., DANE~\cite{vattam_dane:_2011,hummel2003symbolic}) from having a broad coverage of topics and being deployed for realistic use. Conversely, scalable computational approaches such as keyword or citation based search engines have been limited by a dependence on surface or domain similarity. Such search engines aim to maximize similarity to the query which is useful when trying to know what has been done on the problem in the target domain but less useful when trying to find inspiration outside that domain (for example, for Salvador Luria's queries: ``how do bacteria mutate?'' or ``why are bacterial mutation rates so inconsistent?'', similarity maximizing search engines may have found Luria and Delbr{\"u}ck's earlier work on E.coli~\cite{luria1943mutations} but may have failed to recognize more distant sources of inspiration such as slot machines as relevant).

\begin{wrapfigure}{R}{.5\textwidth}
    \begin{center}
    \includegraphics[width=.5\textwidth]{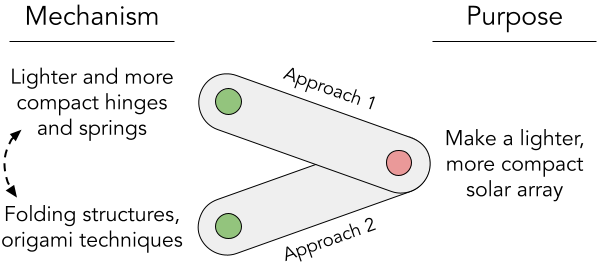}
    \end{center}
    \caption{A diagram of two different yet analogical approaches (dashed arrow) for building lighter and more compact solar arrays, and their representations in purposes and mechanisms.}
    \label{fig:schematic}
\end{wrapfigure}

Recently a novel idea for analogical search was introduced~\cite{hope_kdd17}. In this idea what would otherwise be a complex analogical relation between products is pared down to just two components: purpose (\textit{what problem does it solve?}) and mechanism (\textit{how does it solve that problem?}). Once many such purpose and mechanism pairs are identified, products that solve a similar problem to the query but using diverse mechanisms are searched to help broaden the searcher's perspective on the problem and boost their creativity for coming up with novel mechanism ideas. Anecdotal evidence suggests that this approach may also be applicable to the domain of scientific research. For example, while building lighter and more compact solar panel arrays has been a longstanding challenge for NASA scientists, recognizing how the ancient art form of origami may be applied to create folding structures led to an innovation to use compliant mechanisms to build not just compact but also self-deployable solar arrays~\cite{peraza2014origami,zirbel2013origami,origami_open_innovation} (diagrammatically shown in fig.~\ref{fig:schematic}). \new{The first remaining challenge of analogical search in the scholarly domain is} how we might represent scientific articles as purpose and mechanism pairs at scale and search for those that solve similar purposes using different mechanisms. Recent advances in natural language processing have demonstrated that neural networks that use pre-trained embeddings to encode input text can offer a promising technique to address it. \new{Pre-trained embeddings are} real-valued vectors that represent \new{tokens (\textit{Tokenization} means breaking a piece of text into smaller units; \textit{Tokens} can be words, characters, sub-words, or n-grams.)}, in a high-dimensional space (e.g., typically dimensions of a few dozens to a few thousands) and are shown to capture rich, multi-faceted semantic relations between words~\cite{bengio2003neural,NIPS2014_seq2seq}. Leveraging them, neural networks may be trained to identify purposes and mechanisms from text~\cite{hope_kdd17,hope2021scaling} to enable search-by-analogy (i.e. different mechanisms used for similar purposes). \new{Once candidate papers are retrieved, searchers may use them to come up with novel classes of mechanisms or apply them directly to their own research problems to improve upon the current state. Prior studies in product ideation showed that users of analogical search systems could engage with the results to engender more novel and relevant ideas~\cite{gilon_chi18,chan2018solvent,kittur_pnas19}. Here, we study the remaining open questions as to whether such findings also generalize to the scientific domains of innovation and how they may differ.}

In this paper we present a functioning prototype of an analogical search engine for scientific articles at scale and investigate how such a system can help users explore and adapt distant inspirations. In doing so our system moves beyond manually curated approaches that have limited data (e.g., crowdsourced annotations in~\cite{chan2018solvent} with $\sim${\fnum{2000}} papers) and machine learning approaches that have been limited to simple product descriptions~\cite{hope_kdd17,gilon_chi18,hope2021scaling}. Using the prototypical system, we explore how it enables scientists to interactively search for inspirations for their personalized research problems in a large ($\sim$1.7M) paper corpus. \new{We investigate whether scientists can recognize mapping of analogical relations between the results returned from our analogical search engine and their query problems, and use them to come up with novel ideas. The scale of our corpus} allows us to probe realistic issues including noise, error, and scale as well as how scientists react to a search engine that does not aim to provide only the most similar results to their query.

In order to accomplish these goals we describe how we address several technical issues in the design of an interactive-speed analogical search engine, ranging from \new{developing a machine learning model for extracting purposes and mechanisms in scientific text at a token level granularity}, the pipeline for \new{constructing} a \new{\textit{similarity space}} of purpose embeddings, and enabling these embeddings to be queried at interactive speeds by end users through a search interface. \new{We construct the similarity space by putting semantically related purpose embeddings in close indices from each other such that related purposes can be searched at scale.}

In addition to the technical challenges there are several important questions around the design of analogical search engines that we explore here. A core conceptual difference that distinguishes analogical search engines from other kinds is that the analogs they find for a search query need to maintain some kind of distance from the query, rather than simply maximizing the similarity with it. However, only certain kinds of distance may support generative ideation while others have a detrimental effect. Another question remains as to how much distance is appropriate when it comes to finding analogical inspirations in other domains. While landmark studies of analogical innovation suggest that highly distant domains can provide particularly novel or transformative innovations~\cite{hesse1966models,gickholyoak1980,gick_schema_1983}, recent work suggests the question may be more nuanced and that intermediate levels of distance may be fruitful for finding ideas that are close enough to be relevant but sufficiently distant to be unfamiliar and spur creative adaptation~\cite{fuMeaningFarImpact2013,goncalvesInspirationPeakExploring2013,chanBestDesignIdeas2015}. Using a concrete example from one of our participants who studied ways to facilitate heat transfer in semiconductors, a keyword search engine might find commonly used mechanisms appropriate for direct application (e.g., tweaking the composition of the material) while an analogical search engine might find similar problems in more distant domains which suggest mechanisms that inspire creative adaptation (e.g., nanoscale fins that absorb heat and convert it to mechanical energy). Though more distant conceptual combinations may not {always} lead to immediately feasible or useful ideas, they may result in outsized value after being iterated on~\cite{chanImportanceIterationCreative2015,bergPrimalMarkHow2014,kneeland2020exploring}. 

In the following sections we explore the technical and design challenges for an analogical search engine and how users interact with such a system. First, we describe the development of a human-in-the-loop search engine prototype, in which most elements of the system are functional but human screeners are used to remove obvious noise from the end results in order to maximize our ability to probe how users interact with potentially useful analogical inspirations. Using this prototype we characterize how researchers searching for inspirations for their own problems gain the most benefit from papers that partially match their problem (i.e., match at a high level purpose but mismatch at a lower level specifications of the purpose), and that the benefits are driven not by direct application of the ideas in the paper but by creative adaptation of those ideas to their target domain. Subsequently we describe improvements to the system to enable a fully automated, interactive-speed prototype and case studies with researchers using the system in a realistic way involving reformulation of their queries and self-driven attention to the results. We synthesize the findings of the two studies into design implications for next-generation analogical search engines.

Through extensive in-depth evaluations using an ideation think-aloud protocol~\cite{thinkaloud1,thinkaloud2} with PhD-level researchers working on their own problems, we evaluate the degree to which inspirations spark creative adaptation ideas in a realistic way on scientists' own research problems. Unlike previous work which has often used undergraduate students in the classroom or lab~\cite{vattam_dane:_2011}, and often evaluated systems on pre-determined problems~\cite{fu_chan_design_repo_space}, this study design provides our evaluation with a high degree of external validity and allows us to deeply understand the ways in which encountering our results can engender new ideas. Our final, automated search engine demonstrates how the human-in-the-loop filtering can be removed while achieving a similar accuracy. We conclude with the benefits, design challenges, and opportunities for future analogical search engines from case studies with several researchers. To encourage innovation in this domain, we release our corpus of purpose and mechanism embeddings\footnote{\url{https://github.com/hyeonsuukang/augmenting_tochi22}}.

\section{System Design}
\begin{wraptable}{R}{.6\textwidth}
    \centering
    \begin{tabular}{c r r r r}
    \toprule
    \textbf{Kind (\# of papers)} & \textbf{Avg. length} & \textbf{\# of PP} & \textbf{\# of MN} \\
    \midrule
    Train (\fnum{2021}) & \fnum{196} & \fnum{65261} & \fnum{120586} \\
    Validation (\fnum{50}) & \fnum{170} & \fnum{1510} & \fnum{1988} \\
    \bottomrule
    \end{tabular}
    \caption{Summary statistics of the training and validation datasets: the number of purpose (PP) and mechanism (MN) tokens, the number and avg. token length of paper abstracts.}
    \label{table:training_stats}
    \begin{tabular}{l c c c c c}
    \toprule
    \textbf{Domain} & CS & Eng & BioMed & B \& Eng & \textbf{Total}\\
    \midrule
    \textbf{Count} & 675K & 568K & 336K & 145K & 1.7M\\
    \bottomrule
    \end{tabular}
    \caption{Corpus used in the deployed search engine and its topical distribution: Computer Science (CS), Engineering (Eng), Biomedicine (BioMed), and Business and Engineering (B \& Eng).}
    \label{table:corpus_stats}
\end{wraptable}

\begin{figure}[t]
    \centering
    \includegraphics[width=\textwidth]{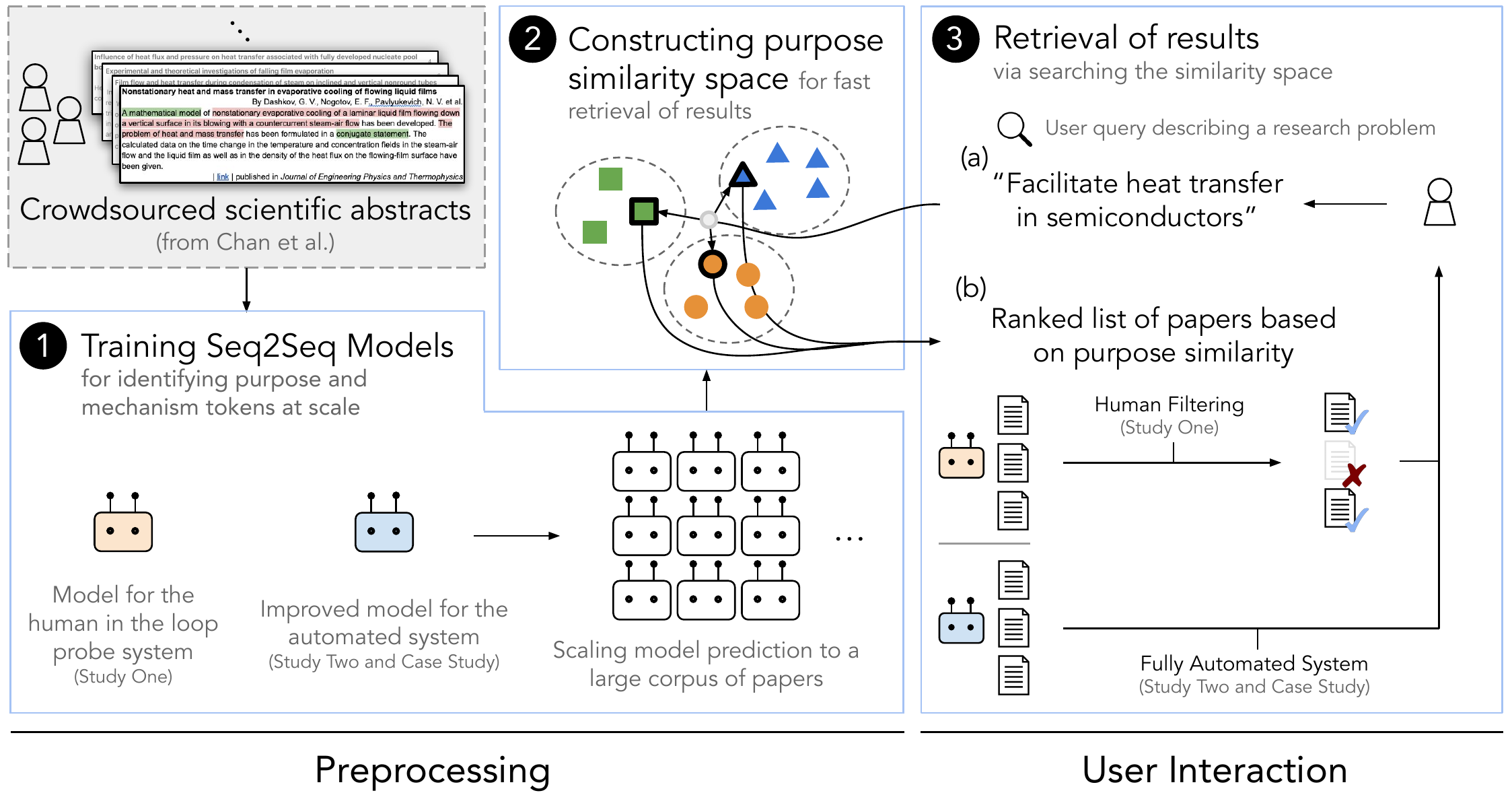}
    \vspace{-1.5em}
    \caption{Components of our system design that address the three core challenges. \cirnum{1} Purpose and mechanism tokens are extracted from paper abstracts at scale. We develop sequence-to-sequence classifiers to classify tokens into purpose, mechanism, or neither, going beyond previous approaches that worked on sentences or relied on crowds. \cirnum{2} We embed the extracted purpose texts using a pre-trained language model (Google's \textsc{Universal Sentence Encoder} (USE)~\cite{universal_sentence_encoder}) and train a tree-based index of vectors to place high semantic similarity vectors in close neighborhoods for efficient lookup. \cirnum{3} When the user query arrives at the system, it is first embedded with USE. This query embedding is then used to lookup the pre-computed tree indices for high similarity purpose embeddings. Paper abstracts for the corresponding purpose embeddings are retrieved from Google Datastore. In the first system, additional human filtering is performed to remove obviously irrelevant results that may have been included due to model errors. Finally, a set of papers with similar purposes to the query but different mechanisms are returned to the users for ideation.}
    \label{fig:system_design}
\end{figure}

The design of our analogical search engine for scientific papers involves three main system requirements. First, a computational pipeline for automatically identifying purposes (\textit{what problems does it solve?}) and mechanisms (\textit{how does it solve those problems}) at scale (e.g., millions of papers), in a token-level granularity from scientific abstracts. Second, an efficient retrieval algorithm for incorporating the identified purpose and mechanism texts into the system to enable search-by-analogy (i.e. paper abstracts that contain similar purposes to a query problem but different mechanisms). Third, end-user interactivity for querying problems of interest \new{(e.g., ``transfer heat in semiconductors,'' ``grow plants using nanoparticle fertilizers'')}. We describe the system design in detail in the following subsections.

\subsection{Stage One. Training Seq2Seq models for identifying purpose and mechanism tokens}
\subsubsection{Overview of Modeling}
In the first stage of the system, purpose and mechanism tokens are identified from paper abstracts (fig.~\ref{fig:system_design}, \cirnum{1}). Research paper abstracts often include descriptions of the most important purpose or \textit{the core problem addressed in a paper} and the proposed mechanism or \textit{the approach taken to address the problem}, making them good candidates for identification and extraction of tokens corresponding to them. For example, for a similar problem of facilitating heat transfer, Paper A may propose an approach that modifies the structure of the material used at the interface between crystalline silicon (semiconductor material) and the substrate, while Paper B may propose a more distant mechanism (due to the mismatch on scale) of fin-based heat sinks commonly used for electronic devices. The goal of this first stage is to automatically identify and extract tokens that correspond to the similar purpose (e.g., `facilitate heat transfer') as well as the mechanisms (e.g., `modifying the structure of the material used at the interface between crystalline silicon' vs. `fin-based heat sinks') from the abstract A and B.

One relevant automated approach for identifying purposes and mechanisms from scientific abstracts is DISA~\cite{disa}, which formulates the task as supervised sentence classification. However, we found that many key sentences in abstracts include both purpose and mechanism, breaking the assumptions of a sentence-level classifier (e.g., ``In this paper, [\textit{a wavelet transforms based method}] for [\textit{filtering noise from images}] is presented.''). To overcome this limitation we follow~\cite{hope2021scaling} and frame purpose and mechanism identification as a sequence-to-sequence (Seq2Seq) learning task~\cite{Seq2SeqICLR,Seq2SeqNIPS} and develop deep neural networks with inductive biases capable of learning token-level patterns in the training dataset. Our dataset consists of crowdsourced annotations from Chan et al. (the dataset is constructed via application of~\cite{chan2018solvent} to a larger corpus of around \fnum{2000} paper abstracts largely in computer science domains) (table.~\ref{table:training_stats}). We train the models to classify input features (tokens or spans of tokens) as either purpose (PP), mechanism (MN) or neither.

We train two deep neural networks (Model 1 and 2), achieving increasing accuracy of classification. The first model is based on a Bi-directional LSTM (BiLSTM) architecture for sequence tagging~\cite{huang2015bidirectional,LSTM_Schmidhuber}, in which the forward (the beginning of the sequence to the end) and the backward passes condition each token position in the text with its left and right context, respectively. A main source of improvement of Model 2 over Model 1 is the ability to more selectively attend to informative \new{tokens} in a sentence rather than treating each \new{token in a sequence as} independent of each other \new{(as a hypothetical example, an extremely effective model based on this approach may assign more weights to the tokens `selectively attend to informative tokens', as they represent the core mechanism described in the previous sentence)} and to leverage the regularities of co-occurrence with surrounding words through {the} self-attention \new{mechanism}~\cite{attention_vaswani}.

\subsubsection{Seq2Seq Model Implementation Details}
We implement the BiLSTM architecture of Model 1 in \textsc{PyTorch}~\cite{pytorch}. We use pre-trained \textsc{GloVe}~\cite{pennington_glove} word embeddings with 300 dimensions, consistent with prior work~\cite{pennington_glove,landauer1997solution,bojanowski2017enriching_subword_info} to represent each token in the sequence as 300-dimensional input vectors for the model. We train the model with a cross entropy loss objective for per-token classification in the three (PP, MN, Neither) token classes.

For Model 2, we adapt the \textsc{SpanRel}~\cite{spanrel} architecture and implement it on \textsc{AllenNLP}~\cite{allennlp}. We implement a self attention mechanism that tunes weights for the core word in each span as well as the boundary words that distinguish the context of use, consistent with~\cite{lee-etal-2017-end}. We use the pre-trained \textsc{ELMo 5.5B}~\cite{elmo} embeddings for token representation following the near state-of-the-art performance reported in~\cite{spanrel} on the scientific Wet Lab Protocol dataset. We train the model using a similar procedure as Model 1. We leave detailed training parameters for Model 1 and 2 to the Appendix.

\subsubsection{\new{Introducing Human-in-the-loop Filtering for Model 1}}
The final classification performance (F1-scores) of Model 1 on the validation set is 0.509 (Purpose), 0.497 (Mechanism), and 0.801 (neither). We found that the limited accuracy contributed to how the system retrieves irrelevant search results. Because reactions to obviously irrelevant results are not useful, we added a human-in-the-loop~\cite{dow2005wizard} filtering stage. The filtering proceeded as follows: members from the research team inputted problem queries received from study participants into the system. Once the model produced matches, they went over from the top of the sorted list and removed only those that are irrelevant to the problem context. They continued filtering until at least 30 papers with reasonable purpose similarity were collected. After Winsorizing at top and bottom 10\%~\cite{winsorizing}, the human filterers reviewed 45 papers per query (SD: 27.6, min: 6, max: 138) for 5 queries (SD: 2.4, min:2, max: 9) to collect 33 (SD: 3.5, min: 30, max: 40) purpose-similar papers (about 12/45 = 26\% error rate). In Study 1 we show that the limited retrieval accuracy of Model 1 is sufficient for use as a probe with this additional human-in-the-loop filtering. In Study 2 and case studies, we demonstrate how this filtering can be removed with Model 2 while achieving a similar accuracy.

\subsubsection{Scaling Model Inference}
In order to have sufficient coverage to return diverse results, we collected an initial corpus of 2.8 million research papers from Springer Nature\footnote{\url{https://dev.springernature.com/}}. After deduplication (based on Digital Object Identifier using BigQuery\footnote{\url{https://cloud.google.com/bigquery}}) and filtering only papers with at least 50 characters in the abstract we were left with 1.7 million papers in four subjects (Table~\ref{table:corpus_stats}). We stored the resulting corpus in Google Cloud storage buckets\footnote{\url{https://cloud.google.com/storage}}. To scale the classification of the Seq2Seq models we used the Apache Beam API\footnote{\url{https://beam.apache.org/}} on Google Cloud Dataflow\footnote{\url{https://cloud.google.com/dataflow/}} to parallelize the operation.

\subsection{Stage Two. Constructing a purpose similarity space}
\subsubsection{Overview} \label{subsubsection:stage2-overview}
In the second stage, the identified purpose texts are incorporated \new{into the system} to enable search-by-analogy of papers that solve similar problems using different mechanisms, at an interactive speed (fig.~\ref{fig:system_design}, \cirnum{2}). Relevant previous approaches include \citet{hope_kdd17} which first clusters similar purposes (through $k$-means with pruning) and subsequently samples within each cluster of similar purposes to maximize the diversity of mechanisms (via a GMM approximation algorithm~\cite{ravi1994heuristic}), or \cite{hope2021scaling} which \new{employs} similarity metrics {to} balance {the} \textit{similarity} to a purpose query and {the} \textit{distance} to a mechanism query (and vice versa). In contrast, from pilot tests in our corpus we discovered that even close purpose matches of scientific papers already had high variance in terms of the mechanisms they propose. We hypothesize that this may be the case due to the enormous span of possible research topics and the relative sparseness of their coverage in our corpus, and/or due to the emphasis on novelty in scientific research that discourages future papers which might contribute relatively small variations to an existing mechanism. We leave exploration of these hypotheses for future work and simplify our sampling of the scientific papers to the one based solely on the similarity of purpose, sufficient for ensuring diversity. 

In order to support fast retrieval (e.g., sub-second response time) of papers with similar purposes at scale (e.g., millions of papers), we pre-train Spotify's \textsc{Annoy}\footnote{\url{https://github.com/spotify/annoy}} indices of nearest neighboring purposes. \new{\textsc{Annoy} trains a neural network to assign an embedding vector corresponding to a purpose an index in the high-dimensional space that brings it close to other indices of purpose vectors that have similar meaning (see \S\ref{subsubsection:stage2_implementation_details} for details of the metric used for the similarity of meaning).}
\textsc{Annoy} uses random projection and tree-building \new{(see~\cite{annoy_readme,random_projection_wiki})} to create read-only, file-based indices. Because it decouples creation of the static index files from lookup, it enables efficient and flexible search by utilizing many parallel processes to quickly load and map indices into memory. 

\subsubsection{\new{Interactive Speed}} \new{Additionally \textsc{Annoy} minimizes its memory footprint in the process. This efficiency, critical for real-time applications such as ours, was further validated during our test of the end-to-end latency on the Web, with the average response taking 2.4s (SD = 0.56s)\footnote{We tested with 20 topically varied search queries that have not previously been entered to the engine to test the latency end-users experience and to exclude the effect of caching from it.}. The level of latency we observed was sufficiently low to enable interactive search by end users (both human-in-the-loop filterers in Study One and researcher participants in case studies).}

\subsubsection{Implementation Details} \label{subsubsection:stage2_implementation_details}
To construct the similarity space, we first encode the purpose texts into high-dimensional embedding vectors which then can be used to compute pairwise semantic similarity. Here, the choice of an encoding algorithm depends on three main constraints. First, the pairwise similarity, when computed, should correlate well with the human-judged semantic similarity between the purposes. Second, similarity calculation between varying lengths of texts should be possible because extracted purposes can differ in length. Third, computationally efficient methods are preferred for scaling. To meet these requirements, we chose \textsc{Universal Sentence Encoder} (\textsc{USE})\footnote{\url{https://tfhub.dev/google/universal-sentence-encoder-large/5}} to encode purposes into fixed 512-dimensional vectors. \textsc{Universal Sentence Encoder} trains a transformer architecture~\cite{attention_vaswani} on a large corpus of both unsupervised (e.g., Wikipedia) and supervised (e.g., Stanford Natural Language Inference dataset~\cite{snli}) data to produce a neural network that can encode text into vectors that meaningfully correlate with human judgment (e.g., evaluated on the semantic textual similarity benchmark~\cite{semeval17}). \textsc{USE} can handle texts of varying lengths (e.g., from short phrases to sentences to paragraphs), and with high efficiency~\cite{universal_sentence_encoder}, thereby making it suitable for our system.

We pre-compute pairwise similarity of the purpose embeddings and store the indices in neighborhoods of high similarity for fast retrieval of similar purposes. As mentioned before, we train the \textsc{Annoy} indices on Google Cloud AI Platform\footnote{\url{https://cloud.google.com/ai-platform}}. We use 1 - the Euclidean distance of normalized vectors (i.e., given two vectors $\vec{u}$ and $\vec{v}$, $\text{distance}(\vec{u}, \vec{v}) = \sqrt{\left(2\left(1 - \text{cos}\left(\vec{u}, \vec{v}\right)\right)\right)}$) as a similarity metric (using a Euclidean distance based metric for nearest neighbor clustering shows good performance, see~\cite{bachrach2014} for a related discussion on the impact of the distance metric on the retrieval performance). We set the hyper-parameter $k$ specifying the number of trees in the forest to 100 (larger $k$'s result in more accurate results but also decreases performance\new{; see~\cite{annoy_readme} for further details}). Empirically, 100 seemed to strike a good balance between the precision-performance trade-off, thus we did not experiment with this parameter further. 

\subsection{Stage Three. Retrieving the results}
In the last stage, the front-end interface interacts with end users and receives problem queries. These queries are then relayed to the back-end for retrieval of papers that solve similar problems using different mechanisms. The retrieved papers are presented on the front-end for users to review (fig.~\ref{fig:system_design}, \cirnum{3}). When a user query is received, the back-end first encodes it using the same encoding algorithm used as the construction method of the purpose similarity space (i.e. \textsc{Universal Sentence Encoder}). Using this query embedding, the back-end searches the pre-trained similarity space for papers with similar purposes. The papers with high purpose similarity are then returned to and displayed on the front-end. We describe the actual interfaces used in the studies in the corresponding design sections (\S\ref{subsubsection:apparatus1}, \S\ref{subsubsection:apparatus2}).

Together the design of our system enabled what is to our knowledge the first functioning prototype of an interactive analogical search engine for scientific papers at scale. In the following sections we report on how such a search engine can help researchers find analogical papers that facilitate creative ideation.

\section{Study 1: Creative Adaptation with a Human-in-the-loop Analogical Search Engine}
In Study 1 we set out to establish the viability of an analogical search engine using a human-in-the-loop probe \new{in the domain of scholarly recommendations}. We investigate whether analogical search returns a distinct and novel set of papers compared to keyword search results, and capture participants' reaction to each result in a randomized order, blind to condition. To deeply understand the process of ideation using analogical papers we ask participants to come up with new ideas for their own research projects after reviewing each paper. Using this data we code ideation outcomes in depth to explore the various ways in which analogical distance can shape ideation outcomes, such as inspiring direct transfer of solutions, or sparking adaptation of ideas into novel combinations.

\subsection{Coding ideation outcomes} \label{section:types-of-brainstorming-ideation-outcomes}
We are interested in studying whether an analogical search engine provides distinctive and complementary value to other commonly used search approaches that rely on surface similarity. In particular, our focus is on the inspirational value rather than the immediate relevance of search results or the direct usefulness of solutions. The highest value of creative inspiration often comes from creatively adapting ideas to reformulate a problem and recognizing new bridges to previously unknown domains that open up entirely new spaces of ideas. For example, recognizing a connection \new{from} the ancient art form of origami to fold intricate structures with paper and building a sufficiently compact, deployable solar panel arrays and radiation shields led NASA to hire origami experts~\cite{peraza2014origami,zirbel2013origami,origami_open_innovation}.

Our approach to measuring ideation outcome is through the use of a quaternary variable categorizing the types of ideation. To capture the inspirational value of analogical search and move beyond the measurements focused on the immediate relevance or the direct usefulness we distinguish the Creative Adaptation and Direct Application types of ideation. In our studies these two types corresponded to think-alouds that result\new{ed} in novel ideas whereas the rest (Background and None) corresponded to think-alouds in which no new ideas {were} produced.

$\sbullet[.75]$~\textbf{Creative Adaptation:} Novel mechanism ideas that involve substantial adaptation of the information provided in the paper. These ideas are typically associated with a higher uncertainty of success due to the less familiar nature of the domains involved.

$\sbullet[.75]$~\textbf{Direct Application:} More directly applicable ideas that involve less adaptation than Creative Adaptation. These ideas are typically associated with a lower uncertainty of success because researchers are more familiar with the domains.

$\sbullet[.75]$~\textbf{Background:} The information provided in the paper is good for background reading (e.g., to learn about other domains).

$\sbullet[.75]$~\textbf{None:} Did not result in new ideas \new{nor was useful for background reading.}

\begin{wrapfigure}{R}{.5\textwidth}
    \begin{center}
    \includegraphics[width=.5\textwidth]{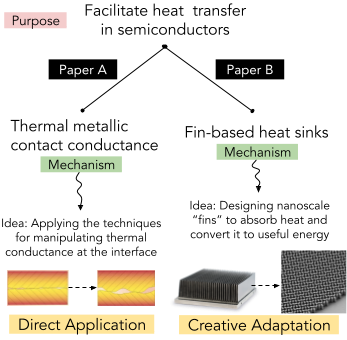}
    \end{center}
    \caption{Example papers for the purpose of facilitating heat transfer heat in semiconductors. (Top) A Direct Application paper involves directly applicable ideas and techniques for manipulating the interface material and structure to control thermal conductance. (Bottom) A Creative Adaptation example involves transferring a distant idea (fin-based design for heat sinks) and creatively adapting it into the target problem context (designing nano-scale fins that could absorb heat and convert it to useful energy). Figure credits: contact configurations and interface resistance from~\cite{interface_resistance}, fin-based heat sink from~\cite{heatfins}, nano-fins from~\cite{nanofins}.}
    \label{fig:ideation_examples}
\end{wrapfigure}

Creative Adaptation ideas generally involve{d} a substantial amount of adaptation, while Direct Application ideas {were} closer to the source domain and more directly applicable. For example, using the data from one of our participants, applying the techniques for manipulating thermal conductance at solid-solid interfaces {wa}s considered a direct application idea for P1 (fig.~\ref{fig:ideation_examples}, left) because he {wa}s familiar with the concept of controlling the interfacial thermal conductivity given the relevant approaches he developed in his current and past research projects. Thus the connections to the source problem {were} directly recognizable. On the other hand, creating a fin-based wall structure for heat transfer {wa}s an example of creative adaptation idea (fig.~\ref{fig:ideation_examples}, right) because of its novelty and the participant's unfamiliarity in relevant domains. The unfamiliarity and uncertainty {wa}s generally more associated with analogs for creative adaptation than direct application. On the other hand, the unfamiliarity also sometimes act{ed} as a barrier to participants' openness and subsequent ideation. Though challenging, in order to recognize novel connections to the source problem the participants may need to suspend their early rejection of a seemingly foreign idea and its surface-level mismatches and engage in deeper processing which could lead to re-imagination and re-formulation of the research problem at hand. \new{To code the Creative Adaptation and Direct Application types of ideation outcomes, the coders took into consideration different linguistic and contextual aspects of the descriptions of the ideas and their think-aloud process (details in \S\ref{subsubsection:data_and_coding}).}

\subsection{Design of the study}

\subsubsection{\new{Participants}}
We recruited eight graduate (four women) researchers in the fields of sciences and engineering via email advertisement at a private R1 U.S. institution. Four were senior PhD students (3rd year or above and one recently defended their thesis) and the rest was 2nd year or below. Disciplinary backgrounds of the participants included: Mechanical (3), Biomedical (2), Environmental (1), Civil (1), and Chemical Engineering (1). Once a participant signed up for the study, we asked them to describe their research problems and send the research team search queries they use to look for inspirations on popular search engines such as Google Scholar\footnote{\url{https://scholar.google.com/}}. Members of the research team screened papers with relevant purposes using these queries on the filtering interface (\new{fig.~\ref{fig:study1_interfaces}}, left). Despite our efforts to collect papers over diverse topical areas, the search engine did not contain enough papers for two of the participants who work on relatively novel fields (e.g., ``machine learning methods of 3D bioprinting''). These participants were interviewed on their current practices for reviewing prior works and coming up with new ideas for research and were not included in the subsequent analyses.

\subsubsection{\new{Study Procedure and Keyword-search Control}} 
\begin{figure}[t]
    \begin{center}
    \includegraphics[width=\textwidth]{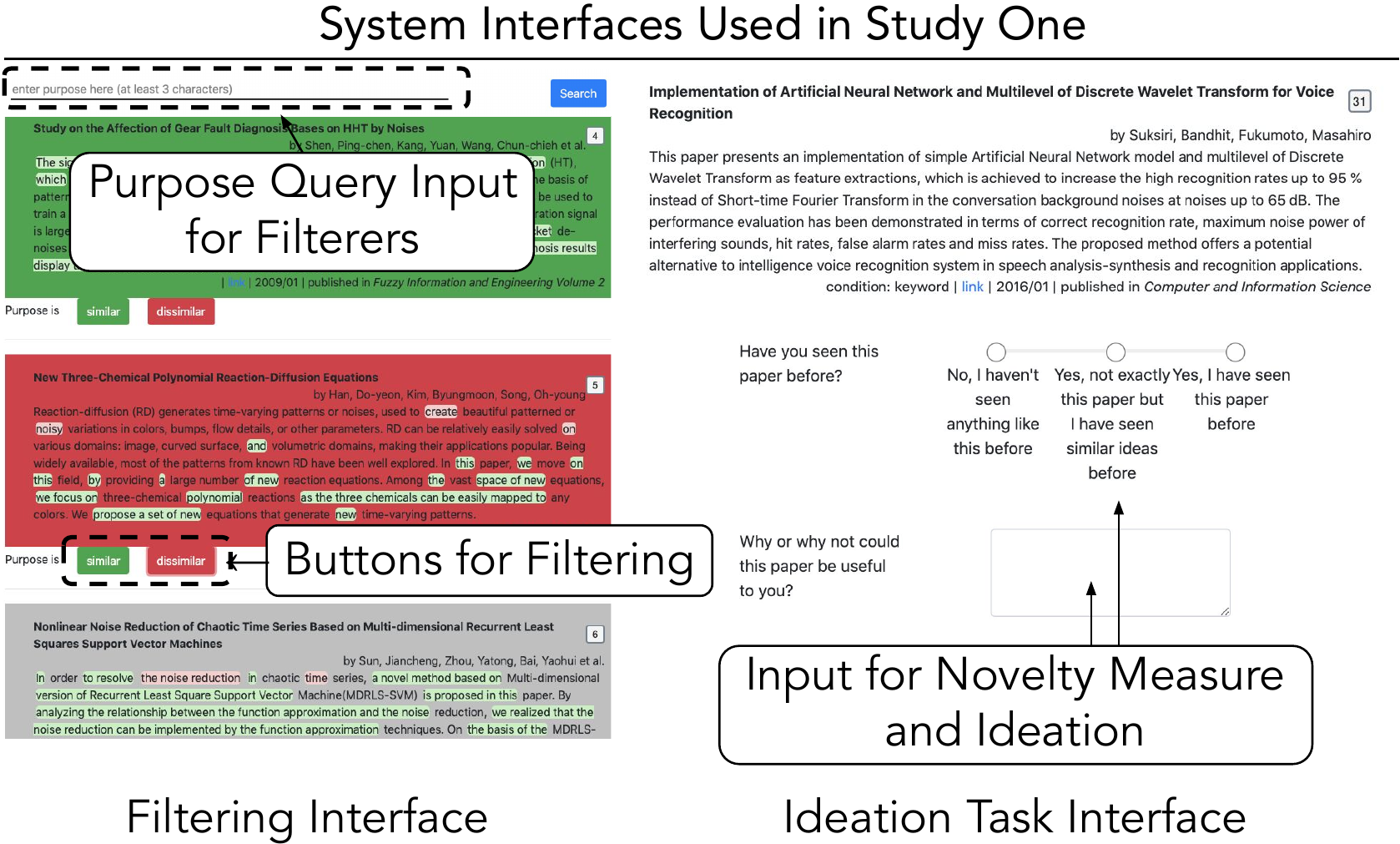}
    \end{center}
    \caption{The front-end interfaces. (Left) Human reviewers used this filtering interface to input search queries received from the participants and remove papers with obviously irrelevant purposes. \new{To assist the reviewers' filtering process, model predicted purpose (e.g., \textit{the noise reduction} and \textit{time}, highlighted in red at the bottom of the filtering interface) and mechanism (highlighted in green) tokens were also provided along with the title and the abstract text.} The background color turned green when the ``Similar'' button is clicked and red when the ``Dissimilar'' button is clicked. (Right) The ideation task interface {wa}s populated with a list of human filtered papers for review by the participants in Study 1 (the order of papers was randomized).}
    \label{fig:study1_interfaces}
\end{figure}

The rest of the participants were then invited to in-person interviews. To ensure that participants would be exposed to a \new{sufficiently} diverse set of analogical mechanisms and to maximize our power to observe the ideation process, we generated a list of top 30 results from the analogical search engine \new{using the search queries provided by the study participants}. As a control condition we also included top 15 results from a keyword-based search engine using the standard \textsc{Okapi BM25} algorithm~\cite{intro_to_IR} ($k_1 = 1.2, b = 0.75$) \new{using the same search queries as the analogical search engine}. The order of results in the list was randomized and participants were blind to condition. \new{To account for the difference in the quantity of exposure in the analysis, we normalized the ideation outcomes by the number of results returned in each condition.} Using this list we employed a think-aloud protocol~\cite{think_aloud_van1994,think_aloud_lewis} in which participants were presented with the title, abstract, and other metadata of papers and asked to think aloud as they read through them with the goal of generating ideas useful for their research using our Web-based interface (fig.~\ref{fig:study1_interfaces}, right). Although time consuming, this approach allowed us to capture rich data on \new{participants' thought process} and how those processes changed and evolved as participants considered how a paper might relate to their own research problems. In addition, we asked the participants to make a judgment on the novelty of each paper on a 3-point Likert-scale. After participants finished reviewing the 45 papers, we interviewed them about their overall thoughts on the results' relevance and novelty and whether there were any surprising or unique results. Each interview lasted about one and a half hours and the participants were compensated \$15/hr for their participation.

\subsubsection{Data and Coding} \label{subsubsection:data_and_coding}
In total, our data consist{ed} of 267 paper recommendations for six participants and their Likert-scale questionnaire responses measuring the content novelty, after removing 3 within-condition duplicates (these papers included cosmetic changes such as different capitalization in the title or abstract). One participant ran out of time towards the end of the interview and only provided novelty measures for the last 17 paper recommendations in the randomized list. Thus, 250 transcripts of participants' think-aloud ideation after reading each paper {were} used for analyzing ideation outcomes.
\new{To code the distance between the Creative Adaptation and Direct Application types of ideation outcomes, the coders took into consideration (1) the verbs used to describe the ideas (e.g., `design', `develop', or `invent' were generally associated more with distant ideas compared to `apply', `use', `adopt'; see Table.~\ref{table:brainstorm-ex}); (2) the context of ideas such as participants' expression of unfamiliarity or uncertainty of the domain involved (e.g., ``I'm not really sure'' vs. ``I'm familiar with this domain''); and (3) participants' perceived immediacy of the idea's applicability (i.e., ideas perceived by participants as more immediately applicable were associated with direct application but not creative adaptation ideas). Two of the authors coded a fraction of the data together (13/250, 5.2\%) and then independently coded the rest blind-to-condition, using the four ideation outcomes types described in \S\ref{section:types-of-brainstorming-ideation-outcomes} and with the following protocol: The coders first judged the existence of an idea. If there was, then its type was further distinguished between Creative Adaptation and Direct Application using the linguistic and contextual descriptions described above (e.g., Creative Adaptation ideas were more frequently associated with the `design' words, higher unfamiliarity and uncertainty of the domains, and less immediate applicability, compared to Direct Application ideas). In case there was no concrete idea in the data, coders further distinguished between the Background vs None cases.}

The agreement between coders was significant, with Cohen's $\kappa = 0.89$ (near perfect agreement) for the four categories of ideation outcome. Given the high level of agreement between the coders, any disagreements were resolved via discussion on a case-by-case basis.
\subsubsection{\new{Apparatus 1}: the human-in-the-loop filtering interface} \label{subsubsection:apparatus1}
In Study 1, members of the research team first received search queries from study participants and reviewed the model-produced purpose matches to filter irrelevant papers using a filtering interface (fig.~\ref{fig:study1_interfaces}, left). This additional step was introduced to ensure that papers with obviously dissimilar purposes are not returned to study participants. Reviewers determined whether each paper contained a clearly irrelevant purpose in which case it was removed by clicking the \textit{Dissimilar} button at the bottom of the paper. On the other hand when the \textit{Similar} button was clicked it turned the background of the paper green in the interface and increased the number of the papers collected so far. Reviewers continued the screening process until at least 30 papers with reasonable purpose similarity were collected.

\subsubsection{\new{Apparatus 2}: the ideation task interface} \label{subsubsection:apparatus2}
The filtered papers were then displayed as a randomized list of papers to study participants (fig.~\ref{fig:study1_interfaces}, right). In addition to the content and metadata of papers (e.g., authors, publication date, venue, etc.), each paper was presented with a Likert-scale question for measuring content novelty and a text input for ideation. 

\subsubsection{Limitations}
To reduce potential biases, our coders were blind to experimental conditions and relied on participants’ statements of \new{ideas' novelty and usefulness} (e.g., ``I've never seen something like this before,'' ``this is not a domain I would've searched if I used Google Scholar''), and achieved a high inter-rater reliability. We believe coders had a reasonable understanding of how participants arrived at specific ideas from descriptions of their current and past research topics, think-alouds, and end-of-experiment discussions. Despite this, \new{we also acknowledge the limitations of this approach and discuss how future research may improve upon it (see \S\ref{subsubsection:exp_validity}).}

\subsubsection{On reporting the results} We report the result of our studies below. To denote statistical significance we use the following notations: $^{*} (\alpha = 0.05)$, $^{**} (\alpha = 0.01)$, $^{***} (\alpha = 0.001)$, $^{****} (\alpha = 0.0001)$. Alpha levels {were} adjusted when appropriate in post-hoc analyses using Bonferroni correction.

\subsection{Result}
\xhdr{Finding novel papers for creative ideas}
Our key measure of success is how paper recommendations from the analogy search engine (hereinafter \textit{analogy papers}) help scientists generate creative ideas for their own research problems. To this end, we investigate a) whether analogy papers are novel and complementary to the papers found from the keyword-search baseline (hereinafter \textit{keyword papers}) and b) whether analogy papers resulted in more creative adaptation ideas than direct application ideas in ideation.

\subsubsection{Analogy papers differed from keyword papers and were judged more novel}
\begin{wrapfigure}{R}{.5\textwidth}
    \begin{center}
    \includegraphics[width=.5\textwidth]{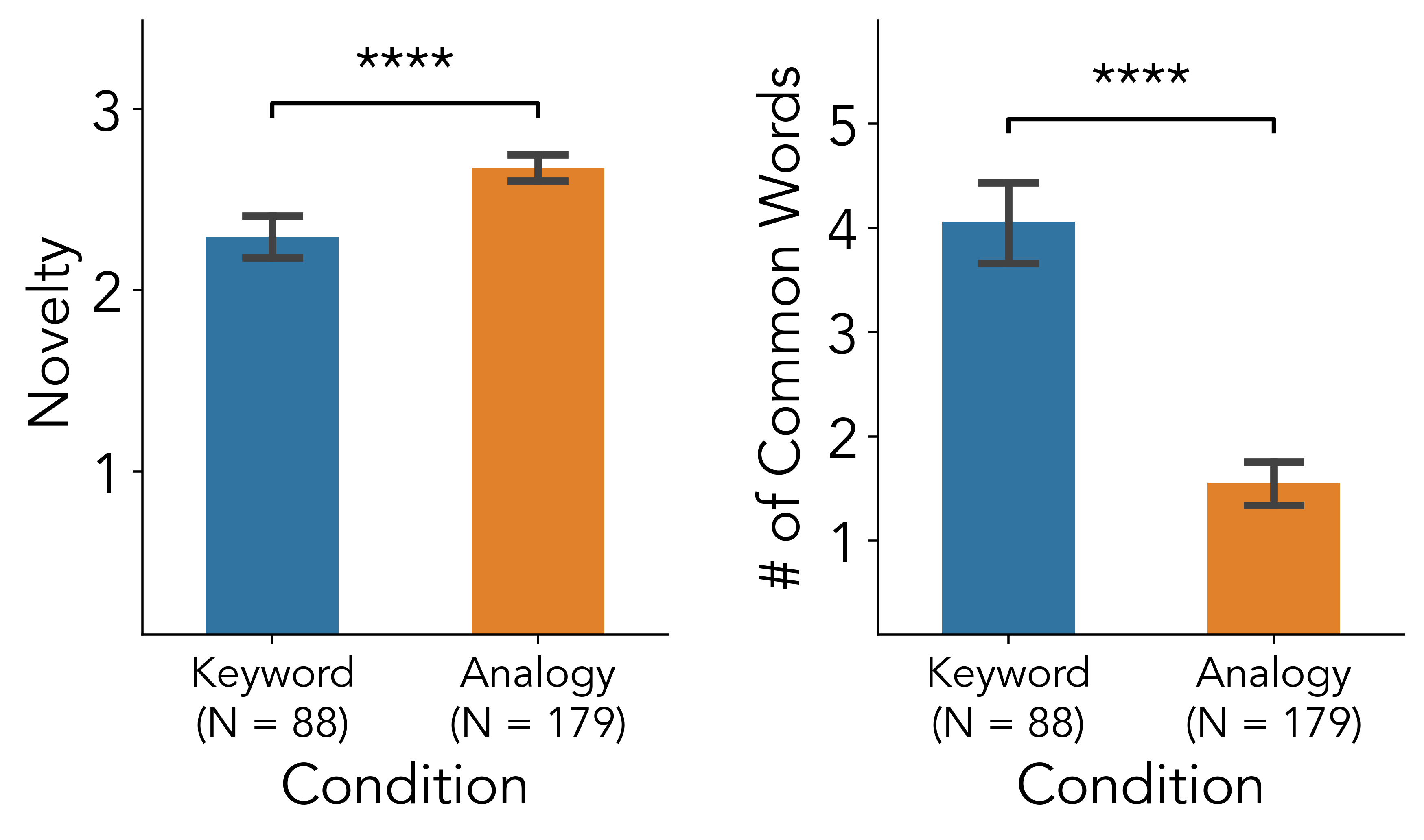}
    \end{center}
    \vspace{-1em}
    \caption{(Left) Participants judged analogy papers significantly more novel. The mean response to the question \textit{"Have you seen this paper before?"} was significantly higher in Analogy: 2.7 (SD: 0.48) than in Keyword: 2.3 (SD: 0.55). (Right) There were significantly more overlapping words between search query terms provided by participants and the title and abstract text of papers: Keyword: 4.1 (SD: 1.74) vs. Analogy: 1.6 (SD: 1.42).}
    \label{fig:novelty_keyword_overlap}
\end{wrapfigure}
The viability of our approach is based on the assumption that the analogy search pipeline returns a different distribution of results than a keyword-based baseline. This assumption appeared to hold true: the keyword-search and analogy-search conditions resulted in almost completely disjoint sets of paper recommendations. Out of the total 267 papers, the overlap between analogy and keyword papers was only one. Analogy papers appeared to represent a complementary set of results users would be unlikely to encounter through keyword-based search.

To further examine this assumption we had participants rate the novelty of the results by asking them ``\textit{have you seen this paper before?}'' on a 3-point Likert scale response options of 1: ``\textit{Yes, I have seen this paper before}'', 2: ``\textit{Yes, not exactly this paper but I have seen similar ideas before}'', and 3: ``\textit{No, I have not seen anything like this before}''. Participants found papers recommended in the analogy condition to contain significantly more novel ideas (2.7, SD: 0.48) compared to the keyword condition (2.3, SD: 0.55) (Welch's two-tailed t-test, $t = -5.53, p = 1.33\times10^{-7}$) (fig.~\ref{fig:novelty_keyword_overlap}, left). Participants thought the ``variance in results is much higher than using other search engines'' (P5) and ``there're a lot of bordering domains... which can be useful if I want to get ideas in them'' (P4).

This difference was also reflected in the content of papers, with keyword papers having significantly more overlapping terms with participant-provided query terms (4.1, SD: 1.74) than analogy papers (1.6, SD: 1.42) (Welch's two-tailed t-test, $t(145.27) = 11.70, p = 1.10\times10^{-22}$) (fig.~\ref{fig:novelty_keyword_overlap}, right)\footnote{We measured the term overlap between participants' queries and the content of papers (title and abstract). To preprocess text, we used \textsc{NLTK}~\cite{nltk} to tokenize papers' content, remove stopwords, digits, and symbols, and lemmatize adjectives, verbs, and adverbs. \new{Finally,} using the processed tokens we constructed a set of unique terms \new{for each paper and the query which was then compared to find overlapping terms}.}. More occurrences of familiar query terms in keyword papers' titles and abstracts may have led participants to perceive them as more familiar.

\subsubsection{Analogy papers resulted in more creative adaptation ideas than direct application ideas}
\begin{figure}[t]
    \centering
    \includegraphics[width=\textwidth]{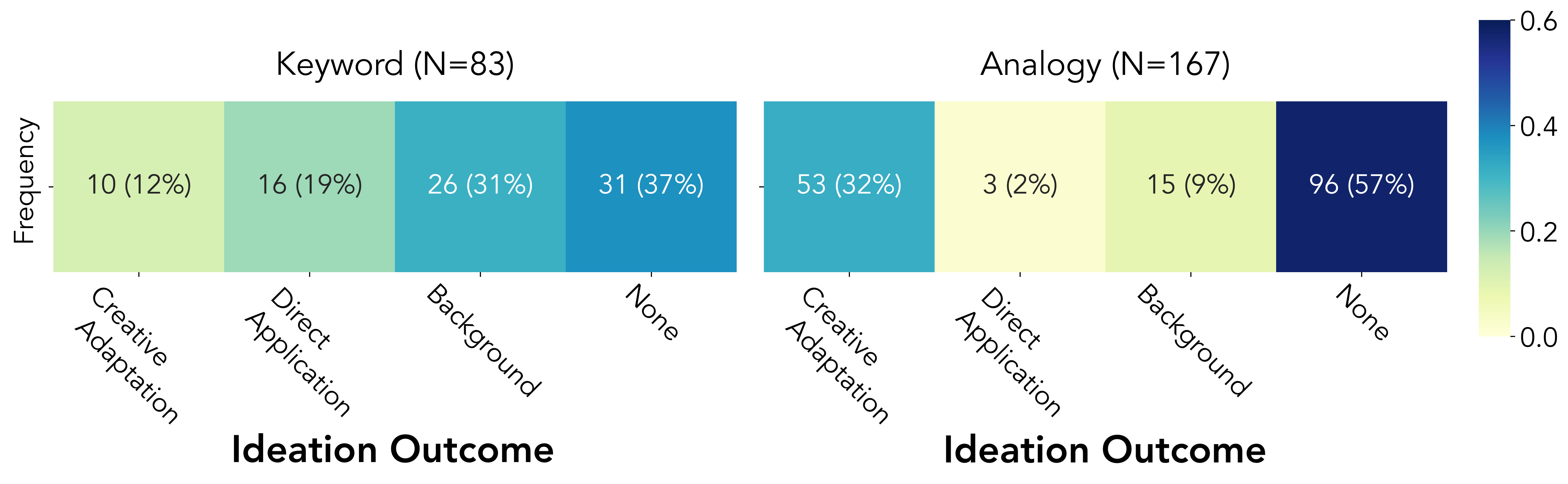}
    \caption{Frequency of the ideation outcome types by condition. Darker colors represent higher rates. Creative adaptation is 5.3 times more frequent among analogy papers (53 in Analogy vs. 10 in Keyword), while most of direct application is from keyword papers (3 in Analogy vs. 16 in Keyword). The distributions differed significantly (chi-squared test, $\chi^2(3) = 52.12, p < 1.0\times10^{-10}$ overall and $\chi^2(1) = 28.41, p = 9.84\times10^{-8}$ for the contrast between the rates of creative adaptation and direct application ideas).}
    \label{fig:ideation-outcome}
\end{figure}
We found that the distribution of ideation outcome types differed significantly between analogy and keyword papers ($\chi^2(3) = 52.12, p < 1.0\times10^{-10}$). Participants came up with more creative adaptation ideas (N = 53; 32\% of total) over direct application ideas (N = 3; 2\%) using analogy papers. In contrast, keyword papers resulted in more direct application ideas (N = 16; 19\%) than creative adaptation ideas (N = 10; 12\%) (fig.~\ref{fig:ideation-outcome}). The difference between creative adaptation and direct application was significant ($\chi^2(1) = 28.41, p = 9.84\times10^{-8}$).

\begin{table*}[t!]
\begin{tabular}{c p{3cm} c p{6cm}}
\toprule
{\textbf{PID}} & {\textbf{Research Problem}} & {\textbf{Type}} & \textbf{Paper Title $\rightarrow$ New Idea} (paraphrased)\\
\midrule
\multirow{10}{*}{1} & \multirow{10}{3cm}{Improve nanoscale heat transfer in semiconductor material} & \multirow{5}{*}{Direct Application} & \textit{Experimental investigation of thermal contact conductance for nominally flat metallic contact} $\rightarrow$ Apply the techniques in the paper to manipulate thermal conductance at the solid-solid interface\\[2cm]
 &  & \multirow{4}{*}{Creative Adaptation} & \textit{Investigation on periodically developed heat transfer in a specially enhanced channel} $\rightarrow$ Design nanoscale ``fins'' to absorb heat and convert it to mechanical energy\\
\midrule
\multirow{10}{*}{2} & \multirow{10}{3cm}{Grow plants better by optimizing entry of nanoparticle fertilizers into the plant} & \multirow{4}{*}{Direct Application} & \textit{Nanoinformatics: Predicting Toxicity Using Computational Modeling} $\rightarrow$ Apply the computational modeling from the paper for predicting toxicity of candidate nanoparticles\\[2cm]
 &  & \multirow{5}{*}{Creative Adaptation} & \textit{Identification of Plant Using Leaf Image Analysis} $\rightarrow$ Invent a hyperspectral 3D imaging mechanism for plants that optically senses, traces, and images plant cells in 3-dimensional structures\\
\midrule
\multirow{12}{*}{3} & \multirow{12}{3cm}{Enhance the evaporation efficiency of thin liquid films in heat pipes and thermosyphones} & \multirow{5}{*}{Direct Application} & \textit{Thin film evaporation effect on heat transport capability in a grooved heat pipe} $\rightarrow$ Adopt the techniques in the paper for manipulating the solid interface's surface properties to balance the film thickness and disjoining pressure\\[3cm]
 &  & \multirow{5}{*}{Creative Adaptation} & \textit{Alkaline treatment kinetics of calcium phosphate by piezoelectric quartz crystal impedance} $\rightarrow$ Design novel liquid film materials for manipulating hydrophobicity to change disjoining pressure\\
\bottomrule
\end{tabular}
\caption{Examples of Direct Application and Creative Adaptation types for three participants (PID). Each participant's research problem is described in the Problem column. While the topics of research problems vary, Creative Adaptation ideas are more distant in terms of content compared to the source problem than Direct Application ideas are, and may be characterized by the use of different sets of verbs (\{\textit{design}, \textit{invent}\} in Creative Adaptation ideas versus \{\textit{apply}, \textit{adopt}\} in Direct Application ideas).}
\label{table:brainstorm-ex}
\end{table*}
To illustrate more concretely the divergent patterns of ideation leading to Creative Adaptation and Direct Application ideas, we describe vignettes from three participants (table~\ref{table:brainstorm-ex}). While Direct Application ideas represented close-knit techniques and mechanisms directly useful for the source problem (described with verbs such as \textit{apply} and \textit{adopt}), Creative Adaptation type ideas were more distant from the source problem and \new{could} be characterized with the use of different verbs associated with significant adaptation (\textit{design} and \textit{invent}). For example, P1's research focused on the methods for improving nanoscale heat transfer in semiconductor materials. Previously he developed mechanisms for manipulating the thermal conductivity at solid-solid interfaces, specifically by adjusting the semiconductor wall structures. Thus, a paper reporting experimental results of manipulating thermal conductance on planar metallic contact points was deemed a directly useful paper that might contain helpful techniques. On the other hand an analogy paper which dealt with the heat transfer phenomenon at a macroscale, using fin-based heat sink designs for electronic devices, gave him a new inspiration: to adapt fins for nanoscale heat transfer in semiconductors to not only transfer heat but also convert it into a useful form of mechanical energy. Despite the mismatch on scale ([macroscale] $\nleftrightarrow$ [microscale]), challenging the assumption of the typical size of a fin-based design engendered an idea to creatively adapt it to convert heat into energy through an array of tiny fins, rather than merely dissipating it into space as in the original formulation of the problem. P1 also found another analogy paper focused on thermal resistance at a liquid-solid interface useful for future ideation because despite its surface dissimilarities, there was a potential mapping that may open up a new space of ideas (e.g., [liquid] $\nleftrightarrow$ [polymer substrate], [solid] $\nleftrightarrow$ [germanium], yet the pairwise relation [liquid:solid] $\leftrightarrow$ [polymer substrate:germanium] may be analogous and interesting)\new{:} ``This is liquid... but it's about liquid-solid interface which can be useful... because for the substrate that sits on top of silicon or germanium you use polymers which have liquid-like properties'' (P1).

In the case of P2, a paper focused on computational methods for toxicity prediction was deemed directly helpful because ``if certain nanomaterials are toxic to certain microorganisms that eat plants or kill them but safe for the plant, we can target these organisms using the nanomaterials as pesticide. Another way this can be helpful is in predicting the chance of toxicity of the nanoparticles in our fertilizers'' (P2). Whereas an analogy paper that uses image analysis for plant identification reminded her of ``hyperspectral imaging in plants, like a CT scan for plants. So making a hyperspectral 3D model using something like this... to optically sense and trace plant cells (such that the entry of fertilizer nanoparticles into plant cells can be monitored, a sub-problem of P2's research problem) would be pretty cool.''

As a third example, P6's research focused on recording and simulating electrical activity using microelectrode arrays. To him, an analogy paper about printing sensors for electrocardiogram (ECG) recording seemed to present an interesting idea despite its mismatch in terms of scale ([nanoscale] $\nleftrightarrow$ [macroscale]) \new{and manufacturing mechanism (}[fabrication] $\nleftrightarrow$ [printing]\new{), because} the pairwise relation between [nanoscale:fabrication] $\leftrightarrow$ [macroscale:printing] engendered a reflection on the relative advantages of different methods and future research directions): ``Interesting idea! Instead of nanoscale fabrication, printing can be a good alternative for example for rapid prototyping. But I think the resolution won't be enough (for use) in nanoscale... works for this particular paper's goal, but an idea for future research is whether we can leverage the benefit of both worlds -- rapid printing and precision of nanoscale fabrication'' (P6).

\subsubsection{The level of purpose-match had different effects on the ideation outcome} \label{section:purpose-match-mediation}
Suggested in these examples is a certain kind of distance the ideas in analogy papers maintain in order to spur creative adaptation. We hypothesize that some amount of difference in purpose facilitates creative adaptation. This process may involve a curvilinear relationship between the degree of purpose mismatch and the resulting ideation outcome, with too much or too little deviation leading to a little-to-no benefit or even an adverse effect on the ideation outcome\new{, a pattern that is consistent with the findings in the literature of creativity and learning outcomes (e.g., Csikszentmihalyi's optimal difficulty~\cite{csikszentmihalyi1990flow}).} For this analysis, we coded each paper based on three levels of purpose-match to the source problem:

$\sbullet[.75]$~\textbf{Full:} Both high- and low-level purposes match

$\sbullet[.75]$~\textbf{Part:} Only the high-level abstract purpose matches. Explicit descriptions of the high-level purpose exist in either title and abstract of the paper. At the same time, certain low-level aspects of the participant's research problem are mismatched as evidenced by relevant comments from the participant

$\sbullet[.75]$~\textbf{None:} Neither high- nor low-level purposes match

\begin{table*}[t]
\begin{tabular}{c c p{10cm}}
\toprule
\textbf{Purpose-Match} & \textbf{PID} & \textbf{Participant Comment} \\
\midrule
\multirow{3}{*}{Full} & \multirow{3}{*}{2} & ``It's a little bit old (from 2010) but I have read papers from that era. I love this... because the paper mentions everything else and especially one word which is `disjoining pressure' -- if I were to publish my current project that's going to be the core topic.''\\
\midrule
\multirow{3}{*}{Part} & \multirow{3}{*}{1} & ``Though I'm not familiar with GFRP-GFRP... but I can see that they're referring to glass fiber reinforced plastic, so this is something not crystalized material... learning about this kind of materials is interesting.''\\
\midrule
None & 3 & ``I don't know what a lot of words mean. I don't typically work with animals cells.''\\
\bottomrule
\end{tabular}
\caption{Examples of different purpose-match types. Purpose-Match shows the level of purpose-match between a recommended paper and each participant's research problem (see table~\ref{table:brainstorm-ex} for descriptions of research problems). Fully matching purposes are those that match at both high- (more abstract) and low-levels (specific details). Partial matches only match at the high-level abstraction and differ in details. The Participant Comment column shows relevant excerpts from the participant.}
\label{table:purpose-match-ex}
\end{table*}
Examples of these types of purpose-match are provided in Table~\ref{table:purpose-match-ex}. \new{High-level match can be considered as a first-order criterion of purpose match and low-level match as a second-order criterion: If the paper does not have overlapping terms in terms of its purpose with the user query cast at a high level (e.g., transfer heat, grow plants) then the low-level match does not matter, but if the paper's purpose matches at the high level, its low-level alignment (e.g., specific aspects of the purpose, such as its scale or materialistic phase) will additionally determine full (i.e., aligned in both high- and low-level aspects of the purpose) vs partial match (i.e., aligned only in the high-level but not low-level aspects of the purpose). Therefore, the coding procedure was symmetrical to the procedure described for coding four types of ideation outcome, with the high-level purpose match deciding between \{Full, Part\} and None match types, while the low-level purpose further distinguishing between Full vs. Partial match.} Following this procedure, two independent coders achieved an inter-rater reliability Cohen's $\kappa = 0.72$ (substantial agreement) and disagreements were resolved \new{with case-by-case discussion}.

We used the \textsc{mediation} package\footnote{\url{https://cran.r-project.org/web/packages/mediation/index.html}}~\cite{tingley2014mediation} to conduct a mediation analysis between the condition, the kind of purpose-match, and the binary Creative Adaptation ideation outcome. The analysis showed that the effect of condition (Keyword vs. Analogy) on the binary outcome of creative adaptation was mediated by the degree of purpose-match\new{, but not by the novelty of content, suggesting that the difference between full vs. partial matching on purpose is much more significant than the variance in the content novelty. We come back to this result in the discussion (\S\ref{subsubsection:control_diversity}).} Table~\ref{table:mediation} presents the result {of} the mediation analys{e}s. The regression coefficient between creative adaptation and condition was significant 
as was the regression coefficient between the degree of purpose match and creative adaptation. 
The indirect effect was $(-.42)\times(.21) = -.09$. We tested the significance of this indirect effect using a bootstrapping procedure~\cite{mediation_bootstrapping} ($p < 2\times 10^{-16}$)\new{, by computing the} unstandardized indirect effects for each of \fnum{1000} boostrapped samples as well as the 95\% confidence interval (CI)\footnote{Alternatively, it is possible that the mediating effect of the degree of purpose-match on the likelihood of creative adaptation outcome is moderated by novelty. However, the result of our analysis showed that this was unlikely: The effect was insignificant using the boostrapping method -.04, ($p = 0.12$, 95\% CI = $[-.09, .01]$).}.

\begin{table*}[t]
\centering  
\begin{tabular}{@{}l *{5}{S[input-symbols=(),table-format=-1.3,table-space-text-post=****,table-align-text-post=false]} @{}}
\toprule
 & \text{Effect of Condition} & \text{Unique Effect} & \text{Indirect Effect} & \multicolumn{2}{c}{\text{CI 95\%}}  \\ \cmidrule(lr){5-6}
 \textit{Mediator} & \text{on Mediator (\textit{a})} & \text{of Mediator (\textit{b})} & \text{(\textit{a$\times$b})} & \text{Lower} & \text{Upper} \\ \midrule
 \multirow{2}{*}{Purpose-match} & -0.42\textsuperscript{****} & 0.21\textsuperscript{****} & -0.09\textsuperscript{****} & -0.14 & -0.05 \\
 & (.08) & (.05) & & & \\ \addlinespace
 \rowcolor{lgrey}
 & 0.40\textsuperscript{****} & -0.06  & -0.02 & -0.07 & 0.02\\
 \rowcolor{lgrey}
 \multirow{-2}{*}{Novelty} & (.07) & (.05) & & & \\ \addlinespace
\multirow{2}{*}{Pid} & -0.02 & 0.03\textsuperscript{*} & -0.001 & -0.02 & 0.02 \\
 & (.22) & (.02) & & & \\
\bottomrule
\end{tabular}
\caption{Regression table of three mediation analyses using \textit{Purpose-match}, \textit{Novelty} and \textit{Pid} (Participant ID) as mediators between Condition and the binary Creative Adaptation outcome variable. Purpose-match was the only significant mediator \new{between Condition and Creative Adaptation} (indirect effect=-.09, significant using a bootstrapping method~\cite{mediation_bootstrapping} with \fnum{1000} iterations, $p < 2\times 10^{-16}$).}
\label{table:mediation}
\end{table*}

\begin{figure}[t]
\begin{minipage}{.33\textwidth}
  \centering
    \includegraphics[width=\textwidth]{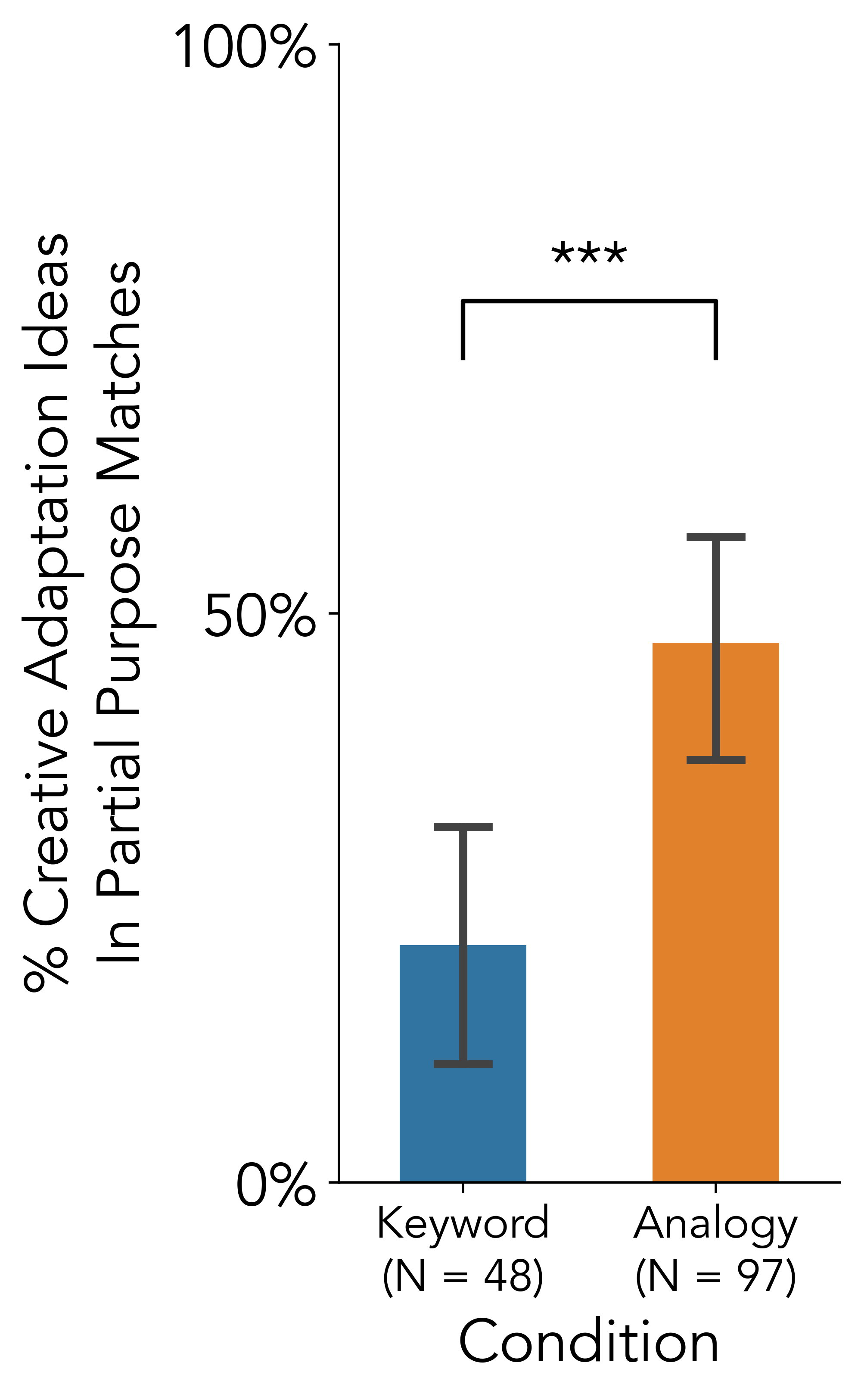}
    \caption{Proportion of creative adaptation ideas among the partial purpose-match papers. Creative Adaptation was significantly more frequent among the analogy papers (47\%) than keyword papers (21\%) (Welch's two-tailed t-test, $p = 9.0\times10^{-4}$.}
    \label{fig:chance_creative_adaptation_mean_ranks}
\end{minipage}%
~\qquad  
\begin{minipage}{.62\textwidth}
  \centering
    \includegraphics[width=\textwidth]{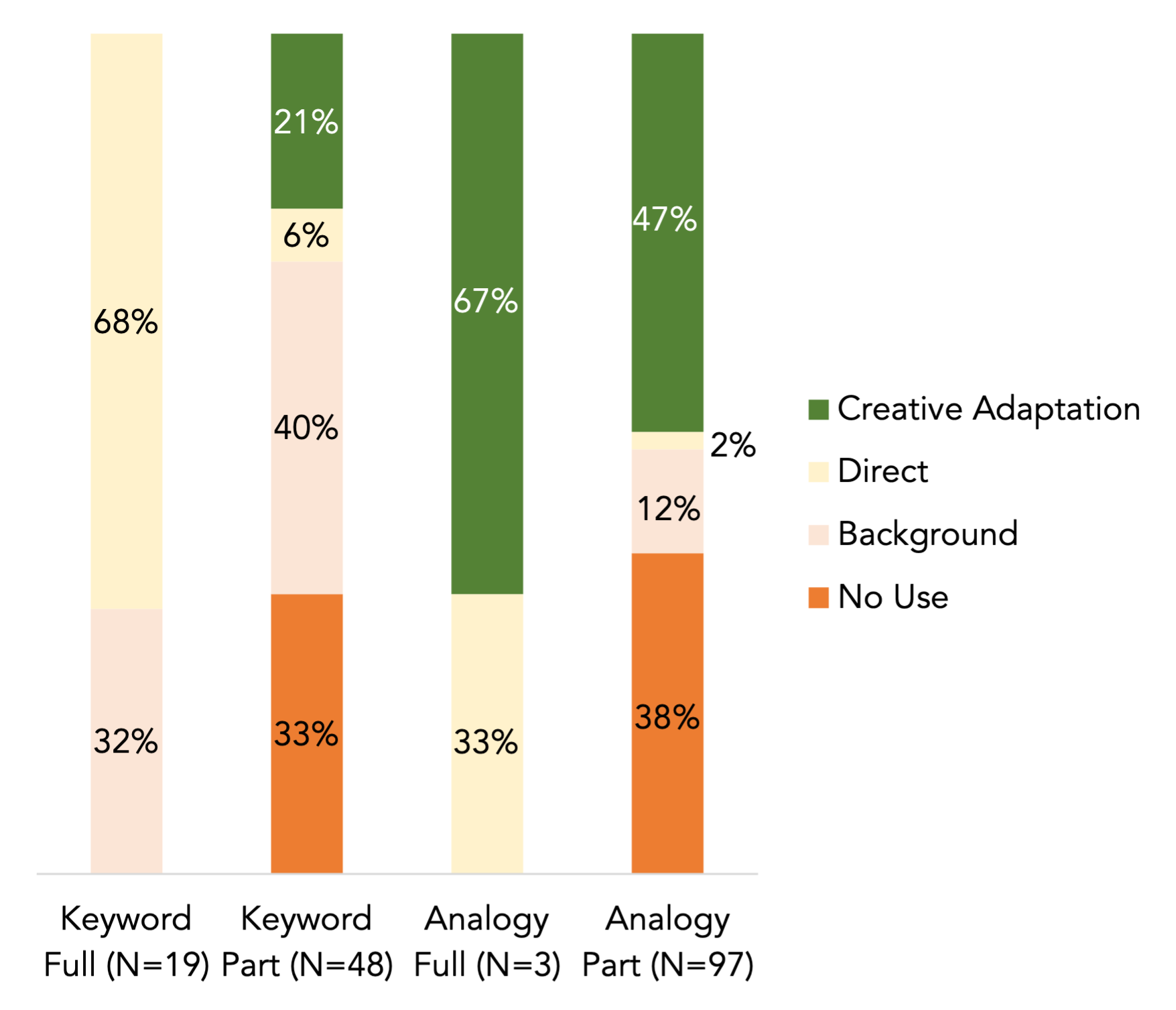}
    \caption{The rate of ideation outcome types in full and partial purpose matches. Among the keyword papers as the purpose mismatch increases, the rate of creative adaptation also increases from 0\% to 21\% (middle). However, this rate is significantly higher among the analogy papers (47\%) than the keyword papers (21\%). Note that while purpose mismatches led to more creative adaptation among analogy papers, a large portion of them also resulted in no ideation outcome (38\%).}
    \label{fig:partial_mismatches_outcome}
\end{minipage}
\end{figure}

Partial purpose matches in both keyword and analogy papers led to creative adaptation, but the rate was significantly higher with analogy papers. As expected, the ratio of direct application decreased from the keyword papers that fully match in purpose (Keyword Full, 68\%) to the keyword papers that partially match in purpose (Keyword Part, 6\%) (fig.~\ref{fig:partial_mismatches_outcome}). At the same time, the rate of creative adaptation increased from the keyword papers that fully match in purpose (Keyword Full, 0\%) to the keyword papers that partially match in purpose (Keyword Part, 21\%). However, the rate of creative adaptation differed significantly between the keyword and analogy papers, with the rate more than doubling among the analogy papers over keyword papers (Analogy Part 47\% vs. Keyword Part 21\%). Homing in on the partial matches, these papers led to creative adaptation ideas significantly more often in analogy search (47\%) than keyword search (21\%) (Welch's two-tailed t-test, $t(112.22) = -3.40, p = 9.0\times10^{-4}$, fig.~\ref{fig:chance_creative_adaptation_mean_ranks}, left). While the partial purpose mismatch was highly associated with creative adaptation ideas, it could be a double-edged sword. Among the analogy papers, 38\% of the partial mismatches resulted in no useful ideation outcome as opposed to the 47\% that resulted in creative adaptation (fig.~\ref{fig:partial_mismatches_outcome}, Analogy Part). Therefore, \textbf{knowing what mismatches are beneficial to creative adaptation} has important implications for facilitating generative misalignment for ideation.

\section{Study 2: Enabling a Fully Automated Analogical Search Engine}
\subsection{Motivation and structure of the study} 
The findings of Study 1 suggest potential benefits of an analogical search engine for scientific research, but a core limitation of interactivity due to the human-in-the-loop system design prevented its use as a more realistic probe for understanding researchers' natural interaction with analogical results. Specifically, the results of Study 1 are limited by the lack of participants' ability to reformulate search queries and the study design that involved returning only a fixed number of papers that blended both keyword and analogy papers in a randomized order. These factors significantly deviate from realistic usage scenarios of a deployed analogical search engine and prevent us from observing the full scope of user interaction. In order to move beyond these limitations, first we need a fully automated pipeline that removes the need for human-in-the-loop filtering, thus allowing us to enable query reformulation and interaction with corresponding search results. To achieve this, we improved the model accuracy on extracting purposes and mechanisms from paper abstracts by training a more sophisticated neural network that leverages more nuanced linguistic patterns. Specifically, we implement{ed} an attention mechanism within a span-based sequence-to-sequence model (Model 2) such that it \new{could} learn words that frequently co-occur to describe coherent purposes or mechanisms in paper abstracts\new{, and as a result, learning more informative words for our purpose} (see Appendix for details of implementation). Through evaluating the system backed by this improved pipeline, we demonstrate how it can remove the human-in-the-loop while maintaining similar levels of accuracy. In the following sections, we report the evaluation results that show 1) an improved token-level prediction accuracy using the span-based Model 2; 2) rankings of the results aligning well with human-judgment of purpose-match from Study 1; and 3) top ranked results of the system maintaining a similar rate of partial purpose matches relative to that of the human-in-the-loop system from Study 1.

The interactivity enabled by the automated analogical search pipeline further allows us to observe its use in more realistic scenarios. To probe how researchers would interact with an analogical search engine and what challenges they might face in the process, we ran case studies with six researchers (\S\ref{section:case studies}). \new{From these studies, w}e uncover potential challenges (\S\ref{section:case studies}) and synthesize design implications for future analogical search engines (\S\ref{section:design implications}).
\subsection{Result}
\begin{minipage}[t]{\textwidth}
    \centering
    \begin{minipage}[tc]{.6\textwidth}
        \centering
        \begin{tabular}{l c c c c}\toprule
        \multirow{2}{*}{\textbf{Model}} & \textbf{Embedding} & \multirow{2}{*}{\textbf{All}} & \multirow{2}{*}{\textbf{PP}} & \multirow{2}{*}{\textbf{MN}} \\
        & (finetuned) & & & \\
        \midrule
        1. Model 2~\cite{spanrel} & ELMo (N) & \textbf{\textcolor{blue}{0.65}} & 0.65 & 0.64 \\
        2. BiLSTM & ELMo (N) & 0.63 & 0.67 & 0.59 \\
        3. BiLSTM & SciBERT (N) & 0.62 & 0.69 & 0.55 \\
        4. BiLSTM-CRF~\cite{elmo} & ELMo (N) & 0.58 & 0.59 & 0.57 \\
        5. BiLSTM & GloVe (Y) & 0.55 & 0.56 & 0.53 \\
        \midrule
        6. Model 1 & GloVe (N) & 0.50 & 0.51 & 0.50 \\
        \bottomrule
        \end{tabular}
        \label{table:models}
        \captionof{table}{F1 scores of different models, sorted by the overall F1 score of Purpose (PP) and Mechanism (MN) detection. The span-based Model 2 gave the best Overall F1 score (blue). In comparison, the average agreement (\%) between two experts' and crowdworkers' annotations was $0.68$ (PP) and $0.72$ (MN)~\cite{chan2018solvent}. We used AllenNLP~\cite{allennlp} to implement the baseline models 1 -- 5.}
    \end{minipage}
    ~\qquad
    \begin{minipage}[tc]{.32\textwidth}
        \centering
        \includegraphics[width=\textwidth]{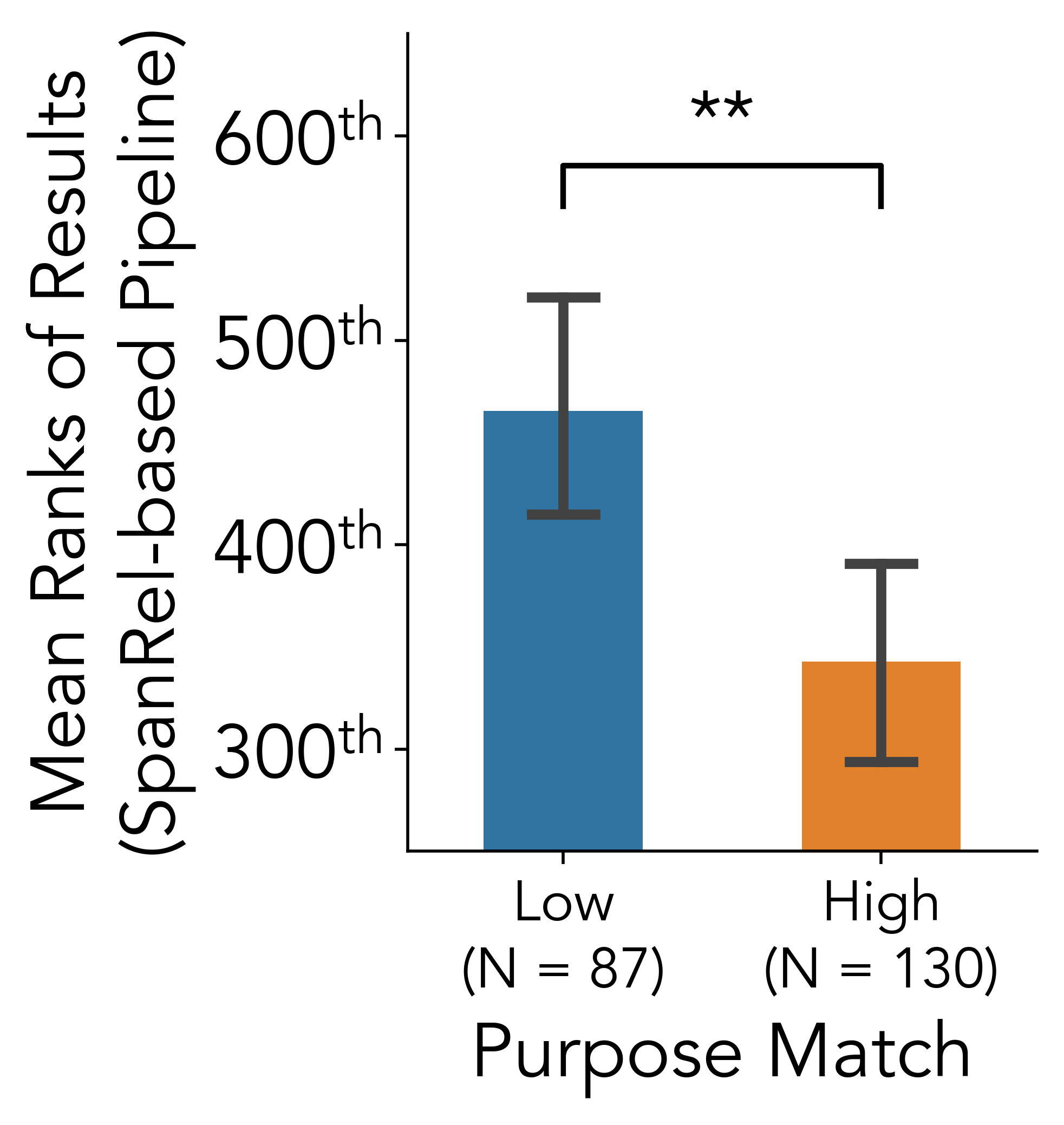}
        \vspace{-1em}
        \captionof{figure}{Mean ranks of human-judged high and low purpose match papers from the span-based pipeline. Low matches were ranked significantly lower (the rank number was higher), on average at $465^{\text{th}}$ (SD: 261.92) than high matches at $343^{\text{th}}$ (SD: 279.48).}
        \label{fig:mean_ranks}
    \end{minipage}
\end{minipage}
\subsubsection{Improved token-level prediction of a span-based model}
First we compared the span-based Model 2 with five other baselines to evaluate the token-level classification performance (Table~7). Model 2's overall F1 score was the highest at 0.65 (Purpose; PP: 0.65, Mechanism; \new{MN: 0.64, an 0.14- and 0.14-absolute-point increase from Model 1, respectively}) on the validation set which represents \new{an overall} 0.15-absolute-point increase from \new{Model 1} used for {the} initial human-in-the-loop analogical search engine.

\subsubsection{Pipeline with a span-based model reflected human judgment for ranking the results}
The improved token-level prediction performance materialized as an increase in the pipeline's ability to judge the degree of purpose match. For this evaluation, we first recorded every query \new{provided by Study 1 participants that human-in-the-loop filterers} used \new{to search and filter the relevant papers}. Then, we simulated the search condition of the filterers for the automated pipeline by providing it input as the exact queries they used. We capped the number of top search results sufficiently large at \fnum{1000} for each query. From these top \fnum{1000} results, we selected papers that also appeared in the human-in-the-loop system and collected the corresponding human-vetted judgments of high or low purpose-match. For each of these papers, we also collected its corresponding rank positions on the new (automated) pipeline's list of results.

We compared the mean ranks of papers that are judged by human filterers as high purpose match to those of low purpose matches. The result showed that the new pipeline indeed was able to distinguish between the two groups of papers; low purpose matches (i.e. papers that were deemed not relevant and subsequently filtered by the judges in Study 1) were placed at significantly lower positions on the list than high purpose matches (i.e. unfiltered papers in Study 1). The mean rank for low purpose matches was 465 while for high purpose matches it was 343 (fig.~\ref{fig:mean_ranks}). This difference was significant ($t(192.49) = 3.29, p = 0.0012$. Welch's two-tailed t-test.).

\subsubsection{\new{Different model performance on finding papers that fully or partially match on purpose}}
\begin{figure}[t]
    \centering
    \includegraphics[width=\textwidth]{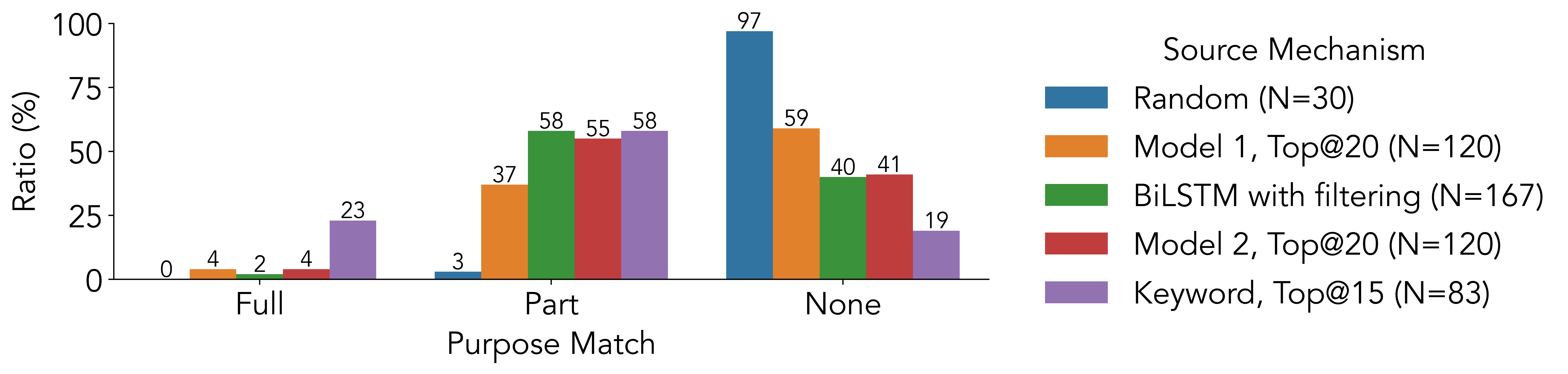}
    \vspace{-1.5em}
    \caption{Distribution of Full, Part, and None purpose matches among the five sourcing mechanisms: \textit{BiLSTM with filtering} represents the human-in-the-loop system (Study 1); \new{\textit{Model 1} represents a} system based on the BiLSTM model \new{alone,} without human-in-the-loop filtering; \textit{Model 2} \new{represents the fully automated system}; \textit{Random} \new{represents randomly sampled papers}; \textit{Keyword} \new{represents} keyword-based search (\new{Control in} Study 1). \textit{Model 2} and \textit{BiLSTM with filtering} showed a similar distribution of purpose matches, \new{and more partial purpose matches than} \textit{BiLSTM} \new{alone}. Random showed mostly no matches. The \textit{Keyword} condition resulted in the highest number of fully matched papers and the lowest number of no matches, suggesting that keyword-based search may be an effective mechanism \new{for direct search tasks, but potentially less effective for inspirational/exploratory search tasks.}}
    \label{fig:model_comparison_pm}
\end{figure}
\xhdr{Data and coding} In addition to the overall rankings reflecting human-vetted judgments we also found that the proportion of partial purpose matches was significant among the top-ranked results. We sourced top 20 results for each participant's research problem with the automated system (Model 2) using the exact queries and order used by the human-in-the-loop filterers in Study 1. We compared this to four other approaches: 1) the human-in-the-loop system in Study 1 (\textit{BiLSTM with filtering}), 2) a BiLSTM-based system excluding the human-in-the-loop from 1 (\textit{BiLSTM}), 3) randomly sampled papers (Random), and 4) a keyword-based search results\new{, which was used as control in} Study 1 (\textit{Keyword}). There were no overlapping papers between Model 2 and other conditions except for {the} Keyword \new{condition} which had 1 overlapping paper. To code the degree of purpose match, we blended the results of Model 2, biLSTM, and Random conditions. Two of the authors coded a fraction of the data together blind-to-condition (7.4\%, $N = 20/270$) following the same procedure used in Study 1. Then they independently coded the rest blind-to-condition achieving an inter-rater agreement of $\kappa = 0.80$ (substantial agreement). We resolved any disagreement through discussion on an individual case basis.
\begin{figure}[t]
\begin{minipage}{.45\textwidth}
    \centering
    \includegraphics[width=\textwidth]{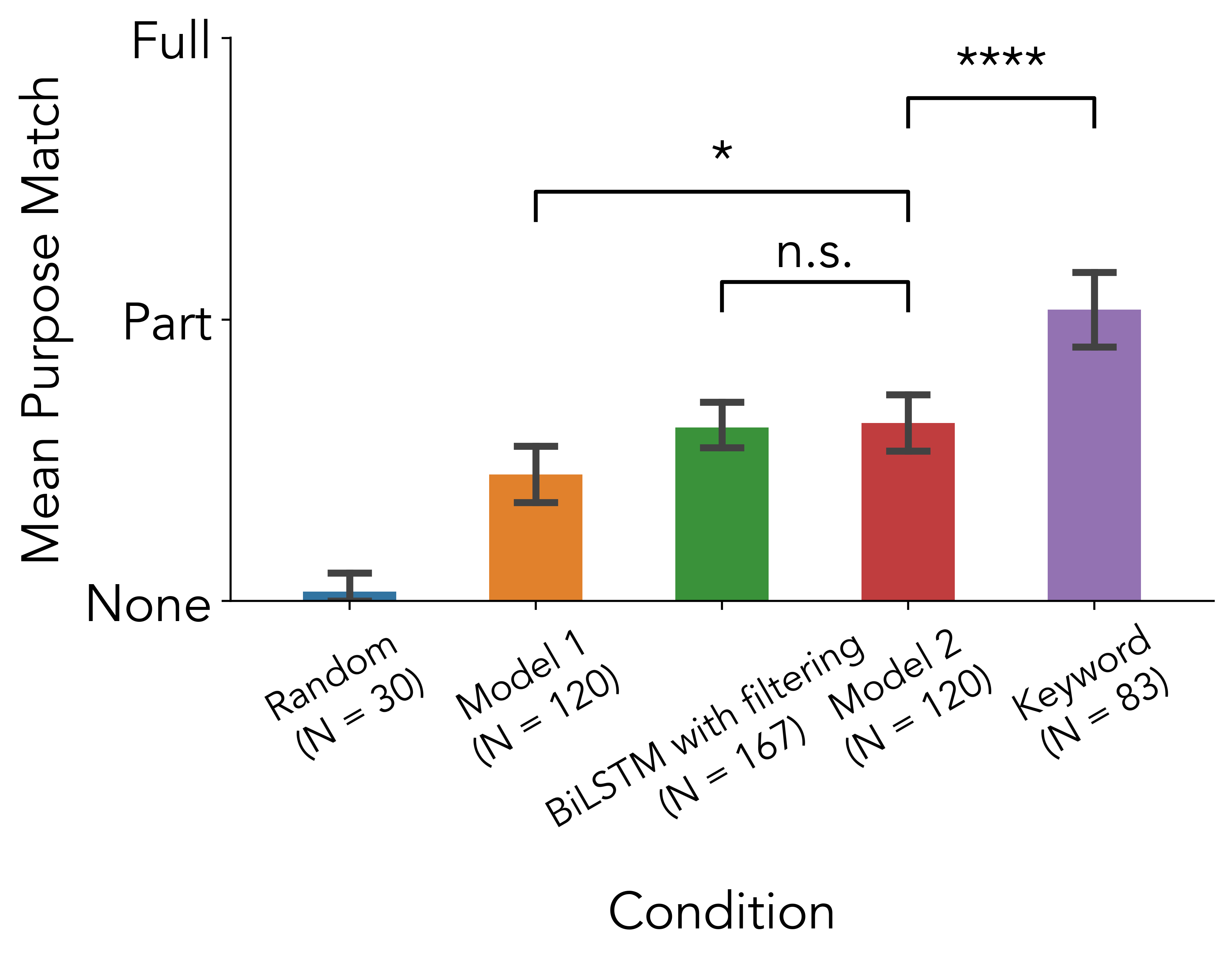}
    \vspace{-1em}
    \caption{The distribution of mean purpose match scores over different conditions (mappings: None $\mapsto$ 0, Part $\mapsto$ 1, and Full $\mapsto$ 2). The mean purpose-match score of the system backed by Model 2 (0.63, SD: 0.56) is significantly higher than that of the system used in Study 1 without the human-in-the-loop (BiLSTM, $\mu=0.45$, SD: 0.58) (Welch's two-tailed t-test, $t(237.87) = 2.49, p = 0.0135$), similar to that of the system with the human-in-the-loop (BiLSTM with filtering, $\mu=0.62$, SD: 0.52) ($t(244.65) = 0.25, p = 0.80$), and significantly lower than that of the keyword-based search (Keyword, $\mu=1.04$, SD:0.65) ($t(159.38) = -4.57, p = 0$).}
    \label{fig:model_comparison_pm_t_test}
\end{minipage}%
~\qquad  
\begin{minipage}{.45\textwidth}
  \centering
    \includegraphics[width=\textwidth]{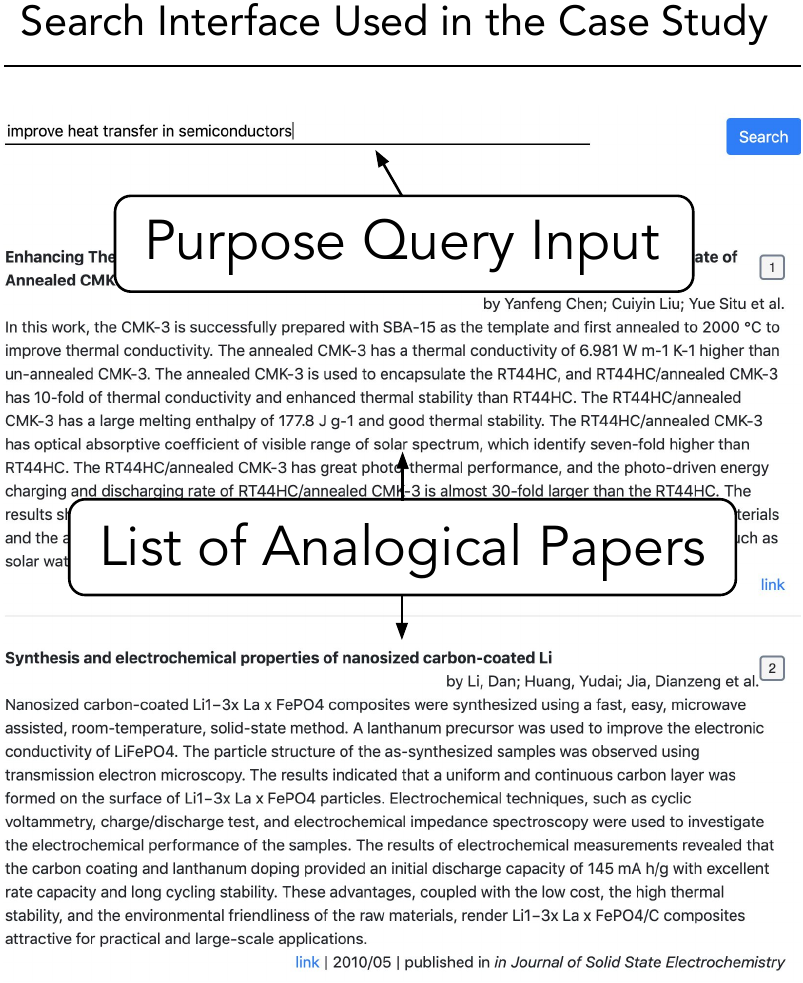}
    \vspace{-1em}
    \caption{The search interface used for case studies featured an input for query reformulation which participants used to iteratively reformulate their queries.}
    \label{fig:case_study_interface}
\end{minipage}
\end{figure}

\xhdr{Result} We found that the Model 2-based system achieved a parity with the human-in-the-loop system (Study 1) for finding purpose matches (fig.~\ref{fig:model_comparison_pm}), with more than half of the system's top 20 results judged to be partial purpose matches. In \new{contrast}, when human-in-the-loop filtering was removed from the BiLSTM-based system, the frequency of partial purpose matches \new{decreased} from 58\% to 37\% \new{while} the frequency of no matches \new{increased} from 40\% to 59\%. Random sampling resulted in mostly irrelevant results, with no alignment on purpose with the source problem. An interesting point of comparison is between the keyword-based and Model 2-based search results. Keyword search mostly outperformed Model 2-based \new{system} by finding full purpose matches at a much higher rate (23\% in keyword search vs. 4\% in the Model 2-based system), \new{with similar rates of} partial purpose matches (58\% vs. 55\%), and \new{significantly less} no purpose matches (19\% vs. 41\%). On average the purpose match score was the highest in keyword-search followed by the Model 2-based and the human-in-the-loop systems (fig.~\ref{fig:model_comparison_pm_t_test}). \new{Combined} with the results of Study 1, this suggests the complementary value of analogical search: The higher rate of full-matches in keyword-search may be good when searchers know what they are looking for, \new{such as in direct search tasks} and foraging from familiar sources of ideas. Nonetheless, because analogy papers were both deemed significantly more novel by the scientists and had little-to-no overlap with keyword-search papers, they augmented keyword-based search results with a complementary set of papers \new{that introduce useful} mistmatches in their purposes. This set of papers may open up new domains of ideas that scientists may not have been aware of, \new{and encourage} creative adaptation.

\section{Case Studies with Researchers} \label{section:case studies}
To further understand what potential interaction challenges prevent future analogical search engines from reaching their full potential, we ran case studies with 6 participants. To this end, we developed a frontend interface that includes a text input for reformulating purpose queries (fig.~\ref{fig:case_study_interface}, right). This frontend interfaced with our automated, Model 2-based backend to display a ranked list of analogical results for a given purpose query. Leveraging the fully automated search engine, we also removed the structure of Study 1 that asked participants to engage with each result they encountered, thus allowing us to observe which results researchers more naturally attend to and engage with. In sum, the design of our case studies differ from Study 1 in three aspects: 1) participants interacted with only the analogical search results ranked in the order of purpose similarity, without blended keyword-based search results; 2) participants reviewed search results returned for their queries and reformulated the queries when needed; and 3) participants looked for papers that interest them and may serve as sources of inspiration for their research problems at their own pace, without being explicitly asked to engage with each result they encounter.

The primary goal of our case studies was to identify generalizable challenges that analogical search engines may face in interactive use, thus providing us insights on how future \new{engines} may be designed and improved. Specifically, we were interested in the challenges related to 1) how researchers recognize relevance of analogical search results and 2) how researchers formulate and reformulate purpose search queries while interacting with analogical search results.

\subsection{Participants and Design}
Participants were asked to formulate purpose queries for their own research problems and interact with the results to find interesting papers. If a paper gave them a new idea relevant to their research project, they were asked to write a short project proposal in a shared Google Doc and explain how the paper helped them to come up with the idea. Interviews were conducted via Zoom and lasted for roughly an hour. Participants were paid \$20 in compensation. One participant was an assistant professor in mechanical engineering at a public R1 U.S. university and five were PhD researchers in the fields of sciences and engineering at a private R1 U.S. university. Two were senior PhD students (3rd year or above) and the rest were 2nd year or below. Disciplinary backgrounds of the participants included Chemical (2), Civil (3), and Mechanical Engineering (1). \new{We note that one participant previously took part in Study 1, whose research focus was the same in terms of its general domain. However, the participant's ideas and the specific papers of interest that led to them did not have overlap between the two studies.} Table~\ref{table:case-studies-research-problems} describes participants' research problems. 
\begin{table*}[t]
\begin{tabular}{c p{13.25cm}}
\toprule
\textbf{PID} & \textbf{Participants' Description of Research Problem} \\
\midrule
\multirow{1}{*}{1} & Improve heat pipe evaporation\\
\rowcolor{lgrey}
\multirow{1}{*}{2} & Computer simulations for fluids in nanoscale and uncovering their heat-transfer properties\\
\multirow{3}{*}{3} & Developing a model to identify complex steps in Nuclear Power Plant (NPP) operation, and understanding what task features and structures cause the complexity and how this influences the operators' performance\\
\rowcolor{lgrey}
\multirow{1}{*}{4} & Designing simulators for training bridge inspectors\\
\multirow{2}{*}{5} & Developing algorithms and extensible frameworks for detecting personal protective equipment (PPE) in construction sites to improve the safety of construction workers\\
\rowcolor{lgrey}
\multirow{1}{*}{6} & Convergence rates of optimization algorithms under multiple initial starting positions\\
\bottomrule
\end{tabular}
\caption{Case study participants' descriptions of own research problems}
\label{table:case-studies-research-problems}
\end{table*}

\textit{Apparatus: Search interface}. The improved \new{performance} of Model 2 \new{backed the} fully automated pipeline without \new{human} filtering. \new{The search interface interacting with this back-end} included a text input for reformulating purpose search queries as well as a \new{list view of search results} that showed a sorted list of papers with similar purposes (fig.~\ref{fig:case_study_interface}).

\subsection{Result}
\subsubsection{Overall impressions} 
Overall participants described their experience with the analogical search engine in a positive light (e.g., ``helps me think at a broad topic or a big picture level'' -- P2; ``find some very interesting and useful ideas, the design is also very simple, good when focusing on key areas of research'' -- P5; and ``very interested now what the future of this engine would look like'' -- P3), but a deeper look suggested that the success of ideation depended on how well searchers were able to engage with analogical results that deviate from their expectations: ``It's surprising that the engine recommends examples like these'' -- P3; ``If I input the same search queries on Google Scholar it'd not normally return these things... this search engine works in a different way'' -- P1.

\subsubsection{``Not the kind of paper I'd look for \textbf{but...}'': The challenge of early rejections} \label{subsubsection:case_studies_early_rejection}
Unlike similarity-maximizing search engines, the diversity in analogical search results can lead to premature rejection of alternative mechanism ideas. \new{One of the factors contributing to premature rejection of alternatives may be the tendency for adherence to a set of existing ideas or concepts, as studied in the literature of design fixation (e.g.,~\cite{jansson1991design}). In our study, the} participants found the variety of domains featured in search results confusing, and it sometimes prevented them from engaging with the ideas therein. For example, P3, whose research studies ways to manage or reduce task complexity for nuclear power plant operators, expected to see results similar to Google Scholar which are typically in the domains of operational and managerial sciences, but was surprised by unfamiliar domains represented in search results: ``These (\textit{distributed networked systems design} or \textit{path planning for automated robots}) are not the kinds of fields that I normally read in, if I found them elsewhere I would've probably thought they're irrelevant and skipped'' (P3). Ranging from unfamiliar terms (P1, P4, P5) to unfamiliar categories of approaches (e.g., ``Not sure what `Gauss-Newton approach for solving constrained optimization' is'' -- P6), or high-level research directions (e.g., ``this is different from my research direction, people who work on this direction might find it interesting, though'' -- P1), participants saw the diversity of results as a challenge for engagement. P1 pointed out a \new{perceived} gap between {the} expectation of least effort and the cognitive processing required when engaging with analogical ideas and adapting them:

\begin{quote}
\textit{``As I understand it, I think this search engine is trying to present papers from related but different fields to let people make connections. But people expect less friction. (The result is) something interesting but I can't directly write it into a project proposal... I think it would be challenging to make people get interested in investing time to read the papers in depth to come up with connections. I wonder what would happen if this was hosted just as an online website (instead of the study context)''} -- P1
\end{quote}

On the other hand, analogs that did get examined more deeply could ultimately lead to creative adaptation. For example, P3 mapped task scheduling among computer processes to task assignment among the nuclear power plant operators, and came up with an idea to adapt algorithmic scheduling used in real-time distributed systems to a scheduling mechanism that could be useful in her research context. Represented symbolically this process was akin to ideating what might best fill in the `?' in the relational structure [scheduling algorithm:processes in distributed systems] $\leftrightarrow$ [?:nuclear power plant operators]: ``I think the algorithms proposed in this paper could be useful for calculating the operator task execution time, the power plant system's response time, and the time margin between the execution time and the system response time... so that the next task assignment can factor in these margins and things related to workers' well-being like rest and time required between switching tasks'' (P3).

Participants seemed to recognize a small number of core relations as kernel for creative adaptation. In the example of P3, \textit{scheduling processes} in the distributed systems paper piqued her interest and led her to connect them with similar concepts in the literature she was already familiar with: ``You need to make that connection... I saw parallels between (distributed systems domain) concepts like [scheduling] and [tasks] and [scheduling tasks for the operators]'' (P3). Similarly, P5 recognized a similarity between [monitoring people's performance] in fitness training and [monitoring whether construction workers are wearing personal protective equipment] in construction sites. He then adapted the idea of tracking heat emission in the fitness context to his {own}: ``I like the idea of [monitoring heat emissions] in fitness training... maybe I can also detect heat emissions from construction workers to see if they are wearing the safety vests or masks while also monitoring the site conditions and worker efficiency. It also gives me an idea to monitor the $\text{CO}_2$ emissions from workers so as to improve the robustness of detection'' (P5). In this case, \textit{monitoring} and the \textit{physical nature} of the activities involved helped P5 see the connection useful for creatively adapting the source idea.
\begin{wrapfigure}{R}{.35\textwidth}
    \begin{center}
    \includegraphics[width=.35\textwidth]{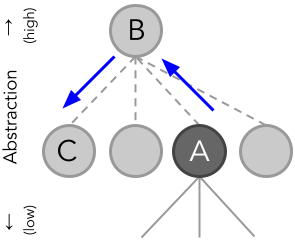}
    \vspace{-1em}
    \caption{Diagram showing different abstraction levels of purposes and their relations. Node \new{\cirnum{A}} \new{corresponds to a more specific query than its} higher-level representation, denoted as \new{\cirnum{B}}. Similarly, node \new{\cirnum{C}} represents a more specific purpose representation \new{of} 
    \new{\cirnum{A}}, accessible via the \new{\cirnum{A}} $\underset{\text{\new{abstraction}}}{\rightarrow}$ \new{\cirnum{B}} $\underset{\text{\new{specification}}}{\rightarrow}$ \new{\cirnum{C}} path.}
    \label{fig:purpose_hierarchy}
    \end{center}
\end{wrapfigure}
\subsubsection{``I don't know what to type in'': The challenge of query (re-)formulation} \label{subsubsection:case_studies_query_reformulation}
Another challenge participants faced was that they were not used to formulating their search queries in terms of high level purposes of their research. On average participants entered 5.2 queries (Min: 1, Max: 18, SD: 5.87), 87\% (27) of which were in the form of a single noun phrase (e.g., ``heat pipe evaporation,'' -- P1, ``task complexity'' -- P3, ``theoretical optimization convergence for non-convex functions'' -- P6) or a comma-separated set of multiple noun phrases (e.g., ``heat transfer, nanoscale, fluid'' -- P2) that represented specific aspects related to research purposes rather than the core purposes themselves. For example, the purpose of `heat pipe evaporation' may be to transfer heat, and the purpose of searching for `theoretical optimization convergence for...' may be to \new{detect} when optimization \new{converges or diverges, or to effectively sample unknown (non-convex) distributions.}

One of the reasons why participants formulated search queries in this way may be \new{wrongly assuming that the search engine used} keyword matching to find results. For example, extensive \new{prior} experience with search engines that highlight matching keywords in abstracts (e.g., Google Scholar) in response to users' search queries can reinforce such assumptions among {the users}. In addition, participants' domain knowledge useful for judging which of the returned papers are relevant may have led them to notice a set of keywords the inclusion of which strongly signifies the relevance of a paper. In contrast, the analogical search results often seemed to not feature such directly similar terms and this contributed to the difficulty of judging whether a result is relevant and how: ``I find these papers not very related to my search query at first. It'd be better if you can use some graph or some pictures to indicate how these papers can relate to my keywords'' (P5); ``I'd not consider... (because) they are totally different, right? They look irrelevant... until I think about it I can realize that it's useful. But if you give me the paper, at first I don't realize that'' (P3).

\new{While} it may not feel as compelling or natural to participants, formulating and abstracting queries at a high level may lead to searching more distant results that are analogous at a higher level. For example, by querying ``detect personal protective equipment'' instead of ``personal protective equipment construction,'' P5 found novel mechanisms of detection, such as general image segmentation algorithms or an approach to monitoring heat in the context of fitness training not specific to construction sites and personal protective equipment but nonetheless useful for creative adaptation. Querying ``scheduling tasks'' instead of ``task complexity'' for P3 resulted in finding scheduling algorithms in distributed computer systems that otherwise P3 would not have encountered, while ``assigning tasks'' led to novel auction mechanisms which made her think about a system in which each power plant operator can bid for a task as opposed to being assigned one. \new{Schematically, fig.~\ref{fig:purpose_hierarchy} shows how} formulating queries at a higher level of abstraction \new{than specifying the problem context in full details} (\new{\cirnum{A} $\rightarrow$ \cirnum{B}}) may lead to \new{discovering} novel mechanisms that are relevant \new{at the high level of abstraction, and in more} distant ways \new{from the original problem formulation} (\new{\cirnum{B} $\rightarrow$ \cirnum{C}}).

\section{Design Implications} \label{section:design implications}
From both the case studies' and Study 1's participants' reflection on the challenges of interacting with analogical search results, common \new{themes} emerged. Here we present three design implications for future analogical search systems synthesized from these results. We use subscripts to denote which study participants participated in when appropriate.

\subsection{Support purpose representation at different levels of abstraction}
Analogical search engines should support users to formulate their purpose queries at different levels of abstraction. Additionally the search engine may prompt users to consider abstracting or specifying their purpose queries in the first place, and explain how it might help bring new insights into their problems. As seen in the case studies (Section~\ref{subsubsection:case_studies_query_reformulation}), scientists recognized their purpose queries may be represented at multiple levels, but prior experiences with similarity maximizing search engines may also have anchored them around pre-existing rigid formulation of purposes. Prompting users to consider their research problems at multiple levels may work against this rigidity, and providing candidate suggestions at varying levels may further reduce the cognitive demand. Moving up on the hierarchy to abstract purpose queries may be possible through removing parts of the query words that correspond to specific constraints, or by replacing them with more general descriptions. For example two participants of Study 1 had an identical purpose representation at a high level (``facilitate heat transfer'') despite the differences in materialistic phases targeted in each purpose: solid material and semiconductors for \p{P1}{Study 1} and liquid thin films for \p{P3}{Study 1}. 

\new{Furthermore}, we also observed that looking for only the exact match of a purpose can lead to missed opportunities. For example, although ``fins represent a different idea for transferring the heat'' and ``they (fins) don't match in terms of the scale -- macro, not nano,'' it nevertheless made \p{P1}{Study 1} wonder ``what if we could design nanoscale wall structures that act like fins that convert heat to mechanical energy?''. A corrollary to th{is} observation is that sometimes the superpositions of misalignment with just the right amount can lead to interesting results. For \p{P4}{Study 1}, a paper presenting experimental techniques for piezoelectric properties was interesting despite its misalignment such as \new{[}\textit{simulation}-based\new{]} (source) $\nleftrightarrow$ \new{[}\textit{experimental}\new{]} (analog) and \new{[}\textit{dielectric properties}\new{]} (source) $\nleftrightarrow$ \new{[}\textit{piezoelectric properties}\new{]} (analog): ``Though it's an experimental study, it's very close in terms of the material and phenomenon so likely to be helpful. Because we might be able to pick up some trends like, if we increased the temperature, the dielectric response gets stronger or weaker, inferred from the experimental piezoelectric responses, which can then be used to corroborate simulation results or help configure its parameters'' (\p{P4}{Study 1}). However, too much deviation seemed detrimental to its potential for inspiration: ``\new{[}Molecular dynamic simulation\new{]} is the same tool, but (this paper studies) \new{[}thermal\new{]} (not \new{[}dielectric\new{]}) properties on \new{[}polymer composites\new{]}... \new{[}polymer composites\new{]} are harder to model'' (\p{P4}{Study 1}). In sum, analogical search engines should support not only the capability to `narrow it down' with specific constraints, but also ways to relax them to broaden the search space when suitable, thus making feasible the sweet spot between too little (i.e. similarity maximization and trivial matches) and too much deviation (i.e. critical misalignment and unusable analogs).

\subsection{Support iterative steering from critical misalignment and towards generative misalignment}
Analogical search engines should recognize that important constraints may be discovered by users only after seeing misaligned analogs, and support this discovery process by presenting effective examples of misalignment to users. Analogs that deviate on some aspects of the source problem but preserve important relations may be particularly conducive to analogical inspiration that opens up not just individual solutions, but entirely new domains of solutions. However at the same time scientists also found it challenging to know how to come up with effective search queries because combinations of misalignment can sometimes lead to an unintended intersection of domains: ``I feel like I'm tricking the machine because \new{[}thin film\new{]} is often used with \new{[}solids\new{]}, and the term \new{[}pressure\new{]} also appears a lot in \new{[}manufacturing\new{]}... so combining them gives a subset of papers concerned with heat transfer in solid materials and in manufacturing'' (\p{P3}{Study 1}); ``on Google Scholar also, I get a lot of polymer strings and get (irrelevant) results like \textit{we use an \new{[}electric\new{]} device to study \new{[}vibration and stress\new{]} of \new{[}polymers\new{]}}... the machine is picking up \new{[}electric\new{]} and \new{[}properties\new{]} such as vibration and stress in the context of studying polymers but what I really want is \new{[}electric properties\new{]} of \new{[}polymers\new{]} \textit{not} \new{[}electronic devices\new{]} to study the \new{[}mechanical properties\new{]} of \new{[}polymers\new{]}'' (\p{P4}{Study 1}). Nonetheless, seeing misaligned analogs can be an effective way of reasoning about salient constraints and reflecting on hidden assumptions. For example, while evaluating papers about designing microelectrode arrays, \p{P6}{Study 1} said:
\textit{``Now I think about this (result), I assumed a lot of things when typing that search query... though impedance and topology are my main focus in microelectrode arrays, the coating, size, interface between a cell membrane and electrodes/sensors, biocompatibility, softness of electrodes, fabrication process, material of the platform: silicon or polymer or graphene, form factor: attaching electrodes to a shank-like structure or a broom-like structure, degree of invasiveness, are all part of the possible areas of research and it makes sense that they showed up -- there is no way the machine would have known that from my query.''}
This excerpt illustrates how knowing what the necessary specifications are and which constraints need to be abstracted to cast a wide-enough net to catch interesting ideas appeared to be a difficult task for scientists, especially when they had to recall important attributes rather than simply recognize them from examples of misalignment. Prior work in cognitive sciences also show how dissimilarity associated with various factors in analogical mappings~\cite{gentner2012analogical} can pressure working memory~\cite{waltz2000role}, increase cognitive load~\cite{sweller1990cognitive}, and increases response time taken to produce correct mappings for analogy problems~\cite{keane1994constraints}. Therefore, analogical search engines should help to reduce the cognitive effort required in the process, for example by proactively retrieving results that are `usefully' misaligned such that searchers can better recognize (as opposed to having to recall) salient constraints and refine their problem representation. This process is deeply exploratory~\cite{white2009exploratory,russell1993cost,CiteSense} in nature, and suggest the importance of both providing end-users a sense of progress over time~\cite{perfect_search_engine} as well as adequate feedback mechanisms for the machine to adjust according to the changing end-user search intent~\cite{schnabel2020doesn,schnabel2019shaping,kelly2003implicit}. For example, while the machine may `correctly' recognize a significant anaogical relevance at a higher level of purpose representation and recommend \textit{macro}-scale mechanisms to a scientist who studies \textit{nano}-scale phenomena (\p{P1}{Study 1}) or solid and semiconductor-based cooling mechanisms to a scientist in liquid and evaporative cooling systems (\p{P3}{Study 1}), these analogs may be critically misaligned on the specific constraints of the problem (i.e. the scale or materialistic phase) and thus considered by end-users as useless and even harmful.

\subsection{\new{Support reflection and explanation of analogical relevance}}
\new{Throughout the process of analogical search, human-AI coordination is critical for success, and an important aspect is how deeply the human users can reflect on the retrieved analogs~\cite{hao2016reflection} and recognize how different notions of relevance may exist for their own problem context, despite potential dissimilarity on the surface.}
\new{Looking at previous examples of the tools and techniques developed for targeted reflection support may be useful to this end. For example, ImageCascade~\cite{koch2020imagesense} provides intelligent support such as automatically generated mood-boards and semantic labels for groups of images to help designers communicate their design intent to others. Another system, Card Mapper, visualizes relative co-occurrences of design concepts using proximity in the design space~\cite{darzentas2019card}. Similarly representing the space of analogical ideas using spatial encoding of similarity between two analogs, or designing information that supports getting a sense of the space of search results --- e.g., semantic category labels similar to ImageCascade's or the distribution of the domains that analogs are pulled from --- may be an avenue for fruitful future research.}
The explanation of relevance is also important especially when there is a risk of early rejection (\S\ref{subsubsection:case_studies_early_rejection}). Using examples from the case studies, one approach to explaining relevance might be to surface a small number of core common features between an analog and a problem query. Such common features were {considered} useful {by} scientists for making analogical connections, and they could creatively adapt them for their own research problem context. When common features are not directly retrieved, generation of more elaborate explanations may be required. \new{We refer to~\cite{bansal2021does,smith2020no,buccinca2021trust,kang2022you} for those interested in future design considerations of automatically generated recommendation explanation.}
\new{Further complementing the direct explanation of relevance approach,} techniques such as prompting or reminding \new{the searchers of previously rejected or overlooked ideas} may also trigger deeper reflection {and} delay premature rejection of the ideas based solely on their surface dissimilarity. Participants from both studies commented that the critical first step towards analogical inspiration may be raising {similarly} enough attention and interest above the initial `hump' of cognitive demand. Gentle reminders (e.g., ``Ask me {later} if this would be interesting and also provide a list of items'' -- \p{P1}{Case Studies}) or resurfacing previously rejected papers in light of new information (\p{P1}{Case Studies}, \p{P3}{Case Studies}) may help with users cross this barrier.

\section{Discussion}
\subsection{\new{Summary of contribution}}
With the exponential growth of research output and the deepening specialization within different fields, encouraging analogical inspiration for scientific innovation that connects distant domains becomes ever more challenging. Our human-in-the-loop and fully automated analogical search engines represent an approach for supporting such analogical inspirations for challenging scientific problems. We have demonstrated in Study 1 that our human-in-the-loop system finds novel results that participants would be unlikely to encounter from keyword-based search, and that these results lead to high levels of creative adaptation. Through a mediation analysis we also showed that this success was driven by the analogical search engine's ability to find \textit{partial} purpose matches (e.g., matching at the high-level purpose but differs at {the} low-level {details}). We saw the nuanced effects of partial purpose alignment on the results' goodness as analogs for inspiration. Through qualitative observations, we described how certain attributes of analogical mapping were perceived as more salient by participants, and that mismatches on them can have either a positive (i.e. generative insights) or a negative impact (i.e. critical misalignment) on creative adaptation. In contrast, keyword-based search resulted in more \textit{full} purpose matches and a higher level of direct application. The value of keyword-based search and analogy-based search thus may complement each other, while keyword-based search can help researchers find `exactly that', analogy-based search can help researchers to switch from a preservative mode (i.e. aiming to find results with maximal similarity to the query) to a generative mode (i.e. aiming to find analogs that are relevant despite the surface dissimilarity) of searching, and ultimately lead them to recognize unusual relations and come up with ways to creatively adapt existing ideas for novel domains.

We also demonstrated how improving the sequence-to-sequence purpose and mechanism identification model can remove the human-in-the-loop but maintain a similar level of accuracy on purpose-match by human judges. This improvement enabled us to develop a fully automated analogical search system to use as a probe to study searchers' more natural interaction with analogical results. Through a series of evaluation we first show that our automated analogical search pipeline can emulate human judgment of purpose match and that it finds partial purpose matches in top ranked results with a similar rate compared to the human-in-the-loop system {used} in Study 1. Then through case studies we find generalizable challenges that future analogical search engines may face, such as early rejection of alternative mechanism ideas and the difficulty of abstracting and representing purposes at the right level. From our studies we synthesize design implications for future analogical search engines, such as supporting purpose representations at different levels of abstraction, supporting the iterative process of steering away from critically misaligned analogs and towards a fertile land of generative misalignment, and providing explanations on why certain analogical search results may be relevant. We envision that future studies will shed light on deeper cognitive sources of the challenges identified here. A fruitful avenue of research may be studying how the dual processing theory~\cite{wason1974dual,kahneman2011thinking} underlies or interacts with analogical search interaction. Studying also how simplification heuristics~\cite{mintzberg1976structure} may \new{negatively bias} interact{ion} with analogical results \new{and how it may be} reduced for expert user populations may be an interesting future direction~\cite{carol2007slowing,lambe2016dual}.

\subsection{\new{Limitations and future work}}
\subsubsection{\new{Experimental design and improving its validity}} \label{subsubsection:exp_validity}
Our findings have several limitations. First the design of our studies may be improved to increase the experimental validity. We believe that our coders of the ideation outcomes had a reasonable understanding of participants’ research context from descriptions of current and past research topics, think-alouds with 45 papers, and end-of-experiment discussions, and that the procedure of coding reduced potential biases (e.g., the coders were blind to experimental conditions, relied on participants’ statements of novelty and distance). Despite this, it is possible that they judged ideas differently from domain experts, for example coding more or fewer ideas as creative adaptation, or pre-filtering useful ideas in the human-in-the-loop stage. In addition, other quality dimensions such as potential for impact or domain-expert-judged idea quality are largely inaccessible within the studies presented here. Future research may improve on these limitations by iterating on the experimental design, collecting data for triangulating the results and capturing \new{other} quality dimension of the generated ideas.

\new{Additionally, future work may add ablation studies to quantify the effects of human filtering in Study 1 on the ideation outcome as well as sensitivity studies to relate how much the increased token-level classification performance of trained models may reduce the burden of human filtering.}
\new{Furthermore, additional experiments with baselines other than keyword-based search using the whole abstract will help pinpoint the potential advantages of representing and matching papers using extracted purposes and mechanisms. For example, Chan et al.~\cite{chan2018solvent} found that embedding all words from an abstract (using GloVe embeddings) resulted in retrieval of fewer analogical items than when extracted purposes and mechanisms were used. Replicating this result with additional approaches such as contextualized word embeddings and pre-trained language models (e.g., ELMo~\cite{elmo}, BERT~\cite{bert}, and SciBERT~\cite{scibert}) will be valuable.}

\subsubsection{\new{Potential sampling bias}}
\new{The sampling strategy in Study 1 was purposefully unbalanced, where analogical papers were sampled twice as much as keyword papers to ensure participants' exposure to sufficiently diverse results. This was crucial for uncovering potential benefits and challenges of our analogical search engine and investigating its viability. This ratio was chosen purposefully, to balance the statistical power for detecting potentially significant differences between the conditions, while also limiting the number of papers that each participant had to review. Given the cognitive burden of reviewing a paper while thinking aloud, we decided on 45 in total with the 2:1 ratio to fit the practical time limits of interviews. However, this may have led to unanticipated effects on ideation outcomes despite having accounted for the difference in sample sizes by measuring the outcomes in ratios. For example, when the results were combined into a single blinded list, the over-representation of analogical results over more purpose-aligned keyword results may have shifted the users' overall perceived value of the list to be more or less positive. Users' perception of diverse results may have been further affected by their relative over-representation. For example, increased cognitive load for processing analogical mapping~\cite{halford1992analogical,sweller1990cognitive,halford1998processing} may suggest that results that fully match on the purpose search query may have been perceived even more favorably than analogical results, due to a negative spill over effect from the rest of the papers in the list, which were less likely matched on the purpose. Investigating whether such factors led to compounding effects beyond our ratio-based measures of usefulness remains an open question for future work.}

\subsubsection{\new{Controlling the diversity of search results}} \label{subsubsection:control_diversity}
\new{Our work is also limited by the lack of controllability in sampling the search results beyond purpose similarity. As described in \S\ref{subsubsection:stage2-overview}, from pilot tests in our corpus we discovered that even close purpose matches of scientific papers already had high variance in terms of the mechanisms they proposed which allowed us to focus our approach to sampling based solely on purpose similarity. 
The simplicity of this approach also means fewer hyper parameters in the sampling mechanism compared to other approaches~\cite{hope_kdd17,hope2021scaling}. However, all the approaches including this work thus far lacked a mechanism for explicitly controlling the diversity in retrieved search results which remains a fruitful avenue for future work. For example, prior research has uncovered the nuanced effects of distance (e.g., near vs. far sources of inspiration~\cite{chan2017semantically,siangliulue2015providing}), suggesting the benefit of targeting analogs at different distance from the source problem for the right context. Future research may also uncover further complexities in the relationship between novelty and purpose-match. The result of our mediation analysis (Table~\ref{table:mediation}) showed that the novelty of content among the search results in Study 1 was not a significant factor to the same extent that the three levels of purpose match was. However, the relationship between novelty and purpose match may be more complex than the levels of manipulation presented in this work. For example,~\cite{diedrich2015creative} suggested a greater importance of novelty than usefulness for predicting creativity scores. Future work may design mechanisms to manipulate the variance in content novelty and alignment in the purpose-mechanism schema to uncover dynamics between the two that go beyond the results from mediation analyses presented here (\S\ref{section:purpose-match-mediation}). Furthermore,} challenges \new{with the abstraction of purposes remain open, for example} how core versus peripheral attributes of research purposes may be identified, and how they may be selectively matched at a specific level of the conceptual hierarchy. \new{Finally, not all query formulations are created equal in terms of their suitability for analogical search. We observed in the case studies that participants wanted to express different query intent via reformulation (\S\ref{subsubsection:case_studies_query_reformulation}). While participants could reformulate their search queries and examine the returned results from our analogical search engine in real-time, it was unclear whether and how specific query formulations may lead to more or less diverse results, and how subsequent queries may be updated after reviewing them. As such, systems that assist users in the potentially tedious process of query reformulation~\cite{white2010predicting} (for example, by way of automatic query expansion~\cite{carpineto2012survey}) in the context of analogical search will be important.}

\subsubsection{\new{Studying the effect of larger context of scientific innovation on analogical innovation}}
Due to our focus on ideation outcomes, our results do not explain how these ideas may be integrated, developed, and shared across the research communities. Studying the lifetime of ideas that goes beyond their inception will deepen our understanding of the factors that currently make analogical innovation such a rare event in sciences (for example, Hofstra et al. suggested that more semantically distant conceptual combinations receive far less uptake~\cite{diversity_innovation_paradox}). Through interviewing our study participants and other colleagues in academia we found emerging structures related to this challenge. Our interviews inform{ed} us that in general the context in which a scientist exists -- such as the scientist's role in a project, the maturity of a project, and the broader academic culture -- can ultimately change how they interact with and seek analogical inspirations. For example a third-year PhD student studying chemical engineering commented ``In the current stage of my project it's more about parameter-tuning -- running multiple experiments at once and comparing which configuration works the best... If I were a first year PhD student maybe I would be in a broader field and exploration.'' In contrast, a PhD in biology who recently defended noted that ``analogical inspirations would perhaps be more useful if you're looking for a postdoc or a faculty position.''

In addition, the underlying career incentive structures in academia may also affect researchers' perception of and openness to analogical inspirations. One of the study participants commented ``since I'm already a third year PhD student and my project is further along and more firmed up, I'm not really looking for really far inspirations... first we push the specific way we have in mind with many iterations on the experiments until, say, publication.'' In addition to the career-wise incentives there are other types of competitive resourcefulness (e.g., social resources such as the advisors' and colleagues' expertise that participants can easily tap into; physical and other forms resources such as tangible artifacts like previously developed code packages or experimental processes and setups). These factors can influence scientists' perception of their advantage and lead them to interpret analogical inspirations as more or less useful, feasible, and directly applicable to their research. This observation is further suggested by survey results that asked our participants: ``\textit{Could this paper be useful to you?},'' their ratings were significantly higher for keyword papers than analogy papers despite them having come up with creative adaptation ideas more often with analogy papers. \new{Therefore,} future work \new{that studies incentive structures, the quality of ideation outcome, their feasibility, the differences in research context e.g., frames of research contribution such as discovery-oriented vs. novel system development-oriented, and taking a longitudinal observation of the variation in such factors will} add \new{a significant} depth to our understanding.

\section{Conclusion}
In this paper we present our novel human-in-the-loop and fully automated analogical search engines for scientific articles. Through a series of evaluations we found that analogous papers that our systems retrieved were novel and useful for sparking creative adaptation ideas. However, significant work is needed to continue this trajectory, including additional understanding of the context and incentives of scientists as well as advances in the data pipeline and interaction methods beyond those described here for a system to maximize its real-world impact. 

We imagine a future in which scholars and designers could find inspirations based on deep analogical similarity that can drive innovation across fields. We hope this work will encourage scientists, designers, and system builders to collaborate across disciplinary boundaries to accelerate the rate of scientific innovation.


\begin{acks}
We thank our study participants for their valuable insights and feedback. This work was supported by Center for Knowledge Acceleration, National Science Foundation (FW-HTF-RL, grant no. 1928631; IIS, grant no. 1816242; SHF, grant no.1814826), the European Research Council (ERC) under the European Union's Horizon 2020 research and innovation programme (grant no. 852686, SIAM) and NSF-BSF grant no. 2017741. This work is also based upon work supported by the Google Cloud Research Credits program with the award GCP19980904.
\end{acks}

\bibliographystyle{ACM-Reference-Format}
\bibliography{main.bib}


\begin{thebibliography}{119}


\ifx \showCODEN    \undefined \def \showCODEN     #1{\unskip}     \fi
\ifx \showDOI      \undefined \def \showDOI       #1{#1}\fi
\ifx \showISBNx    \undefined \def \showISBNx     #1{\unskip}     \fi
\ifx \showISBNxiii \undefined \def \showISBNxiii  #1{\unskip}     \fi
\ifx \showISSN     \undefined \def \showISSN      #1{\unskip}     \fi
\ifx \showLCCN     \undefined \def \showLCCN      #1{\unskip}     \fi
\ifx \shownote     \undefined \def \shownote      #1{#1}          \fi
\ifx \showarticletitle \undefined \def \showarticletitle #1{#1}   \fi
\ifx \showURL      \undefined \def \showURL       {\relax}        \fi
\providecommand\bibfield[2]{#2}
\providecommand\bibinfo[2]{#2}
\providecommand\natexlab[1]{#1}
\providecommand\showeprint[2][]{arXiv:#2}

\bibitem[\protect\citeauthoryear{??}{ran}{[n.d.]}]%
        {random_projection_wiki}
 \bibinfo{year}{[n.d.]}\natexlab{}.
\newblock \bibinfo{title}{Random projection in Locality-sensitive hashing}.
\newblock
\newblock
\urldef\tempurl%
\url{https://en.wikipedia.org/wiki/Locality-sensitive_hashing#Random_projection}
\showURL{%
\tempurl}
\newblock
\shownote{[Online; accessed 23-Jan-2022].}


\bibitem[\protect\citeauthoryear{??}{ann}{[n.d.]}]%
        {annoy_readme}
 \bibinfo{year}{[n.d.]}\natexlab{}.
\newblock \bibinfo{title}{\textsc{Annoy}: How it works}.
\newblock
\newblock
\urldef\tempurl%
\url{https://github.com/spotify/annoy#how-does-it-work}
\showURL{%
\tempurl}
\newblock
\shownote{[Online; accessed 23-Jan-2022].}


\bibitem[\protect\citeauthoryear{Ashley}{Ashley}{1991}]%
        {ashley1991reasoning}
\bibfield{author}{\bibinfo{person}{Kevin~D Ashley}.}
  \bibinfo{year}{1991}\natexlab{}.
\newblock \showarticletitle{Reasoning with cases and hypotheticals in HYPO}.
\newblock \bibinfo{journal}{\emph{International journal of man-machine
  studies}} \bibinfo{volume}{34}, \bibinfo{number}{6} (\bibinfo{year}{1991}),
  \bibinfo{pages}{753--796}.
\newblock


\bibitem[\protect\citeauthoryear{Bachrach, Finkelstein, Gilad-Bachrach, Katzir,
  Koenigstein, Nice, and Paquet}{Bachrach et~al\mbox{.}}{2014}]%
        {bachrach2014}
\bibfield{author}{\bibinfo{person}{Yoram Bachrach}, \bibinfo{person}{Yehuda
  Finkelstein}, \bibinfo{person}{Ran Gilad-Bachrach}, \bibinfo{person}{Liran
  Katzir}, \bibinfo{person}{Noam Koenigstein}, \bibinfo{person}{Nir Nice},
  {and} \bibinfo{person}{Ulrich Paquet}.} \bibinfo{year}{2014}\natexlab{}.
\newblock \showarticletitle{Speeding up the Xbox Recommender System Using a
  Euclidean Transformation for Inner-Product Spaces}. In
  \bibinfo{booktitle}{\emph{Proceedings of the 8th ACM Conference on
  Recommender Systems}} (Foster City, Silicon Valley, California, USA)
  \emph{(\bibinfo{series}{RecSys '14})}. \bibinfo{publisher}{Association for
  Computing Machinery}, \bibinfo{address}{New York, NY, USA},
  \bibinfo{pages}{257–264}.
\newblock
\showISBNx{9781450326681}
\urldef\tempurl%
\url{https://doi.org/10.1145/2645710.2645741}
\showDOI{\tempurl}


\bibitem[\protect\citeauthoryear{Bahdanau, Cho, and Bengio}{Bahdanau
  et~al\mbox{.}}{2016}]%
        {Seq2SeqICLR}
\bibfield{author}{\bibinfo{person}{Dzmitry Bahdanau},
  \bibinfo{person}{Kyunghyun Cho}, {and} \bibinfo{person}{Yoshua Bengio}.}
  \bibinfo{year}{2016}\natexlab{}.
\newblock \bibinfo{title}{Neural Machine Translation by Jointly Learning to
  Align and Translate}.
\newblock
\newblock
\showeprint[arxiv]{1409.0473}~[cs.CL]


\bibitem[\protect\citeauthoryear{Bansal, Wu, Zhou, Fok, Nushi, Kamar, Ribeiro,
  and Weld}{Bansal et~al\mbox{.}}{2021}]%
        {bansal2021does}
\bibfield{author}{\bibinfo{person}{Gagan Bansal}, \bibinfo{person}{Tongshuang
  Wu}, \bibinfo{person}{Joyce Zhou}, \bibinfo{person}{Raymond Fok},
  \bibinfo{person}{Besmira Nushi}, \bibinfo{person}{Ece Kamar},
  \bibinfo{person}{Marco~Tulio Ribeiro}, {and} \bibinfo{person}{Daniel Weld}.}
  \bibinfo{year}{2021}\natexlab{}.
\newblock \showarticletitle{Does the whole exceed its parts? the effect of ai
  explanations on complementary team performance}. In
  \bibinfo{booktitle}{\emph{Proceedings of the 2021 CHI Conference on Human
  Factors in Computing Systems}}. \bibinfo{pages}{1--16}.
\newblock


\bibitem[\protect\citeauthoryear{Beltagy, Lo, and Cohan}{Beltagy
  et~al\mbox{.}}{2019}]%
        {scibert}
\bibfield{author}{\bibinfo{person}{Iz Beltagy}, \bibinfo{person}{Kyle Lo},
  {and} \bibinfo{person}{Arman Cohan}.} \bibinfo{year}{2019}\natexlab{}.
\newblock \showarticletitle{{S}ci{BERT}: A Pretrained Language Model for
  Scientific Text}. In \bibinfo{booktitle}{\emph{Proceedings of the 2019
  Conference on Empirical Methods in Natural Language Processing and the 9th
  International Joint Conference on Natural Language Processing
  (EMNLP-IJCNLP)}}. \bibinfo{publisher}{Association for Computational
  Linguistics}, \bibinfo{address}{Hong Kong, China},
  \bibinfo{pages}{3615--3620}.
\newblock
\urldef\tempurl%
\url{https://doi.org/10.18653/v1/D19-1371}
\showDOI{\tempurl}


\bibitem[\protect\citeauthoryear{Bengio, Ducharme, Vincent, and Janvin}{Bengio
  et~al\mbox{.}}{2003}]%
        {bengio2003neural}
\bibfield{author}{\bibinfo{person}{Yoshua Bengio}, \bibinfo{person}{R{\'e}jean
  Ducharme}, \bibinfo{person}{Pascal Vincent}, {and} \bibinfo{person}{Christian
  Janvin}.} \bibinfo{year}{2003}\natexlab{}.
\newblock \showarticletitle{A neural probabilistic language model}.
\newblock \bibinfo{journal}{\emph{The journal of machine learning research}}
  \bibinfo{volume}{3} (\bibinfo{year}{2003}), \bibinfo{pages}{1137--1155}.
\newblock


\bibitem[\protect\citeauthoryear{Berg}{Berg}{2014}]%
        {bergPrimalMarkHow2014}
\bibfield{author}{\bibinfo{person}{Justin~M. Berg}.}
  \bibinfo{year}{2014}\natexlab{}.
\newblock \showarticletitle{The primal mark: {How} the beginning shapes the end
  in the development of creative ideas}.
\newblock \bibinfo{journal}{\emph{Organizational Behavior and Human Decision
  Processes}} \bibinfo{volume}{125}, \bibinfo{number}{1}
  (\bibinfo{year}{2014}), \bibinfo{pages}{1--17}.
\newblock
\showISSN{07495978}
\urldef\tempurl%
\url{https://doi.org/10.1016/j.obhdp.2014.06.001}
\showDOI{\tempurl}


\bibitem[\protect\citeauthoryear{Bird and Loper}{Bird and Loper}{2004}]%
        {nltk}
\bibfield{author}{\bibinfo{person}{Steven Bird} {and} \bibinfo{person}{Edward
  Loper}.} \bibinfo{year}{2004}\natexlab{}.
\newblock \showarticletitle{{NLTK}: The Natural Language Toolkit}. In
  \bibinfo{booktitle}{\emph{Proceedings of the {ACL} Interactive Poster and
  Demonstration Sessions}}. \bibinfo{publisher}{Association for Computational
  Linguistics}, \bibinfo{address}{Barcelona, Spain}, \bibinfo{pages}{214--217}.
\newblock
\urldef\tempurl%
\url{https://www.aclweb.org/anthology/P04-3031}
\showURL{%
\tempurl}


\bibitem[\protect\citeauthoryear{Bojanowski, Grave, Joulin, and
  Mikolov}{Bojanowski et~al\mbox{.}}{2017}]%
        {bojanowski2017enriching_subword_info}
\bibfield{author}{\bibinfo{person}{Piotr Bojanowski}, \bibinfo{person}{Edouard
  Grave}, \bibinfo{person}{Armand Joulin}, {and} \bibinfo{person}{Tomas
  Mikolov}.} \bibinfo{year}{2017}\natexlab{}.
\newblock \showarticletitle{Enriching Word Vectors with Subword Information}.
\newblock \bibinfo{journal}{\emph{Transactions of the Association for
  Computational Linguistics}}  \bibinfo{volume}{5} (\bibinfo{year}{2017}),
  \bibinfo{pages}{135--146}.
\newblock
\urldef\tempurl%
\url{https://doi.org/10.1162/tacl_a_00051}
\showDOI{\tempurl}


\bibitem[\protect\citeauthoryear{Bornmann and Mutz}{Bornmann and Mutz}{2015}]%
        {bornmann2015growth}
\bibfield{author}{\bibinfo{person}{Lutz Bornmann} {and}
  \bibinfo{person}{R{\"u}diger Mutz}.} \bibinfo{year}{2015}\natexlab{}.
\newblock \showarticletitle{Growth rates of modern science: A bibliometric
  analysis based on the number of publications and cited references}.
\newblock \bibinfo{journal}{\emph{Journal of the Association for Information
  Science and Technology}} \bibinfo{volume}{66}, \bibinfo{number}{11}
  (\bibinfo{year}{2015}), \bibinfo{pages}{2215--2222}.
\newblock


\bibitem[\protect\citeauthoryear{Bowman, Angeli, Potts, and Manning}{Bowman
  et~al\mbox{.}}{2015}]%
        {snli}
\bibfield{author}{\bibinfo{person}{Samuel~R. Bowman}, \bibinfo{person}{Gabor
  Angeli}, \bibinfo{person}{Christopher Potts}, {and}
  \bibinfo{person}{Christopher~D. Manning}.} \bibinfo{year}{2015}\natexlab{}.
\newblock \showarticletitle{A large annotated corpus for learning natural
  language inference}. In \bibinfo{booktitle}{\emph{Proceedings of the 2015
  Conference on Empirical Methods in Natural Language Processing}}.
  \bibinfo{publisher}{Association for Computational Linguistics},
  \bibinfo{address}{Lisbon, Portugal}, \bibinfo{pages}{632--642}.
\newblock
\urldef\tempurl%
\url{https://doi.org/10.18653/v1/D15-1075}
\showDOI{\tempurl}


\bibitem[\protect\citeauthoryear{Bu{\c{c}}inca, Malaya, and
  Gajos}{Bu{\c{c}}inca et~al\mbox{.}}{2021}]%
        {buccinca2021trust}
\bibfield{author}{\bibinfo{person}{Zana Bu{\c{c}}inca},
  \bibinfo{person}{Maja~Barbara Malaya}, {and} \bibinfo{person}{Krzysztof~Z
  Gajos}.} \bibinfo{year}{2021}\natexlab{}.
\newblock \showarticletitle{To trust or to think: cognitive forcing functions
  can reduce overreliance on AI in AI-assisted decision-making}.
\newblock \bibinfo{journal}{\emph{Proceedings of the ACM on Human-Computer
  Interaction}} \bibinfo{volume}{5}, \bibinfo{number}{CSCW1}
  (\bibinfo{year}{2021}), \bibinfo{pages}{1--21}.
\newblock


\bibitem[\protect\citeauthoryear{Carbonell}{Carbonell}{1983}]%
        {carbonell1983learning}
\bibfield{author}{\bibinfo{person}{Jaime~G Carbonell}.}
  \bibinfo{year}{1983}\natexlab{}.
\newblock \showarticletitle{Learning by analogy: Formulating and generalizing
  plans from past experience}.
\newblock In \bibinfo{booktitle}{\emph{Machine learning}}.
  \bibinfo{publisher}{Springer}, \bibinfo{pages}{137--161}.
\newblock


\bibitem[\protect\citeauthoryear{Carbonell}{Carbonell}{1985}]%
        {carbonell1985derivational}
\bibfield{author}{\bibinfo{person}{Jaime~Guillermo Carbonell}.}
  \bibinfo{year}{1985}\natexlab{}.
\newblock \bibinfo{booktitle}{\emph{Derivational analogy: A theory of
  reconstructive problem solving and expertise acquisition.}}
\newblock \bibinfo{type}{{T}echnical {R}eport}.
  \bibinfo{institution}{CARNEGIE-MELLON UNIV PITTSBURGH PA DEPT OF COMPUTER
  SCIENCE}.
\newblock


\bibitem[\protect\citeauthoryear{Carol-anne, Regehr, Mylopoulos, and
  MacRae}{Carol-anne et~al\mbox{.}}{2007}]%
        {carol2007slowing}
\bibfield{author}{\bibinfo{person}{E~Moulton Carol-anne},
  \bibinfo{person}{Glenn Regehr}, \bibinfo{person}{Maria Mylopoulos}, {and}
  \bibinfo{person}{Helen~M MacRae}.} \bibinfo{year}{2007}\natexlab{}.
\newblock \showarticletitle{Slowing down when you should: a new model of expert
  judgment}.
\newblock \bibinfo{journal}{\emph{Academic Medicine}} \bibinfo{volume}{82},
  \bibinfo{number}{10} (\bibinfo{year}{2007}), \bibinfo{pages}{S109--S116}.
\newblock


\bibitem[\protect\citeauthoryear{Carpineto and Romano}{Carpineto and
  Romano}{2012}]%
        {carpineto2012survey}
\bibfield{author}{\bibinfo{person}{Claudio Carpineto} {and}
  \bibinfo{person}{Giovanni Romano}.} \bibinfo{year}{2012}\natexlab{}.
\newblock \showarticletitle{A survey of automatic query expansion in
  information retrieval}.
\newblock \bibinfo{journal}{\emph{Acm Computing Surveys (CSUR)}}
  \bibinfo{volume}{44}, \bibinfo{number}{1} (\bibinfo{year}{2012}),
  \bibinfo{pages}{1--50}.
\newblock


\bibitem[\protect\citeauthoryear{Cer, Diab, Agirre, Lopez-Gazpio, and
  Specia}{Cer et~al\mbox{.}}{2017}]%
        {semeval17}
\bibfield{author}{\bibinfo{person}{Daniel Cer}, \bibinfo{person}{Mona Diab},
  \bibinfo{person}{Eneko Agirre}, \bibinfo{person}{I{\~n}igo Lopez-Gazpio},
  {and} \bibinfo{person}{Lucia Specia}.} \bibinfo{year}{2017}\natexlab{}.
\newblock \showarticletitle{{S}em{E}val-2017 Task 1: Semantic Textual
  Similarity Multilingual and Crosslingual Focused Evaluation}. In
  \bibinfo{booktitle}{\emph{Proceedings of the 11th International Workshop on
  Semantic Evaluation ({S}em{E}val-2017)}}. \bibinfo{publisher}{Association for
  Computational Linguistics}, \bibinfo{address}{Vancouver, Canada},
  \bibinfo{pages}{1--14}.
\newblock
\urldef\tempurl%
\url{https://doi.org/10.18653/v1/S17-2001}
\showDOI{\tempurl}


\bibitem[\protect\citeauthoryear{Cer, Yang, Kong, Hua, Limtiaco, John,
  Constant, Guajardo-Cespedes, Yuan, Tar, et~al\mbox{.}}{Cer
  et~al\mbox{.}}{2018}]%
        {universal_sentence_encoder}
\bibfield{author}{\bibinfo{person}{Daniel Cer}, \bibinfo{person}{Yinfei Yang},
  \bibinfo{person}{Sheng-yi Kong}, \bibinfo{person}{Nan Hua},
  \bibinfo{person}{Nicole Limtiaco}, \bibinfo{person}{Rhomni~St John},
  \bibinfo{person}{Noah Constant}, \bibinfo{person}{Mario Guajardo-Cespedes},
  \bibinfo{person}{Steve Yuan}, \bibinfo{person}{Chris Tar}, {et~al\mbox{.}}}
  \bibinfo{year}{2018}\natexlab{}.
\newblock \showarticletitle{Universal sentence encoder}.
\newblock \bibinfo{journal}{\emph{arXiv preprint arXiv:1803.11175}}
  (\bibinfo{year}{2018}).
\newblock


\bibitem[\protect\citeauthoryear{Chan, Chang, Hope, Shahaf, and Kittur}{Chan
  et~al\mbox{.}}{2018}]%
        {chan2018solvent}
\bibfield{author}{\bibinfo{person}{Joel Chan}, \bibinfo{person}{Joseph~Chee
  Chang}, \bibinfo{person}{Tom Hope}, \bibinfo{person}{Dafna Shahaf}, {and}
  \bibinfo{person}{Aniket Kittur}.} \bibinfo{year}{2018}\natexlab{}.
\newblock \showarticletitle{SOLVENT: A Mixed Initiative System for Finding
  Analogies between Research Papers}.
\newblock \bibinfo{journal}{\emph{Proc. ACM Hum.-Comput. Interact.}}
  \bibinfo{volume}{2}, \bibinfo{number}{CSCW}, Article \bibinfo{articleno}{31}
  (\bibinfo{date}{Nov.} \bibinfo{year}{2018}), \bibinfo{numpages}{21}~pages.
\newblock
\urldef\tempurl%
\url{https://doi.org/10.1145/3274300}
\showDOI{\tempurl}


\bibitem[\protect\citeauthoryear{Chan, Dow, and Schunn}{Chan
  et~al\mbox{.}}{2015}]%
        {chanBestDesignIdeas2015}
\bibfield{author}{\bibinfo{person}{Joel Chan}, \bibinfo{person}{Steven~P. Dow},
  {and} \bibinfo{person}{Christian~D. Schunn}.}
  \bibinfo{year}{2015}\natexlab{}.
\newblock \showarticletitle{Do {The} {Best} {Design} {Ideas} ({Really}) {Come}
  {From} {Conceptually} {Distant} {Sources} {Of} {Inspiration}?}
\newblock \bibinfo{journal}{\emph{Design Studies}}  \bibinfo{volume}{36}
  (\bibinfo{year}{2015}), \bibinfo{pages}{31--58}.
\newblock
\urldef\tempurl%
\url{https://doi.org/10.1016/j.destud.2014.08.001}
\showDOI{\tempurl}


\bibitem[\protect\citeauthoryear{Chan and Schunn}{Chan and Schunn}{2015}]%
        {chanImportanceIterationCreative2015}
\bibfield{author}{\bibinfo{person}{Joel Chan} {and}
  \bibinfo{person}{Christian~D. Schunn}.} \bibinfo{year}{2015}\natexlab{}.
\newblock \showarticletitle{The importance of iteration in creative conceptual
  combination}.
\newblock \bibinfo{journal}{\emph{Cognition}}  \bibinfo{volume}{145}
  (\bibinfo{date}{Dec.} \bibinfo{year}{2015}), \bibinfo{pages}{104--115}.
\newblock
\showISSN{0010-0277}
\urldef\tempurl%
\url{https://doi.org/10.1016/j.cognition.2015.08.008}
\showDOI{\tempurl}


\bibitem[\protect\citeauthoryear{Chan, Siangliulue, Qori~McDonald, Liu,
  Moradinezhad, Aman, Solovey, Gajos, and Dow}{Chan et~al\mbox{.}}{2017}]%
        {chan2017semantically}
\bibfield{author}{\bibinfo{person}{Joel Chan}, \bibinfo{person}{Pao
  Siangliulue}, \bibinfo{person}{Denisa Qori~McDonald}, \bibinfo{person}{Ruixue
  Liu}, \bibinfo{person}{Reza Moradinezhad}, \bibinfo{person}{Safa Aman},
  \bibinfo{person}{Erin~T Solovey}, \bibinfo{person}{Krzysztof~Z Gajos}, {and}
  \bibinfo{person}{Steven~P Dow}.} \bibinfo{year}{2017}\natexlab{}.
\newblock \showarticletitle{Semantically far inspirations considered harmful?
  accounting for cognitive states in collaborative ideation}. In
  \bibinfo{booktitle}{\emph{Proceedings of the 2017 ACM SIGCHI Conference on
  Creativity and Cognition}}. \bibinfo{pages}{93--105}.
\newblock


\bibitem[\protect\citeauthoryear{Csikszentmihalyi and
  Csikzentmihaly}{Csikszentmihalyi and Csikzentmihaly}{1990}]%
        {csikszentmihalyi1990flow}
\bibfield{author}{\bibinfo{person}{Mihaly Csikszentmihalyi} {and}
  \bibinfo{person}{Mihaly Csikzentmihaly}.} \bibinfo{year}{1990}\natexlab{}.
\newblock \bibinfo{booktitle}{\emph{Flow: The psychology of optimal
  experience}}. Vol.~\bibinfo{volume}{1990}.
\newblock \bibinfo{publisher}{Harper \& Row New York}.
\newblock


\bibitem[\protect\citeauthoryear{Darzentas, Velt, Wetzel, Craigon, Wagner,
  Urquhart, and Benford}{Darzentas et~al\mbox{.}}{2019}]%
        {darzentas2019card}
\bibfield{author}{\bibinfo{person}{Dimitrios Darzentas},
  \bibinfo{person}{Raphael Velt}, \bibinfo{person}{Richard Wetzel},
  \bibinfo{person}{Peter~J Craigon}, \bibinfo{person}{Hanne~G Wagner},
  \bibinfo{person}{Lachlan~D Urquhart}, {and} \bibinfo{person}{Steve Benford}.}
  \bibinfo{year}{2019}\natexlab{}.
\newblock \showarticletitle{Card mapper: Enabling data-driven reflections on
  ideation cards}. In \bibinfo{booktitle}{\emph{Proceedings of the 2019 CHI
  Conference on Human Factors in Computing Systems}}. \bibinfo{pages}{1--15}.
\newblock


\bibitem[\protect\citeauthoryear{Davis}{Davis}{2017}]%
        {origami_open_innovation}
\bibfield{author}{\bibinfo{person}{Nicola Davis}.}
  \bibinfo{year}{2017}\natexlab{}.
\newblock \bibinfo{booktitle}{\emph{Nasa needs you: space agency to crowdsource
  origami designs for shield}}.
\newblock
\urldef\tempurl%
\url{https://www.theguardian.com/science/2017/jul/20/nasa-needs-you-space-agency-to-crowdsource-origami-designs-for-shield}
\showURL{%
\tempurl}


\bibitem[\protect\citeauthoryear{de~Solla~Price}{de~Solla~Price}{1965}]%
        {de_Solla_Price510}
\bibfield{author}{\bibinfo{person}{Derek~J. de Solla~Price}.}
  \bibinfo{year}{1965}\natexlab{}.
\newblock \showarticletitle{Networks of Scientific Papers}.
\newblock \bibinfo{journal}{\emph{Science}} \bibinfo{volume}{149},
  \bibinfo{number}{3683} (\bibinfo{year}{1965}), \bibinfo{pages}{510--515}.
\newblock
\showISSN{0036-8075}
\urldef\tempurl%
\url{https://doi.org/10.1126/science.149.3683.510}
\showDOI{\tempurl}
\showeprint{https://science.sciencemag.org/content/149/3683/510.full.pdf}


\bibitem[\protect\citeauthoryear{Devlin, Chang, Lee, and Toutanova}{Devlin
  et~al\mbox{.}}{2019}]%
        {bert}
\bibfield{author}{\bibinfo{person}{Jacob Devlin}, \bibinfo{person}{Ming-Wei
  Chang}, \bibinfo{person}{Kenton Lee}, {and} \bibinfo{person}{Kristina
  Toutanova}.} \bibinfo{year}{2019}\natexlab{}.
\newblock \showarticletitle{{BERT}: Pre-training of Deep Bidirectional
  Transformers for Language Understanding}. In
  \bibinfo{booktitle}{\emph{Proceedings of the 2019 Conference of the North
  {A}merican Chapter of the Association for Computational Linguistics: Human
  Language Technologies, Volume 1 (Long and Short Papers)}}.
  \bibinfo{publisher}{Association for Computational Linguistics},
  \bibinfo{address}{Minneapolis, Minnesota}, \bibinfo{pages}{4171--4186}.
\newblock
\urldef\tempurl%
\url{https://doi.org/10.18653/v1/N19-1423}
\showDOI{\tempurl}


\bibitem[\protect\citeauthoryear{Diedrich, Benedek, Jauk, and
  Neubauer}{Diedrich et~al\mbox{.}}{2015}]%
        {diedrich2015creative}
\bibfield{author}{\bibinfo{person}{Jennifer Diedrich}, \bibinfo{person}{Mathias
  Benedek}, \bibinfo{person}{Emanuel Jauk}, {and} \bibinfo{person}{Aljoscha~C
  Neubauer}.} \bibinfo{year}{2015}\natexlab{}.
\newblock \showarticletitle{Are creative ideas novel and useful?}
\newblock \bibinfo{journal}{\emph{Psychology of Aesthetics, Creativity, and the
  Arts}} \bibinfo{volume}{9}, \bibinfo{number}{1} (\bibinfo{year}{2015}),
  \bibinfo{pages}{35}.
\newblock


\bibitem[\protect\citeauthoryear{Dow, MacIntyre, Lee, Oezbek, Bolter, and
  Gandy}{Dow et~al\mbox{.}}{2005}]%
        {dow2005wizard}
\bibfield{author}{\bibinfo{person}{Steven Dow}, \bibinfo{person}{Blair
  MacIntyre}, \bibinfo{person}{Jaemin Lee}, \bibinfo{person}{Christopher
  Oezbek}, \bibinfo{person}{Jay~David Bolter}, {and} \bibinfo{person}{Maribeth
  Gandy}.} \bibinfo{year}{2005}\natexlab{}.
\newblock \showarticletitle{Wizard of Oz support throughout an iterative design
  process}.
\newblock \bibinfo{journal}{\emph{IEEE Pervasive Computing}}
  \bibinfo{volume}{4}, \bibinfo{number}{4} (\bibinfo{year}{2005}),
  \bibinfo{pages}{18--26}.
\newblock


\bibitem[\protect\citeauthoryear{Dunbar}{Dunbar}{1997}]%
        {dunbarHowScientistsThink1997}
\bibfield{author}{\bibinfo{person}{K.~N. Dunbar}.}
  \bibinfo{year}{1997}\natexlab{}.
\newblock \showarticletitle{How scientists think: {On}-line creativity and
  conceptual change in science}.
\newblock In \bibinfo{booktitle}{\emph{Creative thought: {An} investigation of
  conceptual structures and processes}},
  \bibfield{editor}{\bibinfo{person}{T.~B. Ward}, \bibinfo{person}{S.~M.
  Smith}, {and} \bibinfo{person}{J.~Vaid}} (Eds.). \bibinfo{address}{Washington
  D.C.}, \bibinfo{pages}{461--493}.
\newblock


\bibitem[\protect\citeauthoryear{Eliasmith and Thagard}{Eliasmith and
  Thagard}{2001}]%
        {eliasmith2001integrating}
\bibfield{author}{\bibinfo{person}{Chris Eliasmith} {and} \bibinfo{person}{Paul
  Thagard}.} \bibinfo{year}{2001}\natexlab{}.
\newblock \showarticletitle{Integrating structure and meaning: A distributed
  model of analogical mapping}.
\newblock \bibinfo{journal}{\emph{Cognitive Science}} \bibinfo{volume}{25},
  \bibinfo{number}{2} (\bibinfo{year}{2001}), \bibinfo{pages}{245--286}.
\newblock


\bibitem[\protect\citeauthoryear{Fazel-Zarandi and Yu}{Fazel-Zarandi and
  Yu}{2008}]%
        {ontology_based_expertise_finding}
\bibfield{author}{\bibinfo{person}{Maryam Fazel-Zarandi} {and}
  \bibinfo{person}{Eric Yu}.} \bibinfo{year}{2008}\natexlab{}.
\newblock \showarticletitle{Ontology-Based Expertise Finding}. In
  \bibinfo{booktitle}{\emph{Practical Aspects of Knowledge Management}},
  \bibfield{editor}{\bibinfo{person}{Takahira Yamaguchi}} (Ed.).
  \bibinfo{publisher}{Springer Berlin Heidelberg}, \bibinfo{address}{Berlin,
  Heidelberg}, \bibinfo{pages}{232--243}.
\newblock
\showISBNx{978-3-540-89447-6}


\bibitem[\protect\citeauthoryear{Fonteyn, Kuipers, and Grobe}{Fonteyn
  et~al\mbox{.}}{1993}]%
        {thinkaloud1}
\bibfield{author}{\bibinfo{person}{Marsha~E Fonteyn}, \bibinfo{person}{Benjamin
  Kuipers}, {and} \bibinfo{person}{Susan~J Grobe}.}
  \bibinfo{year}{1993}\natexlab{}.
\newblock \showarticletitle{A description of think aloud method and protocol
  analysis}.
\newblock \bibinfo{journal}{\emph{Qualitative health research}}
  \bibinfo{volume}{3}, \bibinfo{number}{4} (\bibinfo{year}{1993}),
  \bibinfo{pages}{430--441}.
\newblock


\bibitem[\protect\citeauthoryear{Forbus}{Forbus}{2001}]%
        {forbus2001exploring}
\bibfield{author}{\bibinfo{person}{Kenneth Forbus}.}
  \bibinfo{year}{2001}\natexlab{}.
\newblock \bibinfo{booktitle}{\emph{Exploring analogy in the large}}.
\newblock \bibinfo{publisher}{MIT Press}.
\newblock


\bibitem[\protect\citeauthoryear{Forbus, Ferguson, and Gentner}{Forbus
  et~al\mbox{.}}{1994}]%
        {forbus1994incremental}
\bibfield{author}{\bibinfo{person}{Kenneth~D Forbus}, \bibinfo{person}{Ronald~W
  Ferguson}, {and} \bibinfo{person}{Dedre Gentner}.}
  \bibinfo{year}{1994}\natexlab{}.
\newblock \showarticletitle{Incremental structure-mapping}. In
  \bibinfo{booktitle}{\emph{Proceedings of the sixteenth annual conference of
  the Cognitive Science Society}}. \bibinfo{pages}{313--318}.
\newblock


\bibitem[\protect\citeauthoryear{Forbus, Ferguson, Lovett, and Gentner}{Forbus
  et~al\mbox{.}}{2017}]%
        {forbus2017extending}
\bibfield{author}{\bibinfo{person}{Kenneth~D Forbus}, \bibinfo{person}{Ronald~W
  Ferguson}, \bibinfo{person}{Andrew Lovett}, {and} \bibinfo{person}{Dedre
  Gentner}.} \bibinfo{year}{2017}\natexlab{}.
\newblock \showarticletitle{Extending SME to handle large-scale cognitive
  modeling}.
\newblock \bibinfo{journal}{\emph{Cognitive Science}} \bibinfo{volume}{41},
  \bibinfo{number}{5} (\bibinfo{year}{2017}), \bibinfo{pages}{1152--1201}.
\newblock


\bibitem[\protect\citeauthoryear{Fu, Chan, Cagan, Kotovsky, Schunn, and
  Wood}{Fu et~al\mbox{.}}{2013a}]%
        {fuMeaningFarImpact2013}
\bibfield{author}{\bibinfo{person}{Katherine Fu}, \bibinfo{person}{Joel Chan},
  \bibinfo{person}{Jonathan Cagan}, \bibinfo{person}{Kenneth Kotovsky},
  \bibinfo{person}{Christian Schunn}, {and} \bibinfo{person}{Kristin Wood}.}
  \bibinfo{year}{2013}\natexlab{a}.
\newblock \showarticletitle{The meaning of ``near'' and ``far'': the impact of
  structuring design databases and the effect of distance of analogy on design
  output}.
\newblock \bibinfo{journal}{\emph{Journal of Mechanical Design}}
  \bibinfo{volume}{135}, \bibinfo{number}{2} (\bibinfo{year}{2013}),
  \bibinfo{pages}{021007}.
\newblock


\bibitem[\protect\citeauthoryear{Fu, Chan, Schunn, Cagan, and Kotovsky}{Fu
  et~al\mbox{.}}{2013b}]%
        {fu_chan_design_repo_space}
\bibfield{author}{\bibinfo{person}{Katherine Fu}, \bibinfo{person}{Joel Chan},
  \bibinfo{person}{Christian Schunn}, \bibinfo{person}{Jonathan Cagan}, {and}
  \bibinfo{person}{Kenneth Kotovsky}.} \bibinfo{year}{2013}\natexlab{b}.
\newblock \showarticletitle{Expert representation of design repository space: A
  comparison to and validation of algorithmic output}.
\newblock \bibinfo{journal}{\emph{Design Studies}} \bibinfo{volume}{34},
  \bibinfo{number}{6} (\bibinfo{year}{2013}), \bibinfo{pages}{729 -- 762}.
\newblock
\showISSN{0142-694X}
\urldef\tempurl%
\url{https://doi.org/10.1016/j.destud.2013.06.002}
\showDOI{\tempurl}


\bibitem[\protect\citeauthoryear{Gardner, Grus, Neumann, Tafjord, Dasigi, Liu,
  Peters, Schmitz, and Zettlemoyer}{Gardner et~al\mbox{.}}{2018}]%
        {allennlp}
\bibfield{author}{\bibinfo{person}{Matt Gardner}, \bibinfo{person}{Joel Grus},
  \bibinfo{person}{Mark Neumann}, \bibinfo{person}{Oyvind Tafjord},
  \bibinfo{person}{Pradeep Dasigi}, \bibinfo{person}{Nelson~F. Liu},
  \bibinfo{person}{Matthew Peters}, \bibinfo{person}{Michael Schmitz}, {and}
  \bibinfo{person}{Luke Zettlemoyer}.} \bibinfo{year}{2018}\natexlab{}.
\newblock \showarticletitle{{A}llen{NLP}: A Deep Semantic Natural Language
  Processing Platform}. In \bibinfo{booktitle}{\emph{Proceedings of Workshop
  for {NLP} Open Source Software ({NLP}-{OSS})}}.
  \bibinfo{publisher}{Association for Computational Linguistics},
  \bibinfo{address}{Melbourne, Australia}, \bibinfo{pages}{1--6}.
\newblock
\urldef\tempurl%
\url{https://doi.org/10.18653/v1/W18-2501}
\showDOI{\tempurl}


\bibitem[\protect\citeauthoryear{Gentner}{Gentner}{1983}]%
        {gentner1983structure}
\bibfield{author}{\bibinfo{person}{Dedre Gentner}.}
  \bibinfo{year}{1983}\natexlab{}.
\newblock \showarticletitle{Structure-mapping: A theoretical framework for
  analogy}.
\newblock \bibinfo{journal}{\emph{Cognitive science}} \bibinfo{volume}{7},
  \bibinfo{number}{2} (\bibinfo{year}{1983}), \bibinfo{pages}{155--170}.
\newblock


\bibitem[\protect\citeauthoryear{Gentner, Brem, Ferguson, Wolff, Markman, and
  Forbus}{Gentner et~al\mbox{.}}{1997}]%
        {gentnerAnalogyCreativityWorks1997}
\bibfield{author}{\bibinfo{person}{D. Gentner}, \bibinfo{person}{S. Brem},
  \bibinfo{person}{R.~W. Ferguson}, \bibinfo{person}{P. Wolff},
  \bibinfo{person}{A.~B. Markman}, {and} \bibinfo{person}{K.~D. Forbus}.}
  \bibinfo{year}{1997}\natexlab{}.
\newblock \showarticletitle{Analogy and {Creativity} in the {Works} of
  {Johannes} {Kepler}}.
\newblock In \bibinfo{booktitle}{\emph{Creative thought: {An} investigation of
  conceptual structures and processes}},
  \bibfield{editor}{\bibinfo{person}{T.~B. Ward}, \bibinfo{person}{J.~Vaid},
  {and} \bibinfo{person}{S.~M. Smith}} (Eds.). \bibinfo{publisher}{American
  Psychological Association}, \bibinfo{address}{Washington D.C.},
  \bibinfo{pages}{403--459}.
\newblock


\bibitem[\protect\citeauthoryear{Gentner and Landers}{Gentner and
  Landers}{1985}]%
        {analogical_reminding}
\bibfield{author}{\bibinfo{person}{Dedre Gentner} {and}
  \bibinfo{person}{Russell Landers}.} \bibinfo{year}{1985}\natexlab{}.
\newblock \showarticletitle{ANALOGICAL REMINDING: A GOOD MATCH IS HARD TO
  FIND.}. In \bibinfo{booktitle}{\emph{Unknown Host Publication Title}}.
  \bibinfo{publisher}{IEEE}, \bibinfo{pages}{607--613}.
\newblock


\bibitem[\protect\citeauthoryear{Gentner and Smith}{Gentner and Smith}{2012}]%
        {gentner2012analogical}
\bibfield{author}{\bibinfo{person}{Dedre Gentner} {and} \bibinfo{person}{Linsey
  Smith}.} \bibinfo{year}{2012}\natexlab{}.
\newblock \showarticletitle{Analogical reasoning}.
\newblock \bibinfo{journal}{\emph{Encyclopedia of human behavior}}
  \bibinfo{volume}{2} (\bibinfo{year}{2012}), \bibinfo{pages}{130--136}.
\newblock


\bibitem[\protect\citeauthoryear{Gick and Holyoak}{Gick and Holyoak}{1980}]%
        {gickholyoak1980}
\bibfield{author}{\bibinfo{person}{Mary~L Gick} {and} \bibinfo{person}{Keith~J
  Holyoak}.} \bibinfo{year}{1980}\natexlab{}.
\newblock \showarticletitle{Analogical problem solving}.
\newblock \bibinfo{journal}{\emph{Cognitive psychology}} \bibinfo{volume}{12},
  \bibinfo{number}{3} (\bibinfo{year}{1980}), \bibinfo{pages}{306--355}.
\newblock


\bibitem[\protect\citeauthoryear{Gick and Holyoak}{Gick and Holyoak}{1983}]%
        {gick_schema_1983}
\bibfield{author}{\bibinfo{person}{Mary~L. Gick} {and}
  \bibinfo{person}{Keith~J. Holyoak}.} \bibinfo{year}{1983}\natexlab{}.
\newblock \showarticletitle{Schema induction and analogical transfer}.
\newblock \bibinfo{journal}{\emph{Cognitive Psychology}} \bibinfo{volume}{15},
  \bibinfo{number}{1} (\bibinfo{year}{1983}), \bibinfo{pages}{1 -- 38}.
\newblock
\showISSN{0010-0285}
\urldef\tempurl%
\url{https://doi.org/10.1016/0010-0285(83)90002-6}
\showDOI{\tempurl}


\bibitem[\protect\citeauthoryear{Gilon, Chan, Ng, Liifshitz-Assaf, Kittur, and
  Shahaf}{Gilon et~al\mbox{.}}{2018}]%
        {gilon_chi18}
\bibfield{author}{\bibinfo{person}{Karni Gilon}, \bibinfo{person}{Joel Chan},
  \bibinfo{person}{Felicia~Y. Ng}, \bibinfo{person}{Hila Liifshitz-Assaf},
  \bibinfo{person}{Aniket Kittur}, {and} \bibinfo{person}{Dafna Shahaf}.}
  \bibinfo{year}{2018}\natexlab{}.
\newblock \showarticletitle{Analogy Mining for Specific Design Needs}. In
  \bibinfo{booktitle}{\emph{Proceedings of the 2018 CHI Conference on Human
  Factors in Computing Systems}} (Montreal QC, Canada)
  \emph{(\bibinfo{series}{CHI '18})}. \bibinfo{publisher}{ACM},
  \bibinfo{address}{New York, NY, USA}, Article \bibinfo{articleno}{121},
  \bibinfo{numpages}{11}~pages.
\newblock
\showISBNx{978-1-4503-5620-6}
\urldef\tempurl%
\url{https://doi.org/10.1145/3173574.3173695}
\showDOI{\tempurl}


\bibitem[\protect\citeauthoryear{Gonçalves, Cardoso, and
  Badke-Schaub}{Gonçalves et~al\mbox{.}}{2013}]%
        {goncalvesInspirationPeakExploring2013}
\bibfield{author}{\bibinfo{person}{Milene Gonçalves}, \bibinfo{person}{Carlos
  Cardoso}, {and} \bibinfo{person}{Petra Badke-Schaub}.}
  \bibinfo{year}{2013}\natexlab{}.
\newblock \showarticletitle{Inspiration peak: exploring the semantic distance
  between design problem and textual inspirational stimuli}.
\newblock \bibinfo{journal}{\emph{International Journal of Design Creativity
  and Innovation}} \bibinfo{volume}{1}, \bibinfo{number}{4}
  (\bibinfo{year}{2013}), \bibinfo{pages}{215--232}.
\newblock


\bibitem[\protect\citeauthoryear{Gruber and Barrett}{Gruber and
  Barrett}{1974}]%
        {gruberDarwinManPsychological1974}
\bibfield{author}{\bibinfo{person}{Howard~E. Gruber} {and}
  \bibinfo{person}{Paul~H. Barrett}.} \bibinfo{year}{1974}\natexlab{}.
\newblock \bibinfo{booktitle}{\emph{Darwin on man: {A} psychological study of
  scientific creativity}}.
\newblock \bibinfo{publisher}{E. P. Dutton}, \bibinfo{address}{New York, NY,
  England}.
\newblock
\showISBNx{978-0-525-08877-6}
\newblock
\shownote{Pages: xxv, 495.}


\bibitem[\protect\citeauthoryear{Halford}{Halford}{1992}]%
        {halford1992analogical}
\bibfield{author}{\bibinfo{person}{Graeme~S Halford}.}
  \bibinfo{year}{1992}\natexlab{}.
\newblock \showarticletitle{Analogical reasoning and conceptual complexity in
  cognitive development}.
\newblock \bibinfo{journal}{\emph{Human Development}} \bibinfo{volume}{35},
  \bibinfo{number}{4} (\bibinfo{year}{1992}), \bibinfo{pages}{193--217}.
\newblock


\bibitem[\protect\citeauthoryear{Halford, Wilson, and Phillips}{Halford
  et~al\mbox{.}}{1998}]%
        {halford1998processing}
\bibfield{author}{\bibinfo{person}{Graeme~S Halford},
  \bibinfo{person}{William~H Wilson}, {and} \bibinfo{person}{Steven Phillips}.}
  \bibinfo{year}{1998}\natexlab{}.
\newblock \showarticletitle{Processing capacity defined by relational
  complexity: Implications for comparative, developmental, and cognitive
  psychology}.
\newblock \bibinfo{journal}{\emph{Behavioral and brain sciences}}
  \bibinfo{volume}{21}, \bibinfo{number}{6} (\bibinfo{year}{1998}),
  \bibinfo{pages}{803--831}.
\newblock


\bibitem[\protect\citeauthoryear{Hao, Ku, Liu, Hu, Bodner, Grabner, and
  Fink}{Hao et~al\mbox{.}}{2016}]%
        {hao2016reflection}
\bibfield{author}{\bibinfo{person}{Ning Hao}, \bibinfo{person}{Yixuan Ku},
  \bibinfo{person}{Meigui Liu}, \bibinfo{person}{Yi Hu}, \bibinfo{person}{Mark
  Bodner}, \bibinfo{person}{Roland~H Grabner}, {and} \bibinfo{person}{Andreas
  Fink}.} \bibinfo{year}{2016}\natexlab{}.
\newblock \showarticletitle{Reflection enhances creativity: Beneficial effects
  of idea evaluation on idea generation}.
\newblock \bibinfo{journal}{\emph{Brain and cognition}}  \bibinfo{volume}{103}
  (\bibinfo{year}{2016}), \bibinfo{pages}{30--37}.
\newblock


\bibitem[\protect\citeauthoryear{Hesse}{Hesse}{1966a}]%
        {hesseModelsAnalogiesScience1966}
\bibfield{author}{\bibinfo{person}{M. Hesse}.}
  \bibinfo{year}{1966}\natexlab{a}.
\newblock \bibinfo{booktitle}{\emph{Models and analogies in science}}.
\newblock \bibinfo{address}{Notre Dame, IN}.
\newblock


\bibitem[\protect\citeauthoryear{Hesse}{Hesse}{1966b}]%
        {hesse1966models}
\bibfield{author}{\bibinfo{person}{Mary~B Hesse}.}
  \bibinfo{year}{1966}\natexlab{b}.
\newblock \showarticletitle{Models and analogies in science}.
\newblock  (\bibinfo{year}{1966}).
\newblock


\bibitem[\protect\citeauthoryear{Hochreiter and Schmidhuber}{Hochreiter and
  Schmidhuber}{1997}]%
        {LSTM_Schmidhuber}
\bibfield{author}{\bibinfo{person}{Sepp Hochreiter} {and}
  \bibinfo{person}{J{\"u}rgen Schmidhuber}.} \bibinfo{year}{1997}\natexlab{}.
\newblock \showarticletitle{Long short-term memory}.
\newblock \bibinfo{journal}{\emph{Neural computation}} \bibinfo{volume}{9},
  \bibinfo{number}{8} (\bibinfo{year}{1997}), \bibinfo{pages}{1735--1780}.
\newblock


\bibitem[\protect\citeauthoryear{Hofstadter, Mitchell,
  et~al\mbox{.}}{Hofstadter et~al\mbox{.}}{1995}]%
        {hofstadter1995copycat}
\bibfield{author}{\bibinfo{person}{Douglas~R Hofstadter},
  \bibinfo{person}{Melanie Mitchell}, {et~al\mbox{.}}}
  \bibinfo{year}{1995}\natexlab{}.
\newblock \showarticletitle{The copycat project: A model of mental fluidity and
  analogy-making}.
\newblock \bibinfo{journal}{\emph{Advances in connectionist and neural
  computation theory}}  \bibinfo{volume}{2} (\bibinfo{year}{1995}),
  \bibinfo{pages}{205--267}.
\newblock


\bibitem[\protect\citeauthoryear{Hofstra, Kulkarni, Galvez, He, Jurafsky, and
  McFarland}{Hofstra et~al\mbox{.}}{2020}]%
        {diversity_innovation_paradox}
\bibfield{author}{\bibinfo{person}{Bas Hofstra}, \bibinfo{person}{Vivek~V
  Kulkarni}, \bibinfo{person}{Sebastian Munoz-Najar Galvez},
  \bibinfo{person}{Bryan He}, \bibinfo{person}{Dan Jurafsky}, {and}
  \bibinfo{person}{Daniel~A McFarland}.} \bibinfo{year}{2020}\natexlab{}.
\newblock \showarticletitle{The diversity--innovation paradox in science}.
\newblock \bibinfo{journal}{\emph{Proceedings of the National Academy of
  Sciences}} \bibinfo{volume}{117}, \bibinfo{number}{17}
  (\bibinfo{year}{2020}), \bibinfo{pages}{9284--9291}.
\newblock


\bibitem[\protect\citeauthoryear{Holyoak and Thagard}{Holyoak and
  Thagard}{1989}]%
        {holyoak1989analogical}
\bibfield{author}{\bibinfo{person}{Keith~J Holyoak} {and} \bibinfo{person}{Paul
  Thagard}.} \bibinfo{year}{1989}\natexlab{}.
\newblock \showarticletitle{Analogical mapping by constraint satisfaction}.
\newblock \bibinfo{journal}{\emph{Cognitive science}} \bibinfo{volume}{13},
  \bibinfo{number}{3} (\bibinfo{year}{1989}), \bibinfo{pages}{295--355}.
\newblock


\bibitem[\protect\citeauthoryear{Holyoak and Thagard}{Holyoak and
  Thagard}{1996}]%
        {holyoakAnalogicalScientist1996}
\bibfield{author}{\bibinfo{person}{K.~J. Holyoak} {and} \bibinfo{person}{P.
  Thagard}.} \bibinfo{year}{1996}\natexlab{}.
\newblock \showarticletitle{The analogical scientist}.
\newblock In \bibinfo{booktitle}{\emph{Mental {Leaps}: {Analogy} in {Creative}
  {Thought}}}, \bibfield{editor}{\bibinfo{person}{K.~J. Holyoak} {and}
  \bibinfo{person}{P.~Thagard}} (Eds.). \bibinfo{address}{Cambridge, MA},
  \bibinfo{pages}{185--209}.
\newblock


\bibitem[\protect\citeauthoryear{Hope, Chan, Kittur, and Shahaf}{Hope
  et~al\mbox{.}}{2017}]%
        {hope_kdd17}
\bibfield{author}{\bibinfo{person}{Tom Hope}, \bibinfo{person}{Joel Chan},
  \bibinfo{person}{Aniket Kittur}, {and} \bibinfo{person}{Dafna Shahaf}.}
  \bibinfo{year}{2017}\natexlab{}.
\newblock \showarticletitle{Accelerating Innovation Through Analogy Mining}. In
  \bibinfo{booktitle}{\emph{Proceedings of the 23rd ACM SIGKDD International
  Conference on Knowledge Discovery and Data Mining}} (Halifax, NS, Canada)
  \emph{(\bibinfo{series}{KDD '17})}. \bibinfo{publisher}{ACM},
  \bibinfo{address}{New York, NY, USA}, \bibinfo{pages}{235--243}.
\newblock
\showISBNx{978-1-4503-4887-4}
\urldef\tempurl%
\url{https://doi.org/10.1145/3097983.3098038}
\showDOI{\tempurl}


\bibitem[\protect\citeauthoryear{Hope, Tamari, Kang, Hershcovich, Chan, Kittur,
  and Shahaf}{Hope et~al\mbox{.}}{2022}]%
        {hope2021scaling}
\bibfield{author}{\bibinfo{person}{Tom Hope}, \bibinfo{person}{Ronen Tamari},
  \bibinfo{person}{Hyeonsu Kang}, \bibinfo{person}{Daniel Hershcovich},
  \bibinfo{person}{Joel Chan}, \bibinfo{person}{Aniket Kittur}, {and}
  \bibinfo{person}{Dafna Shahaf}.} \bibinfo{year}{2022}\natexlab{}.
\newblock \showarticletitle{Scaling Creative Inspiration with Fine-Grained
  Functional Facets of Product Ideas}. In \bibinfo{booktitle}{\emph{Proceedings
  of the SIGCHI Conference on Human Factors in Computing Systems}} (New
  Orleans, LA, USA) \emph{(\bibinfo{series}{CHI '22})}.
  \bibinfo{publisher}{Association for Computing Machinery},
  \bibinfo{address}{New York, NY, USA}, \bibinfo{numpages}{15}~pages.
\newblock
\urldef\tempurl%
\url{https://doi.org/10.1145/3491102.3517434}
\showDOI{\tempurl}


\bibitem[\protect\citeauthoryear{Huang and Chen}{Huang and Chen}{2017}]%
        {disa}
\bibfield{author}{\bibinfo{person}{Hen-Hsen Huang} {and}
  \bibinfo{person}{Hsin-Hsi Chen}.} \bibinfo{year}{2017}\natexlab{}.
\newblock \showarticletitle{DISA: A Scientific Writing Advisor with Deep
  Information Structure Analysis}. In \bibinfo{booktitle}{\emph{Proceedings of
  the Twenty-Sixth International Joint Conference on Artificial Intelligence,
  {IJCAI-17}}}. \bibinfo{pages}{5229--5231}.
\newblock
\urldef\tempurl%
\url{https://doi.org/10.24963/ijcai.2017/773}
\showDOI{\tempurl}


\bibitem[\protect\citeauthoryear{Huang, Xu, and Yu}{Huang
  et~al\mbox{.}}{2015}]%
        {huang2015bidirectional}
\bibfield{author}{\bibinfo{person}{Zhiheng Huang}, \bibinfo{person}{Wei Xu},
  {and} \bibinfo{person}{Kai Yu}.} \bibinfo{year}{2015}\natexlab{}.
\newblock \showarticletitle{Bidirectional LSTM-CRF models for sequence
  tagging}.
\newblock \bibinfo{journal}{\emph{arXiv preprint arXiv:1508.01991}}
  (\bibinfo{year}{2015}).
\newblock


\bibitem[\protect\citeauthoryear{Hummel and Holyoak}{Hummel and
  Holyoak}{2003}]%
        {hummel2003symbolic}
\bibfield{author}{\bibinfo{person}{John~E Hummel} {and}
  \bibinfo{person}{Keith~J Holyoak}.} \bibinfo{year}{2003}\natexlab{}.
\newblock \showarticletitle{A symbolic-connectionist theory of relational
  inference and generalization.}
\newblock \bibinfo{journal}{\emph{Psychological review}} \bibinfo{volume}{110},
  \bibinfo{number}{2} (\bibinfo{year}{2003}), \bibinfo{pages}{220}.
\newblock


\bibitem[\protect\citeauthoryear{Jansson and Smith}{Jansson and Smith}{1991}]%
        {jansson1991design}
\bibfield{author}{\bibinfo{person}{David~G Jansson} {and}
  \bibinfo{person}{Steven~M Smith}.} \bibinfo{year}{1991}\natexlab{}.
\newblock \showarticletitle{Design fixation}.
\newblock \bibinfo{journal}{\emph{Design studies}} \bibinfo{volume}{12},
  \bibinfo{number}{1} (\bibinfo{year}{1991}), \bibinfo{pages}{3--11}.
\newblock


\bibitem[\protect\citeauthoryear{Jiang, Xu, Araki, and Neubig}{Jiang
  et~al\mbox{.}}{2020}]%
        {spanrel}
\bibfield{author}{\bibinfo{person}{Zhengbao Jiang}, \bibinfo{person}{Wei Xu},
  \bibinfo{person}{Jun Araki}, {and} \bibinfo{person}{Graham Neubig}.}
  \bibinfo{year}{2020}\natexlab{}.
\newblock \showarticletitle{Generalizing Natural Language Analysis through
  Span-relation Representations}. In \bibinfo{booktitle}{\emph{Proceedings of
  the 58th Annual Meeting of the Association for Computational Linguistics}}.
  \bibinfo{publisher}{Association for Computational Linguistics},
  \bibinfo{address}{Online}, \bibinfo{pages}{2120--2133}.
\newblock
\urldef\tempurl%
\url{https://doi.org/10.18653/v1/2020.acl-main.192}
\showDOI{\tempurl}


\bibitem[\protect\citeauthoryear{Jinha}{Jinha}{2010}]%
        {jinha2010article}
\bibfield{author}{\bibinfo{person}{Arif~E Jinha}.}
  \bibinfo{year}{2010}\natexlab{}.
\newblock \showarticletitle{Article 50 million: an estimate of the number of
  scholarly articles in existence}.
\newblock \bibinfo{journal}{\emph{Learned Publishing}} \bibinfo{volume}{23},
  \bibinfo{number}{3} (\bibinfo{year}{2010}), \bibinfo{pages}{258--263}.
\newblock


\bibitem[\protect\citeauthoryear{Kahneman}{Kahneman}{2011}]%
        {kahneman2011thinking}
\bibfield{author}{\bibinfo{person}{Daniel Kahneman}.}
  \bibinfo{year}{2011}\natexlab{}.
\newblock \bibinfo{booktitle}{\emph{Thinking, fast and slow}}.
\newblock \bibinfo{publisher}{Macmillan}.
\newblock


\bibitem[\protect\citeauthoryear{Kang, Kocielnik, Head, Yang, Latzke, Kittur,
  Weld, Downey, and Bragg}{Kang et~al\mbox{.}}{2022}]%
        {kang2022you}
\bibfield{author}{\bibinfo{person}{Hyeonsu~B Kang}, \bibinfo{person}{Rafal
  Kocielnik}, \bibinfo{person}{Andrew Head}, \bibinfo{person}{Jiangjiang Yang},
  \bibinfo{person}{Matt Latzke}, \bibinfo{person}{Aniket Kittur},
  \bibinfo{person}{Daniel~S Weld}, \bibinfo{person}{Doug Downey}, {and}
  \bibinfo{person}{Jonathan Bragg}.} \bibinfo{year}{2022}\natexlab{}.
\newblock \showarticletitle{From Who You Know to What You Read: Augmenting
  Scientific Recommendations with Implicit Social Networks}. In
  \bibinfo{booktitle}{\emph{Proceedings of the SIGCHI Conference on Human
  Factors in Computing Systems}} (New Orleans, LA, USA)
  \emph{(\bibinfo{series}{CHI '22})}. \bibinfo{publisher}{Association for
  Computing Machinery}, \bibinfo{address}{New York, NY, USA},
  \bibinfo{numpages}{23}~pages.
\newblock
\urldef\tempurl%
\url{https://doi.org/10.1145/3491102.3517470}
\showDOI{\tempurl}


\bibitem[\protect\citeauthoryear{Keane, Ledgeway, and Duff}{Keane
  et~al\mbox{.}}{1994}]%
        {keane1994constraints}
\bibfield{author}{\bibinfo{person}{Mark~T Keane}, \bibinfo{person}{Tim
  Ledgeway}, {and} \bibinfo{person}{Stuart Duff}.}
  \bibinfo{year}{1994}\natexlab{}.
\newblock \showarticletitle{Constraints on analogical mapping: A comparison of
  three models}.
\newblock \bibinfo{journal}{\emph{Cognitive Science}} \bibinfo{volume}{18},
  \bibinfo{number}{3} (\bibinfo{year}{1994}), \bibinfo{pages}{387--438}.
\newblock


\bibitem[\protect\citeauthoryear{Kelly and Teevan}{Kelly and Teevan}{2003}]%
        {kelly2003implicit}
\bibfield{author}{\bibinfo{person}{Diane Kelly} {and} \bibinfo{person}{Jaime
  Teevan}.} \bibinfo{year}{2003}\natexlab{}.
\newblock \showarticletitle{Implicit feedback for inferring user preference: a
  bibliography}. In \bibinfo{booktitle}{\emph{Acm Sigir Forum}},
  Vol.~\bibinfo{volume}{37}. ACM New York, NY, USA, \bibinfo{pages}{18--28}.
\newblock


\bibitem[\protect\citeauthoryear{Kingma and Ba}{Kingma and Ba}{2014}]%
        {kingma2014adam}
\bibfield{author}{\bibinfo{person}{Diederik~P. Kingma} {and}
  \bibinfo{person}{Jimmy Ba}.} \bibinfo{year}{2014}\natexlab{}.
\newblock \bibinfo{title}{Adam: A Method for Stochastic Optimization}.
\newblock
\newblock
\showeprint[arxiv]{1412.6980}~[cs.LG]


\bibitem[\protect\citeauthoryear{Kittur, Yu, Hope, Chan, Lifshitz-Assaf, Gilon,
  Ng, Kraut, and Shahaf}{Kittur et~al\mbox{.}}{2019}]%
        {kittur_pnas19}
\bibfield{author}{\bibinfo{person}{Aniket Kittur}, \bibinfo{person}{Lixiu Yu},
  \bibinfo{person}{Tom Hope}, \bibinfo{person}{Joel Chan},
  \bibinfo{person}{Hila Lifshitz-Assaf}, \bibinfo{person}{Karni Gilon},
  \bibinfo{person}{Felicia Ng}, \bibinfo{person}{Robert~E Kraut}, {and}
  \bibinfo{person}{Dafna Shahaf}.} \bibinfo{year}{2019}\natexlab{}.
\newblock \showarticletitle{Scaling up analogical innovation with crowds and
  AI}.
\newblock \bibinfo{journal}{\emph{Proceedings of the National Academy of
  Sciences}} \bibinfo{volume}{116}, \bibinfo{number}{6} (\bibinfo{year}{2019}),
  \bibinfo{pages}{1870--1877}.
\newblock


\bibitem[\protect\citeauthoryear{Kneeland, Schilling, and Aharonson}{Kneeland
  et~al\mbox{.}}{2020}]%
        {kneeland2020exploring}
\bibfield{author}{\bibinfo{person}{Madeline~K Kneeland},
  \bibinfo{person}{Melissa~A Schilling}, {and} \bibinfo{person}{Barak~S
  Aharonson}.} \bibinfo{year}{2020}\natexlab{}.
\newblock \showarticletitle{Exploring uncharted territory: Knowledge search
  processes in the origination of outlier innovation}.
\newblock \bibinfo{journal}{\emph{Organization Science}} \bibinfo{volume}{31},
  \bibinfo{number}{3} (\bibinfo{year}{2020}), \bibinfo{pages}{535--557}.
\newblock


\bibitem[\protect\citeauthoryear{Koch, Taffin, Beaudouin-Lafon, Laine, Lucero,
  and MacKay}{Koch et~al\mbox{.}}{2020}]%
        {koch2020imagesense}
\bibfield{author}{\bibinfo{person}{Janin Koch}, \bibinfo{person}{Nicolas
  Taffin}, \bibinfo{person}{Michel Beaudouin-Lafon}, \bibinfo{person}{Markku
  Laine}, \bibinfo{person}{Andr{\'e}s Lucero}, {and} \bibinfo{person}{Wendy~E
  MacKay}.} \bibinfo{year}{2020}\natexlab{}.
\newblock \showarticletitle{ImageSense: An Intelligent Collaborative Ideation
  Tool to Support Diverse Human-Computer Partnerships}.
\newblock \bibinfo{journal}{\emph{Proceedings of the ACM on Human-Computer
  Interaction}} \bibinfo{volume}{4}, \bibinfo{number}{CSCW1}
  (\bibinfo{year}{2020}), \bibinfo{pages}{1--27}.
\newblock


\bibitem[\protect\citeauthoryear{Lambe, O'Reilly, Kelly, and Curristan}{Lambe
  et~al\mbox{.}}{2016}]%
        {lambe2016dual}
\bibfield{author}{\bibinfo{person}{Kathryn~Ann Lambe}, \bibinfo{person}{Gary
  O'Reilly}, \bibinfo{person}{Brendan~D Kelly}, {and} \bibinfo{person}{Sarah
  Curristan}.} \bibinfo{year}{2016}\natexlab{}.
\newblock \showarticletitle{Dual-process cognitive interventions to enhance
  diagnostic reasoning: a systematic review}.
\newblock \bibinfo{journal}{\emph{BMJ quality \& safety}} \bibinfo{volume}{25},
  \bibinfo{number}{10} (\bibinfo{year}{2016}), \bibinfo{pages}{808--820}.
\newblock


\bibitem[\protect\citeauthoryear{Landauer and Dumais}{Landauer and
  Dumais}{1997}]%
        {landauer1997solution}
\bibfield{author}{\bibinfo{person}{Thomas~K Landauer} {and}
  \bibinfo{person}{Susan~T Dumais}.} \bibinfo{year}{1997}\natexlab{}.
\newblock \showarticletitle{A solution to Plato's problem: The latent semantic
  analysis theory of acquisition, induction, and representation of knowledge.}
\newblock \bibinfo{journal}{\emph{Psychological review}} \bibinfo{volume}{104},
  \bibinfo{number}{2} (\bibinfo{year}{1997}), \bibinfo{pages}{211}.
\newblock


\bibitem[\protect\citeauthoryear{Lee, He, Lewis, and Zettlemoyer}{Lee
  et~al\mbox{.}}{2017}]%
        {lee-etal-2017-end}
\bibfield{author}{\bibinfo{person}{Kenton Lee}, \bibinfo{person}{Luheng He},
  \bibinfo{person}{Mike Lewis}, {and} \bibinfo{person}{Luke Zettlemoyer}.}
  \bibinfo{year}{2017}\natexlab{}.
\newblock \showarticletitle{End-to-end Neural Coreference Resolution}. In
  \bibinfo{booktitle}{\emph{Proceedings of the 2017 Conference on Empirical
  Methods in Natural Language Processing}}. \bibinfo{publisher}{Association for
  Computational Linguistics}, \bibinfo{address}{Copenhagen, Denmark},
  \bibinfo{pages}{188--197}.
\newblock
\urldef\tempurl%
\url{https://doi.org/10.18653/v1/D17-1018}
\showDOI{\tempurl}


\bibitem[\protect\citeauthoryear{Lewis}{Lewis}{1982}]%
        {think_aloud_lewis}
\bibfield{author}{\bibinfo{person}{Clayton Lewis}.}
  \bibinfo{year}{1982}\natexlab{}.
\newblock \bibinfo{booktitle}{\emph{Using the" thinking-aloud" method in
  cognitive interface design}}.
\newblock \bibinfo{publisher}{IBM TJ Watson Research Center Yorktown Heights,
  NY}.
\newblock


\bibitem[\protect\citeauthoryear{Luria and Delbr{\"u}ck}{Luria and
  Delbr{\"u}ck}{1943}]%
        {luria1943mutations}
\bibfield{author}{\bibinfo{person}{Salvador~E Luria} {and} \bibinfo{person}{Max
  Delbr{\"u}ck}.} \bibinfo{year}{1943}\natexlab{}.
\newblock \showarticletitle{Mutations of bacteria from virus sensitivity to
  virus resistance}.
\newblock \bibinfo{journal}{\emph{Genetics}} \bibinfo{volume}{28},
  \bibinfo{number}{6} (\bibinfo{year}{1943}), \bibinfo{pages}{491}.
\newblock


\bibitem[\protect\citeauthoryear{Manning, Raghavan, and Sch\"{u}tze}{Manning
  et~al\mbox{.}}{2008}]%
        {intro_to_IR}
\bibfield{author}{\bibinfo{person}{Christopher~D. Manning},
  \bibinfo{person}{Prabhakar Raghavan}, {and} \bibinfo{person}{Hinrich
  Sch\"{u}tze}.} \bibinfo{year}{2008}\natexlab{}.
\newblock \bibinfo{booktitle}{\emph{Introduction to Information Retrieval}}.
\newblock \bibinfo{publisher}{Cambridge University Press},
  \bibinfo{address}{USA}.
\newblock
\showISBNx{0521865719}


\bibitem[\protect\citeauthoryear{McDonald and Ackerman}{McDonald and
  Ackerman}{2000}]%
        {ackerman_expertise_recommender}
\bibfield{author}{\bibinfo{person}{David~W. McDonald} {and}
  \bibinfo{person}{Mark~S. Ackerman}.} \bibinfo{year}{2000}\natexlab{}.
\newblock \showarticletitle{Expertise Recommender: A Flexible Recommendation
  System and Architecture}. In \bibinfo{booktitle}{\emph{Proceedings of the
  2000 ACM Conference on Computer Supported Cooperative Work}} (Philadelphia,
  Pennsylvania, USA) \emph{(\bibinfo{series}{CSCW '00})}.
  \bibinfo{publisher}{Association for Computing Machinery},
  \bibinfo{address}{New York, NY, USA}, \bibinfo{pages}{231–240}.
\newblock
\showISBNx{1581132220}
\urldef\tempurl%
\url{https://doi.org/10.1145/358916.358994}
\showDOI{\tempurl}


\bibitem[\protect\citeauthoryear{Mintzberg, Raisinghani, and Theoret}{Mintzberg
  et~al\mbox{.}}{1976}]%
        {mintzberg1976structure}
\bibfield{author}{\bibinfo{person}{Henry Mintzberg}, \bibinfo{person}{Duru
  Raisinghani}, {and} \bibinfo{person}{Andre Theoret}.}
  \bibinfo{year}{1976}\natexlab{}.
\newblock \showarticletitle{The structure of" unstructured" decision
  processes}.
\newblock \bibinfo{journal}{\emph{Administrative science quarterly}}
  (\bibinfo{year}{1976}), \bibinfo{pages}{246--275}.
\newblock


\bibitem[\protect\citeauthoryear{Noorden}{Noorden}{2014}]%
        {noorden_2014}
\bibfield{author}{\bibinfo{person}{Richar~Van Noorden}.}
  \bibinfo{year}{2014}\natexlab{}.
\newblock \bibinfo{title}{Global scientific output doubles every nine years}.
\newblock
\newblock
\urldef\tempurl%
\url{http://blogs.nature.com/news/2014/05/global-scientific-output-doubles-every-nine-years.html}
\showURL{%
\tempurl}


\bibitem[\protect\citeauthoryear{Oppenheimer}{Oppenheimer}{1956}]%
        {oppenheimerAnalogyScience1956}
\bibfield{author}{\bibinfo{person}{R. Oppenheimer}.}
  \bibinfo{year}{1956}\natexlab{}.
\newblock \showarticletitle{Analogy in science}.
\newblock \bibinfo{journal}{\emph{American Psychologist}} \bibinfo{volume}{11},
  \bibinfo{number}{3} (\bibinfo{year}{1956}), \bibinfo{pages}{127--135}.
\newblock
\showISSN{0003-066X}


\bibitem[\protect\citeauthoryear{Paszke, Gross, Massa, Lerer, Bradbury, Chanan,
  Killeen, Lin, Gimelshein, Antiga, et~al\mbox{.}}{Paszke
  et~al\mbox{.}}{2019}]%
        {pytorch}
\bibfield{author}{\bibinfo{person}{Adam Paszke}, \bibinfo{person}{Sam Gross},
  \bibinfo{person}{Francisco Massa}, \bibinfo{person}{Adam Lerer},
  \bibinfo{person}{James Bradbury}, \bibinfo{person}{Gregory Chanan},
  \bibinfo{person}{Trevor Killeen}, \bibinfo{person}{Zeming Lin},
  \bibinfo{person}{Natalia Gimelshein}, \bibinfo{person}{Luca Antiga},
  {et~al\mbox{.}}} \bibinfo{year}{2019}\natexlab{}.
\newblock \showarticletitle{Pytorch: An imperative style, high-performance deep
  learning library}.
\newblock \bibinfo{journal}{\emph{arXiv preprint arXiv:1912.01703}}
  (\bibinfo{year}{2019}).
\newblock


\bibitem[\protect\citeauthoryear{Pennington, Socher, and Manning}{Pennington
  et~al\mbox{.}}{2014}]%
        {pennington_glove}
\bibfield{author}{\bibinfo{person}{Jeffrey Pennington},
  \bibinfo{person}{Richard Socher}, {and} \bibinfo{person}{Christopher
  Manning}.} \bibinfo{year}{2014}\natexlab{}.
\newblock \showarticletitle{{G}love: Global Vectors for Word Representation}.
  In \bibinfo{booktitle}{\emph{Proceedings of the 2014 conference on empirical
  methods in natural language processing (EMNLP '14)}}.
  \bibinfo{publisher}{Association for Computational Linguistics},
  \bibinfo{address}{Doha, Qatar}, \bibinfo{pages}{1532--1543}.
\newblock
\urldef\tempurl%
\url{https://doi.org/10.3115/v1/D14-1162}
\showDOI{\tempurl}


\bibitem[\protect\citeauthoryear{Peraza-Hernandez, Hartl, Malak~Jr, and
  Lagoudas}{Peraza-Hernandez et~al\mbox{.}}{2014}]%
        {peraza2014origami}
\bibfield{author}{\bibinfo{person}{Edwin~A Peraza-Hernandez},
  \bibinfo{person}{Darren~J Hartl}, \bibinfo{person}{Richard~J Malak~Jr}, {and}
  \bibinfo{person}{Dimitris~C Lagoudas}.} \bibinfo{year}{2014}\natexlab{}.
\newblock \showarticletitle{Origami-inspired active structures: a synthesis and
  review}.
\newblock \bibinfo{journal}{\emph{Smart Materials and Structures}}
  \bibinfo{volume}{23}, \bibinfo{number}{9} (\bibinfo{year}{2014}),
  \bibinfo{pages}{094001}.
\newblock


\bibitem[\protect\citeauthoryear{Peters, Neumann, Iyyer, Gardner, Clark, Lee,
  and Zettlemoyer}{Peters et~al\mbox{.}}{2018}]%
        {elmo}
\bibfield{author}{\bibinfo{person}{Matthew Peters}, \bibinfo{person}{Mark
  Neumann}, \bibinfo{person}{Mohit Iyyer}, \bibinfo{person}{Matt Gardner},
  \bibinfo{person}{Christopher Clark}, \bibinfo{person}{Kenton Lee}, {and}
  \bibinfo{person}{Luke Zettlemoyer}.} \bibinfo{year}{2018}\natexlab{}.
\newblock \showarticletitle{Deep Contextualized Word Representations}. In
  \bibinfo{booktitle}{\emph{Proceedings of the 2018 Conference of the North
  {A}merican Chapter of the Association for Computational Linguistics: Human
  Language Technologies, Volume 1 (Long Papers)}}.
  \bibinfo{publisher}{Association for Computational Linguistics},
  \bibinfo{address}{New Orleans, Louisiana}, \bibinfo{pages}{2227--2237}.
\newblock
\urldef\tempurl%
\url{https://doi.org/10.18653/v1/N18-1202}
\showDOI{\tempurl}


\bibitem[\protect\citeauthoryear{Preacher and Hayes}{Preacher and
  Hayes}{2004}]%
        {mediation_bootstrapping}
\bibfield{author}{\bibinfo{person}{Kristopher~J. Preacher} {and}
  \bibinfo{person}{Andrew~F. Hayes}.} \bibinfo{year}{2004}\natexlab{}.
\newblock \showarticletitle{SPSS and SAS procedures for estimating indirect
  effects in simple mediation models}.
\newblock \bibinfo{journal}{\emph{Behavior Research Methods, Instruments, {\&}
  Computers}} \bibinfo{volume}{36}, \bibinfo{number}{4} (\bibinfo{date}{01 Nov}
  \bibinfo{year}{2004}), \bibinfo{pages}{717--731}.
\newblock
\showISSN{1532-5970}
\urldef\tempurl%
\url{https://doi.org/10.3758/BF03206553}
\showDOI{\tempurl}


\bibitem[\protect\citeauthoryear{Ravi, Rosenkrantz, and Tayi}{Ravi
  et~al\mbox{.}}{1994}]%
        {ravi1994heuristic}
\bibfield{author}{\bibinfo{person}{Sekharipuram~S Ravi},
  \bibinfo{person}{Daniel~J Rosenkrantz}, {and} \bibinfo{person}{Giri~Kumar
  Tayi}.} \bibinfo{year}{1994}\natexlab{}.
\newblock \showarticletitle{Heuristic and special case algorithms for
  dispersion problems}.
\newblock \bibinfo{journal}{\emph{Operations Research}} \bibinfo{volume}{42},
  \bibinfo{number}{2} (\bibinfo{year}{1994}), \bibinfo{pages}{299--310}.
\newblock


\bibitem[\protect\citeauthoryear{Russell, Stefik, Pirolli, and Card}{Russell
  et~al\mbox{.}}{1993}]%
        {russell1993cost}
\bibfield{author}{\bibinfo{person}{Daniel~M Russell}, \bibinfo{person}{Mark~J
  Stefik}, \bibinfo{person}{Peter Pirolli}, {and} \bibinfo{person}{Stuart~K
  Card}.} \bibinfo{year}{1993}\natexlab{}.
\newblock \showarticletitle{The cost structure of sensemaking}. In
  \bibinfo{booktitle}{\emph{Proceedings of the INTERACT'93 and CHI'93
  conference on Human factors in computing systems}}. ACM,
  \bibinfo{pages}{269--276}.
\newblock


\bibitem[\protect\citeauthoryear{Savage}{Savage}{2016}]%
        {nanofins}
\bibfield{author}{\bibinfo{person}{Neil Savage}.}
  \bibinfo{year}{2016}\natexlab{}.
\newblock \bibinfo{booktitle}{\emph{Nanofins Make a Better Hologram}}.
\newblock
\urldef\tempurl%
\url{https://spectrum.ieee.org/tech-talk/semiconductors/optoelectronics/nanofins-make-a-better-hologram}
\showURL{%
\tempurl}


\bibitem[\protect\citeauthoryear{Schnabel, Bennett, and Joachims}{Schnabel
  et~al\mbox{.}}{2019}]%
        {schnabel2019shaping}
\bibfield{author}{\bibinfo{person}{Tobias Schnabel}, \bibinfo{person}{Paul~N
  Bennett}, {and} \bibinfo{person}{Thorsten Joachims}.}
  \bibinfo{year}{2019}\natexlab{}.
\newblock \showarticletitle{Shaping feedback data in recommender systems with
  interventions based on information foraging theory}. In
  \bibinfo{booktitle}{\emph{Proceedings of the Twelfth ACM International
  Conference on Web Search and Data Mining}}. \bibinfo{pages}{546--554}.
\newblock


\bibitem[\protect\citeauthoryear{Schnabel, Ramos, and Amershi}{Schnabel
  et~al\mbox{.}}{2020}]%
        {schnabel2020doesn}
\bibfield{author}{\bibinfo{person}{Tobias Schnabel}, \bibinfo{person}{Gonzalo
  Ramos}, {and} \bibinfo{person}{Saleema Amershi}.}
  \bibinfo{year}{2020}\natexlab{}.
\newblock \showarticletitle{“Who doesn’t like dinosaurs?” Finding and
  Eliciting Richer Preferences for Recommendation}. In
  \bibinfo{booktitle}{\emph{Fourteenth ACM Conference on Recommender Systems}}.
  \bibinfo{pages}{398--407}.
\newblock


\bibitem[\protect\citeauthoryear{Siangliulue, Chan, Gajos, and Dow}{Siangliulue
  et~al\mbox{.}}{2015}]%
        {siangliulue2015providing}
\bibfield{author}{\bibinfo{person}{Pao Siangliulue}, \bibinfo{person}{Joel
  Chan}, \bibinfo{person}{Krzysztof~Z Gajos}, {and} \bibinfo{person}{Steven~P
  Dow}.} \bibinfo{year}{2015}\natexlab{}.
\newblock \showarticletitle{Providing timely examples improves the quantity and
  quality of generated ideas}. In \bibinfo{booktitle}{\emph{Proceedings of the
  2015 ACM SIGCHI Conference on Creativity and Cognition}}.
  \bibinfo{pages}{83--92}.
\newblock


\bibitem[\protect\citeauthoryear{Smith-Renner, Fan, Birchfield, Wu,
  Boyd-Graber, Weld, and Findlater}{Smith-Renner et~al\mbox{.}}{2020}]%
        {smith2020no}
\bibfield{author}{\bibinfo{person}{Alison Smith-Renner}, \bibinfo{person}{Ron
  Fan}, \bibinfo{person}{Melissa Birchfield}, \bibinfo{person}{Tongshuang Wu},
  \bibinfo{person}{Jordan Boyd-Graber}, \bibinfo{person}{Daniel~S Weld}, {and}
  \bibinfo{person}{Leah Findlater}.} \bibinfo{year}{2020}\natexlab{}.
\newblock \showarticletitle{No explainability without accountability: An
  empirical study of explanations and feedback in interactive ml}. In
  \bibinfo{booktitle}{\emph{Proceedings of the 2020 CHI Conference on Human
  Factors in Computing Systems}}. \bibinfo{pages}{1--13}.
\newblock


\bibitem[\protect\citeauthoryear{Streeter and Lochbaum}{Streeter and
  Lochbaum}{1988}]%
        {automatic_representation_semantic_structure}
\bibfield{author}{\bibinfo{person}{L. Streeter} {and} \bibinfo{person}{K.
  Lochbaum}.} \bibinfo{year}{1988}\natexlab{}.
\newblock \showarticletitle{An expert/expert-locating system based on automatic
  representation of semantic structure}. In
  \bibinfo{booktitle}{\emph{Proceedings. The Fourth Conference on Artificial
  Intelligence Applications}}. \bibinfo{publisher}{IEEE Computer Society},
  \bibinfo{address}{Los Alamitos, CA, USA},
  \bibinfo{pages}{345,346,347,348,349,350}.
\newblock
\urldef\tempurl%
\url{https://doi.org/10.1109/CAIA.1988.196129}
\showDOI{\tempurl}


\bibitem[\protect\citeauthoryear{Sutskever, Vinyals, and Le}{Sutskever
  et~al\mbox{.}}{2014a}]%
        {NIPS2014_seq2seq}
\bibfield{author}{\bibinfo{person}{Ilya Sutskever}, \bibinfo{person}{Oriol
  Vinyals}, {and} \bibinfo{person}{Quoc~V Le}.}
  \bibinfo{year}{2014}\natexlab{a}.
\newblock \showarticletitle{Sequence to Sequence Learning with Neural
  Networks}. In \bibinfo{booktitle}{\emph{Advances in Neural Information
  Processing Systems}}, \bibfield{editor}{\bibinfo{person}{Z.~Ghahramani},
  \bibinfo{person}{M.~Welling}, \bibinfo{person}{C.~Cortes},
  \bibinfo{person}{N.~Lawrence}, {and} \bibinfo{person}{K.~Q. Weinberger}}
  (Eds.), Vol.~\bibinfo{volume}{27}. \bibinfo{publisher}{Curran Associates,
  Inc.}
\newblock
\urldef\tempurl%
\url{https://proceedings.neurips.cc/paper/2014/file/a14ac55a4f27472c5d894ec1c3c743d2-Paper.pdf}
\showURL{%
\tempurl}


\bibitem[\protect\citeauthoryear{Sutskever, Vinyals, and Le}{Sutskever
  et~al\mbox{.}}{2014b}]%
        {Seq2SeqNIPS}
\bibfield{author}{\bibinfo{person}{Ilya Sutskever}, \bibinfo{person}{Oriol
  Vinyals}, {and} \bibinfo{person}{Quoc~V. Le}.}
  \bibinfo{year}{2014}\natexlab{b}.
\newblock \showarticletitle{Sequence to Sequence Learning with Neural
  Networks}. In \bibinfo{booktitle}{\emph{Proceedings of the 27th International
  Conference on Neural Information Processing Systems - Volume 2}} (Montreal,
  Canada) \emph{(\bibinfo{series}{NIPS'14})}. \bibinfo{publisher}{MIT Press},
  \bibinfo{address}{Cambridge, MA, USA}, \bibinfo{pages}{3104–3112}.
\newblock


\bibitem[\protect\citeauthoryear{Sweller, Chandler, Tierney, and
  Cooper}{Sweller et~al\mbox{.}}{1990}]%
        {sweller1990cognitive}
\bibfield{author}{\bibinfo{person}{John Sweller}, \bibinfo{person}{Paul
  Chandler}, \bibinfo{person}{Paul Tierney}, {and} \bibinfo{person}{Martin
  Cooper}.} \bibinfo{year}{1990}\natexlab{}.
\newblock \showarticletitle{Cognitive load as a factor in the structuring of
  technical material.}
\newblock \bibinfo{journal}{\emph{Journal of experimental psychology: general}}
  \bibinfo{volume}{119}, \bibinfo{number}{2} (\bibinfo{year}{1990}),
  \bibinfo{pages}{176}.
\newblock


\bibitem[\protect\citeauthoryear{Teevan, Alvarado, Ackerman, and Karger}{Teevan
  et~al\mbox{.}}{2004}]%
        {perfect_search_engine}
\bibfield{author}{\bibinfo{person}{Jaime Teevan}, \bibinfo{person}{Christine
  Alvarado}, \bibinfo{person}{Mark~S. Ackerman}, {and}
  \bibinfo{person}{David~R. Karger}.} \bibinfo{year}{2004}\natexlab{}.
\newblock \showarticletitle{The Perfect Search Engine is Not Enough: A Study of
  Orienteering Behavior in Directed Search}. In
  \bibinfo{booktitle}{\emph{Proceedings of the SIGCHI Conference on Human
  Factors in Computing Systems}} (Vienna, Austria) \emph{(\bibinfo{series}{CHI
  ’04})}. \bibinfo{publisher}{Association for Computing Machinery},
  \bibinfo{address}{New York, NY, USA}, \bibinfo{pages}{415–422}.
\newblock
\showISBNx{1581137028}
\urldef\tempurl%
\url{https://doi.org/10.1145/985692.985745}
\showDOI{\tempurl}


\bibitem[\protect\citeauthoryear{ThermoCool}{ThermoCool}{2021}]%
        {heatfins}
\bibfield{author}{\bibinfo{person}{ThermoCool}.}
  \bibinfo{year}{2021}\natexlab{}.
\newblock \bibinfo{booktitle}{\emph{SKIVED FIN HEAT SINKS}}.
\newblock
\urldef\tempurl%
\url{https://thermocoolcorp.com/project/skived-fins/}
\showURL{%
\tempurl}


\bibitem[\protect\citeauthoryear{Tingley, Yamamoto, Hirose, Keele, and
  Imai}{Tingley et~al\mbox{.}}{2014}]%
        {tingley2014mediation}
\bibfield{author}{\bibinfo{person}{Dustin Tingley}, \bibinfo{person}{Teppei
  Yamamoto}, \bibinfo{person}{Kentaro Hirose}, \bibinfo{person}{Luke Keele},
  {and} \bibinfo{person}{Kosuke Imai}.} \bibinfo{year}{2014}\natexlab{}.
\newblock \showarticletitle{Mediation: R package for causal mediation
  analysis}.
\newblock  (\bibinfo{year}{2014}).
\newblock


\bibitem[\protect\citeauthoryear{Turney}{Turney}{2008}]%
        {turney2008latent}
\bibfield{author}{\bibinfo{person}{Peter~D Turney}.}
  \bibinfo{year}{2008}\natexlab{}.
\newblock \showarticletitle{The latent relation mapping engine: Algorithm and
  experiments}.
\newblock \bibinfo{journal}{\emph{Journal of Artificial Intelligence Research}}
   \bibinfo{volume}{33} (\bibinfo{year}{2008}), \bibinfo{pages}{615--655}.
\newblock


\bibitem[\protect\citeauthoryear{Van~Someren, Barnard, and
  Sandberg}{Van~Someren et~al\mbox{.}}{1994a}]%
        {thinkaloud2}
\bibfield{author}{\bibinfo{person}{MW Van~Someren}, \bibinfo{person}{YF
  Barnard}, {and} \bibinfo{person}{JAC Sandberg}.}
  \bibinfo{year}{1994}\natexlab{a}.
\newblock \showarticletitle{The think aloud method: a practical approach to
  modelling cognitive}.
\newblock \bibinfo{journal}{\emph{London: AcademicPress}}
  (\bibinfo{year}{1994}).
\newblock


\bibitem[\protect\citeauthoryear{Van~Someren, Barnard, and
  Sandberg}{Van~Someren et~al\mbox{.}}{1994b}]%
        {think_aloud_van1994}
\bibfield{author}{\bibinfo{person}{MW Van~Someren}, \bibinfo{person}{YF
  Barnard}, {and} \bibinfo{person}{JAC Sandberg}.}
  \bibinfo{year}{1994}\natexlab{b}.
\newblock \showarticletitle{The think aloud method: a practical approach to
  modelling cognitive}.
\newblock \bibinfo{journal}{\emph{London: AcademicPress}}
  (\bibinfo{year}{1994}).
\newblock


\bibitem[\protect\citeauthoryear{Vaswani, Shazeer, Parmar, Uszkoreit, Jones,
  Gomez, Kaiser, and Polosukhin}{Vaswani et~al\mbox{.}}{2017}]%
        {attention_vaswani}
\bibfield{author}{\bibinfo{person}{Ashish Vaswani}, \bibinfo{person}{Noam
  Shazeer}, \bibinfo{person}{Niki Parmar}, \bibinfo{person}{Jakob Uszkoreit},
  \bibinfo{person}{Llion Jones}, \bibinfo{person}{Aidan~N. Gomez},
  \bibinfo{person}{undefinedukasz Kaiser}, {and} \bibinfo{person}{Illia
  Polosukhin}.} \bibinfo{year}{2017}\natexlab{}.
\newblock \showarticletitle{Attention is All You Need}. In
  \bibinfo{booktitle}{\emph{Proceedings of the 31st International Conference on
  Neural Information Processing Systems}} (Long Beach, California, USA)
  \emph{(\bibinfo{series}{NIPS'17})}. \bibinfo{publisher}{Curran Associates
  Inc.}, \bibinfo{address}{Red Hook, NY, USA}, \bibinfo{pages}{6000–6010}.
\newblock
\showISBNx{9781510860964}


\bibitem[\protect\citeauthoryear{Vattam, Wiltgen, Helms, Goel, and Yen}{Vattam
  et~al\mbox{.}}{2011}]%
        {vattam_dane:_2011}
\bibfield{author}{\bibinfo{person}{Swaroop Vattam}, \bibinfo{person}{Bryan
  Wiltgen}, \bibinfo{person}{Michael Helms}, \bibinfo{person}{Ashok~K Goel},
  {and} \bibinfo{person}{Jeannette Yen}.} \bibinfo{year}{2011}\natexlab{}.
\newblock \showarticletitle{DANE: fostering creativity in and through
  biologically inspired design}.
\newblock In \bibinfo{booktitle}{\emph{Design Creativity 2010}}.
  \bibinfo{publisher}{Springer}, \bibinfo{pages}{115--122}.
\newblock


\bibitem[\protect\citeauthoryear{Veloso and Carbonell}{Veloso and
  Carbonell}{1993}]%
        {veloso1993derivational}
\bibfield{author}{\bibinfo{person}{Manuela~M Veloso} {and}
  \bibinfo{person}{Jaime~G Carbonell}.} \bibinfo{year}{1993}\natexlab{}.
\newblock \showarticletitle{Derivational analogy in PRODIGY: Automating case
  acquisition, storage, and utilization}.
\newblock In \bibinfo{booktitle}{\emph{Case-Based Learning}}.
  \bibinfo{publisher}{Springer}, \bibinfo{pages}{55--84}.
\newblock


\bibitem[\protect\citeauthoryear{Waltz, Lau, Grewal, and Holyoak}{Waltz
  et~al\mbox{.}}{2000}]%
        {waltz2000role}
\bibfield{author}{\bibinfo{person}{James~A Waltz}, \bibinfo{person}{Albert
  Lau}, \bibinfo{person}{Sara~K Grewal}, {and} \bibinfo{person}{Keith~J
  Holyoak}.} \bibinfo{year}{2000}\natexlab{}.
\newblock \showarticletitle{The role of working memory in analogical mapping}.
\newblock \bibinfo{journal}{\emph{Memory \& Cognition}} \bibinfo{volume}{28},
  \bibinfo{number}{7} (\bibinfo{year}{2000}), \bibinfo{pages}{1205--1212}.
\newblock


\bibitem[\protect\citeauthoryear{Wason and Evans}{Wason and Evans}{1974}]%
        {wason1974dual}
\bibfield{author}{\bibinfo{person}{Peter~C Wason} {and}
  \bibinfo{person}{J~St~BT Evans}.} \bibinfo{year}{1974}\natexlab{}.
\newblock \showarticletitle{Dual processes in reasoning?}
\newblock \bibinfo{journal}{\emph{Cognition}} \bibinfo{volume}{3},
  \bibinfo{number}{2} (\bibinfo{year}{1974}), \bibinfo{pages}{141--154}.
\newblock


\bibitem[\protect\citeauthoryear{White, Bennett, and Dumais}{White
  et~al\mbox{.}}{2010}]%
        {white2010predicting}
\bibfield{author}{\bibinfo{person}{Ryen~W White}, \bibinfo{person}{Paul~N
  Bennett}, {and} \bibinfo{person}{Susan~T Dumais}.}
  \bibinfo{year}{2010}\natexlab{}.
\newblock \showarticletitle{Predicting short-term interests using
  activity-based search context}. In \bibinfo{booktitle}{\emph{Proceedings of
  the 19th ACM international conference on Information and knowledge
  management}}. \bibinfo{pages}{1009--1018}.
\newblock


\bibitem[\protect\citeauthoryear{White and Roth}{White and Roth}{2009}]%
        {white2009exploratory}
\bibfield{author}{\bibinfo{person}{Ryen~W White} {and} \bibinfo{person}{Resa~A
  Roth}.} \bibinfo{year}{2009}\natexlab{}.
\newblock \showarticletitle{Exploratory search: Beyond the query-response
  paradigm}.
\newblock \bibinfo{journal}{\emph{Synthesis lectures on information concepts,
  retrieval, and services}} \bibinfo{volume}{1}, \bibinfo{number}{1}
  (\bibinfo{year}{2009}), \bibinfo{pages}{1--98}.
\newblock


\bibitem[\protect\citeauthoryear{Wilcox and Keselman}{Wilcox and
  Keselman}{2003}]%
        {winsorizing}
\bibfield{author}{\bibinfo{person}{Rand~R Wilcox} {and} \bibinfo{person}{HJ
  Keselman}.} \bibinfo{year}{2003}\natexlab{}.
\newblock \showarticletitle{Modern robust data analysis methods: measures of
  central tendency.}
\newblock \bibinfo{journal}{\emph{Psychological methods}} \bibinfo{volume}{8},
  \bibinfo{number}{3} (\bibinfo{year}{2003}), \bibinfo{pages}{254}.
\newblock


\bibitem[\protect\citeauthoryear{Yovanovich, Culham, and Teertstra}{Yovanovich
  et~al\mbox{.}}{2004}]%
        {interface_resistance}
\bibfield{author}{\bibinfo{person}{Michael Yovanovich},
  \bibinfo{person}{Richard Culham}, {and} \bibinfo{person}{Peter Teertstra}.}
  \bibinfo{year}{2004}\natexlab{}.
\newblock \bibinfo{booktitle}{\emph{Calculating interface resistance}}.
\newblock
\urldef\tempurl%
\url{http://www.thermalengineer.com/library/calculating_interface_resistance.htm}
\showURL{%
\tempurl}


\bibitem[\protect\citeauthoryear{Zhang, Qu, Giles, and Song}{Zhang
  et~al\mbox{.}}{2008}]%
        {CiteSense}
\bibfield{author}{\bibinfo{person}{Xiaolong Zhang}, \bibinfo{person}{Yan Qu},
  \bibinfo{person}{C.~Lee Giles}, {and} \bibinfo{person}{Piyou Song}.}
  \bibinfo{year}{2008}\natexlab{}.
\newblock \showarticletitle{CiteSense: Supporting Sensemaking of Research
  Literature}. In \bibinfo{booktitle}{\emph{Proceedings of the SIGCHI
  Conference on Human Factors in Computing Systems}} (Florence, Italy)
  \emph{(\bibinfo{series}{CHI ’08})}. \bibinfo{publisher}{Association for
  Computing Machinery}, \bibinfo{address}{New York, NY, USA},
  \bibinfo{pages}{677–680}.
\newblock
\showISBNx{9781605580111}
\urldef\tempurl%
\url{https://doi.org/10.1145/1357054.1357161}
\showDOI{\tempurl}


\bibitem[\protect\citeauthoryear{Zirbel, Wilson, Magleby, and Howell}{Zirbel
  et~al\mbox{.}}{2013}]%
        {zirbel2013origami}
\bibfield{author}{\bibinfo{person}{Shannon~A Zirbel}, \bibinfo{person}{Mary~E
  Wilson}, \bibinfo{person}{Spencer~P Magleby}, {and} \bibinfo{person}{Larry~L
  Howell}.} \bibinfo{year}{2013}\natexlab{}.
\newblock \showarticletitle{An origami-inspired self-deployable array}. In
  \bibinfo{booktitle}{\emph{ASME 2013 Conference on Smart Materials, Adaptive
  Structures and Intelligent Systems}}. American Society of Mechanical
  Engineers Digital Collection.
\newblock


\end{thebibliography}

\section*{Appendix A. Reproducibility} \label{appendix}

\xhdr{Training and validation datasets} \label{appendix:dataset} The original annotation dataset from~\cite{chan2018solvent} also includes Background and Findings annotations which we exclude due to their relatively high confusion rates among the annotators with the Purpose and Mechanism classes and to balance the number of available training examples per annotation class. 

\xhdr{Model parameter selection} \label{appendix:model_parameters} We experimented with changing the model capacity relative to the signal present in the training dataset by tuning the number of hidden layers and the nodes used in each model architecture. For Model 1 we found a hidden layer of 100 nodes was sufficient. We optimized this model using the cross-entropy loss and the Adam optimizer~\cite{kingma2014adam} with a 0.0001 learning rate. For Model 2, we found three hidden layers with 256 nodes led to an improved accuracy on the validation set. We trained this model with an L2 regularizer ($\alpha = 0.01$), dropouts with the rate of 0.3, and the Adam optimizer with a 0.001 learning rate.

\xhdr{Span-based model architecture} \label{appendix:spanrel_implementation}
We adapt SpanRel~\cite{spanrel} as architecture for the span-based Model 2. SpanRel combines the boundary representation (BiLSTM) and the content representation with a self-attention mechanism for finding the core words. More specifically, given a sentence $\bm{x}$ $=$ $[e_1,$ $e_2,$ $\cdots,$ $e_n]$, of $n$ token embeddings, a span $s_i = [\omega_{s_i}, \omega_{s_i + 1}, \cdots, \omega_{f_i}]$ is a concatenation of the \textit{content representation} ${\bm{z_i}}^c$ (weighted average across all token embeddings in the span; SelfAttn) and the \textit{boundary representation} ${\bm{z_i}}^b$ of the start ($s_i$) and end positions ($f_i$) of the span:
\begin{align*}
    \bm{u_1}, \bm{u_2}, \cdots, \bm{u_n} &= \text{BiLSTM}(\bm{e_1}, \bm{e_2}, \cdots, \bm{e_n}) \\
    \bm{z_i}^c &= \text{SelfAttn}(\bm{e_{s_i}}, \bm{e_{s_i + 1}}, \cdots, \bm{e_{f_i}}) \\
    \bm{z_i}^b &= [\bm{u_{s_i}}; \bm{u_{f_i}}] \\
    \bm{z_i} &= [\bm{z_i}^c; \bm{z_i}^b]
\end{align*}
We use the contextualized ELMo 5.5B embeddings\footnote{\url{https://allennlp.org/elmo}} for token representation, following the near state-of-the-art performance reported on the named entity recognition task on the Wet Lab Protocol dataset in~\cite{spanrel}. We refer to~\cite{spanrel,lee-etal-2017-end} for further details.

\xhdr{Other parameters} We use GloVe vectors for input feature representation for Model 1 with 300 dimensions, consistent with the prior work~\cite{pennington_glove,landauer1997solution,bojanowski2017enriching_subword_info}. For Model 2, we use the contextualized ELMo 5.5B embeddings as described above which have pre-determined \fnum{1024} dimensions. We use Universal Sentence Encoder (USE)~\cite{universal_sentence_encoder} for encoding purposes. 
A USE embedding vector has pre-determined 512 dimensions.

\end{document}